\documentclass[bookmarks=true]{aastex6}

\usepackage{amssymb,amsmath,eurosym}
\usepackage{mathrsfs}
\usepackage{enumerate}

\usepackage{natbib}
\usepackage{cancel}

\newcommand{\figref}[1]{Figure \ref{#1}}
\newcommand{\vect}[1]{\mathbf{#1}}

\newcommand{\fnc}[1]{\textsf{#1}} 
\newcommand{\unit}[1]{\ensuremath{\, \mathrm{#1}}}
\newcommand{\del}{\nabla}
\newcommand{\bigo}{\ensuremath{\mathcal{O}}}
\providecommand{\abs}[1]{\lvert#1\rvert}
\DeclareMathOperator{\atan}{atan}
\DeclareMathOperator{\erf}{erf}
\newcommand{\change}[1]{{#1}}

\begin{document}
\title{Magnetoacoustic Waves in a Stratified Atmosphere with a Magnetic Null Point}
\author{Lucas A. Tarr\altaffilmark{1,2}, Mark Linton, and James Leake}
\affil{U.S. Naval Research Laboratory 4555 Overlook Ave SW, Washington, DC 20375, USA}
\altaffiltext{1}{National Research Council Research Associate}
\altaffiltext{2}{\url{lucas.tarr.ctr@nrl.navy.mil}}

\begin{abstract}
  We perform nonlinear MHD simulations to study the propagation of magnetoacoustic waves from the photosphere to the low corona.  We focus on a 2D system with a gravitationally stratified atmosphere and three photospheric concentrations of magnetic flux that produce a magnetic null point with a magnetic dome topology.  We find that a single wavepacket introduced at the lower boundary splits into multiple secondary wavepackets.  A portion of the packet refracts towards the null due to the varying Alfv\'en speed.  Waves incident on the equipartition contour surrounding the null, where the sound and Alfv\'en speeds coincide, partially transmit, reflect, and mode convert between branches of the local dispersion relation.  Approximately $15.5\%$ of the wavepacket's initial energy ($E_{input}$) converges on the null, mostly as a fast magnetoacoustic wave.  Conversion is very efficient: $70\%$ of the energy incident on the null is converted to slow modes propagating away from the null, $7\%$ leaves as a fast wave, and the remaining $23\%$ (0.036$E_{input}$) is locally dissipated.  The acoustic energy leaving the null is strongly concentrated along field lines near each of the null's four separatrices.  The portion of the wavepacket that refracts towards the null, and the amount of current accumulation, depends on the vertical and horizontal wavenumbers and the centroid position of the wavepacket as it crosses the photosphere.  Regions that refract towards or away from the null do not simply coincide with regions of open versus closed magnetic field or regions of particular field orientation.  We also model wavepacket propagation using a WKB method and find that it agrees qualitatively, though not quantitatively, with the results of the numerical simulation.
\end{abstract}

\section{Introduction}

 \change{In this paper we connect the study of magnetohydrodynamic (MHD) waves between two mostly separate fields of inquiry: propagation through stellar atmospheres and propagation near magnetic null points.  MHD waves serve as both a source of observed phenomena \citep[e.g., periodic shocks, acoustic halos:][]{Vecchio:2009, Rajaguru:2013} and an effect of other phenomena \citep[e.g., magnetic reconnection, convective buffeting:][]{Longcope:2012, Stangalini:2013a}.  Understanding how waves propagate through the highly inhomogeneous solar atmosphere is therefore essential both for interpreting solar observations and for predicting the consequences of processes we wish to study on the Sun.  A detailed study of wave propagation can also be used as a diagnostic for determining the properties of the plasma, for instance through coronal seismology \citep[see review by][\S5.2]{Jess:2015} or modeling that reproduces temporal and spectral characteristics of spectral lines \citep{Vigeesh:2011}.}

In a homogeneous plasma, MHD waves come in three basic types, Alfv\'en, slow, and fast \citep{Cowling:1957}.  These correspond to the six roots, one positive and one negative for each wave type, of a single wave equation derived by combining the linearized MHD equations \citep{Ferraro:1958, Stix:1992, Goedbloed:2004}.  The Alfv\'en wave is always incompressible; the fast and slow modes may be compressible or incompressible, depending on the situation, and are generically termed magnetoacoustic waves.  \change{In an inhomogeneous plasma, three types of waves still typically exist in some modified forms which reduce to the three basic types in appropriate limits.  The modes may also be degenerate in certain locations throughout the plasma \citep{Zhugzhda:1984a} and may therefore exchange wave energy from one type to another in a process termed mode conversion \citep{Cally:2001}.}

\change{Gravity creates a natural inhomogeneity by stratifying the density in the direction of the gravitation force.  Decades of studies of MHD waves in atmospheres have used parameters with gradients only in that direction \citep{Ferraro:1958,Osterbrock:1961,McLellan:1968, Nye:1976, Zhugzhda:1984a, Hasan:1992, Cally:2001}.  More recently, numerical solutions to the MHD equations have made it much easier to study systems containing gradients in two or three dimensions, usually through a spatially varying magnetic field \citep[][to name a few]{Rosenthal:2002, Bogdan:2003, DeMoortel:2004,Carlsson:2006,Cally:2008,Fedun:2011a, Felipe:2012, Nutto:2012,Russell:2013,Santamaria:2015}.  However, most of this work has focused on regions where gradients in the magnetic field are small compared to other length scales, for example, in simplified sunspot models \citep{Felipe:2012}.  Rapidly varying (in space) magnetic fields have received less attention in the context of stratified atmospheres.}

\change{In contrast, near a magnetic null point (where the magnetic field $\vect{B}=\vect{0}$) the field necessarily has strong gradients, and this greatly affects the behavior of waves propagating nearby.  Because the phase speed of MHD waves is proportional to the magnetic field strength and traveling waves refract towards regions of low phase speed, MHD waves will tend to be guided towards nulls as they propagate.  The plasma properties are such that the MHD waves become degenerate near nulls, and this allows for mode conversion between the wave types at these locations.  Null points are expected to be rather common in the low solar atmosphere \citep[roughly one per supergranual cell, by multiple estimates:][]{Close:2004, Regnier:2008, Longcope:2009, Freed:2015}, so it is vital to understand this fundamental plasma process in the solar context.}

\change{MHD wave propagation near nulls has been extensively studied, especially over the past decade in a series of papers by McLaughlin and coauthors \citep[][where the last two are recent review articles]{Craig:1991, Hassam:1992, Craig:1993, McLaughlin:2004, McLaughlin:2005, McLaughlin:2006a, McLaughlin:2006b,Longcope:2007, McLaughlin:2008, McLaughlin:2009, Afanasyev:2012, Longcope:2012, McLaughlin:2011, Pontin:2012}.  Waves exhibit complex behavior near nulls, and many approximations may be used to reduce that complexity: a cold plasma limit ($\beta=0$, where $\beta = 2P\mu_0/B^2$ is the ratio of plasma $(P)$ to magnetic pressure), uniform density and temperature backgrounds, linear and/or symmetric nulls, and solving for the linearized instead of full MHD equations, are all common.  The references above have each used combinations of these approximations, and this has greatly informed our understanding of how waves travel through a strongly inhomogeneous plasma.  With that background, we are in a position to analyze the effects of gravitational stratification and of a magnetic null topology in combination in a single simulation that will adhere much more closely to the inhomogeneous environment spanning from the photosphere to low corona on the Sun.}

\begin{figure}
  \includegraphics[width=0.49\textwidth]{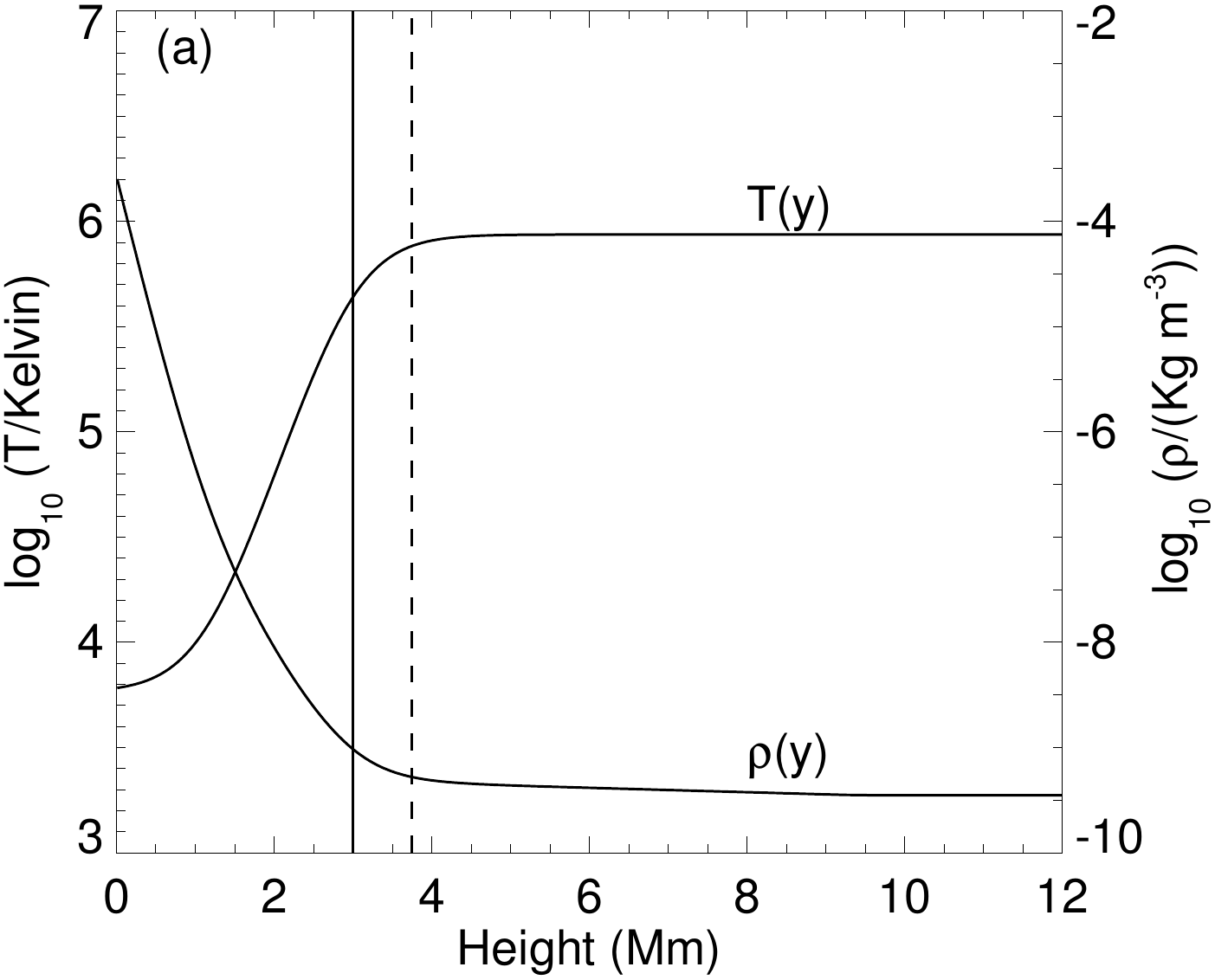}
  \includegraphics[width=0.49\textwidth]{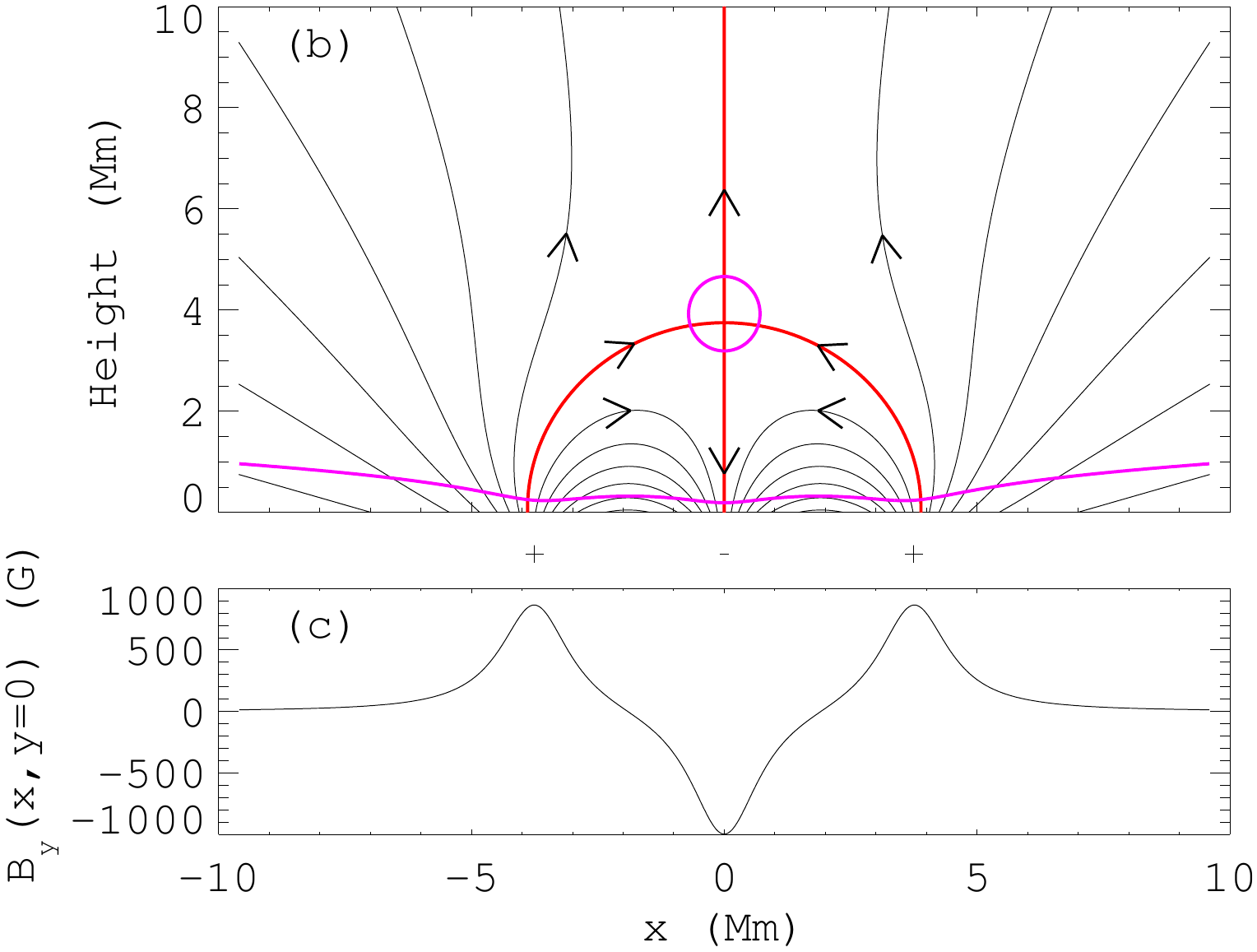}
  \caption{\label{fig:initial-atmo} (a) Temperature and density profiles for the initial stratified atmosphere.  Dashed line designates the height of the null point, solid line the height of the transition region.  (b) Top: initial magnetic field configuration, with a selection of field lines (contours of $A_z$).  Arrows indicate magnetic direction, and the thick red lines are separatrix field lines connected to the null point at $\vect{r}=(0,3.75)\unit{Mm}$.  Equipartition contours, along which $c_s=v_A$, are shown in magenta.  The minus and plus symbols indicate the location of the negative and positive flux sources.  (b) Bottom: vertical field strength along $y=0.$}
\end{figure}

\change{The stratified atmosphere with magnetic field that we will study is described in detail in \S\ref{sec:strat} and \S\ref{sec:Bini} and illustrated in \figref{fig:initial-atmo}.  The atmosphere contains a single null point.  Important for now is the fact that the sound speed and Alfv\'en speed ($c_s$ and $v_A$, defined in \S\ref{sec:initial}) both vary throughout the domain.  The ratio of the two is a key parameter in both stellar atmosphere and finite $\beta$ null point investigations.  As explained in \S\ref{sec:ray}, at locations where $c_s/v_A\approx 1$ there is near equipartition between pressure and magnetic forces, and it is possible for mode conversion to take place (note that $c_s/v_A=1$ for $\beta = 2c_s^2/\gamma v_A^2=1.2$).}

\change{There are two equipartition contours in our simulation domain, shown in magenta in \figref{fig:initial-atmo}(b): the first running horizontally near the lower boundary, the second roughly circular surrounding the null.  The region around the first of these contours has been studied extensively in the case of an isothermally stratified atmosphere and uniform, arbitrarily directed magnetic field \citep[][see also references in the last of these]{Zhugzhda:1984a, Hassam:1992, Cally:2001, Cally:2007, Hansen:2009}.  In that case, the MHD wave equation can be solved analytically.  For a concise presentation of the analytic theory, see \citet{Hansen:2009} and references therein.  The angle of attack between the wavevector $\vect{k}$ and the inclined magnetic field direction determines the coefficients of transmission, reflection, and mode conversion at this layer.  According to Equation (26) of \citet{Schunker:2006}, the transmission coefficient $T$ is
\begin{equation}\label{eq:transmission}
  T = \exp(-\pi k H_e\sin^2\theta),
\end{equation}
where $k=\abs{\vect{k}}$ is the total wavenumber, $\theta$ is the attack angle between the wavevector and the (uniform) magnetic field direction, and $H_e$ is the scale height of the equipartition region, defined as the derivative along a path $\tau$ of the ratio $c_s/v_A$, evaluated at the equipartition point: $H_e^{-1} = \partial_\tau(c_s^2/v_A^2)\rvert_{c_s=v_A}$.  Transmission peaks for $\theta=0$ and decreases for increasing attack angles.}

\change{Equation \eqref{eq:transmission} shows how observations of atmospheric oscillations should have a strong dependence on the local magnetic field orientation.  Despite the simplifications, the analytic theory \citep{Hansen:2009} captures the essential wave behavior, as verified by numerical simulations and recent observations of acoustic shocks, active region halos, sunspot oscillations, small fluxtube oscillations, running penumbral waves, and many other phenomena \citep{Vecchio:2009, Felipe:2010a,Fedun:2011b, Stangalini:2011, Rajaguru:2013, Jess:2013, Kontogiannis:2014,Lohner-bottcher:2015, Khomenko:2013}.  \citet{Khomenko:2009a} in particular was able to explain the enhanced power around sunspots known as acoustic halos in terms of mode conversion at the $c_s=v_A$ layer (their model contains only a single such layer).  Recent numerical simulations show how the dependence on formation height of spectral lines, peak frequency of oscillatory power at each location, the surface of equipartition, and the magnetic field direction all combine to fit the conclusions drawn from the analytic models \citep{Rijs:2015, Przybylski:2015}.}

\change{Although the magnetic field in our model varies all along the lower equipartition curve of \figref{fig:initial-atmo}(b), and substantially more than in the slowly varying sunspot models of, e.g., \citet{Felipe:2012} and \citet{Rijs:2015}, the major difference between it and these other models is the presence of the null.  Surrounding the null point in \figref{fig:initial-atmo}(b) is the second equipartition curve along which $c_s=v_A$.  Wave dynamics at nulls have been studied before, but are not well studied in the context of the low solar atmosphere, where finite plasma $\beta$, partial ionization, and stratification are all important.}

\change{Linear nulls with $\beta=0$ have received the most attention \citep{Craig:1991, Hassam:1992, McLaughlin:2004, McLaughlin:2006a, Longcope:2007}.  For a linear null, $\abs{B}\propto r$, the radial distance from the null.  As the pressure forces are set to zero the equipartition curve surrounding the null shrinks to zero, and the slow magnetoacoustic wave is no longer a solution to the MHD wave equation; mode conversion is thus precluded for $\beta=0$.  For linearized MHD, the fast mode velocity decreases to zero at the null, so that the fast mode cannot pass through the null and instead becomes trapped.  This can be visualized using a Wentzel--Kramers--Brillouin \citep[WKB][]{Weinberg:1962} method to trace rays for the MHD waves, in analogy to geometric optics (see \S\ref{sec:ray} and Appendix \ref{sec:wkb}).  The radial increase in the phase speed (and hence index of refraction) causes incident rays to spiral inwards \citep{McLaughlin:2004, Longcope:2012, Afanasyev:2012}.  The wavelength of the incoming waveform decreases, causing an exponential increase in the current density at the null in the limit of no resistivity.  These results generalize to linear nulls in three dimension \citep{McLaughlin:2008}.}

\change{Departures from (i) zero resistivity, (ii) linearized MHD, (iii) a linear null, or (iv) zero $\beta$ will substantially alter the behavior described above.  (i) Finite resistivity causes partial reflection of an incident wave from the null point, even in a $\beta=0$ plasma \citep{Craig:1991, Longcope:2007, Longcope:2012}: wave energy is neither perfectly Ohmically dissipated nor perfectly reflected at the null.  (ii) Nonlinear MHD effects create shocks, which allows the fast wave to pass through the null \citep{McLaughlin:2009, Afanasyev:2012}.  (iii) For more realistic magnetic fields, at some radius the linear null approximation is no longer valid.  Examples are the quadrupolar fields studied by \citet{McLaughlin:2006a} and \citet{Longcope:2012}, the double null system of \citet{McLaughlin:2005}, or any field extrapolated from a photospheric magnetogram \citep{Longcope:2009}.  In these cases, inflection points in the spatial variation of the phase speed will cause a wave front to split, so that a portion of the wave refracts towards the null while other portions refract away.  Refraction is thus very important for the propagation of fast waves because they travel nearly isotropically relative to the magnetic field, and this is true even for $\beta=0$.}

\change{(iv) A finite $\beta$ produces more dramatic changes than relaxing assumputions (i), (ii), and (iii) by reintroducing the slow mode and allowing a coupling between the fast and slow waves \citep{McLaughlin:2006b, McLaughlin:2009, McLaughlin:2012a}.  The phase speeds of the fast and slow waves are now everywhere nonzero and it is possible for linear waves (or rays in the WKB approximation) to pass through a null.  The wave speed still varies substantially near the null and causes a focusing effect.  This can lead to shock formation and the collapse of the null into a current sheet, both of which dissipate to heat the plasma.  The general conclusion is that including pressure forces does not simply allow wave energy to pass smoothly through a null \citep{McLaughlin:2006b, McLaughlin:2009, Gruszecki:2011, Afanasyev:2012}.}

\change{The finite $\beta$ wave--null interaction studies referenced above have all used ad hoc initial conditions, such as an initially circular pulse that fully surrounds the null.  We will study the more realistic situation where a wave is introduced by boundary driving, mimicking a convective process, and is allowed to propagate self--consistently into the null point region.  We will focus our study on the properties of the waves in the vicinity of the null point, with particular emphasis on how they are affected by the equipartition region surrounding the null and the nontrivial topology of the null.  We will answer several questions: How much of the initial energy of the wave packet is incident upon the null?  What is the efficiency of mode conversion around the null?  How much of the initial energy makes it to the coronal portion of the domain?  Answering these questions will serve two purposes.  First, to advance the study of MHD wave behavior around magnetic null points, which is a fundamental process in plasma physics; and second, to understand how an inhomogeneous magnetic field affects wave propagation in stellar atmospheres.}

We use a combination of ray tracing, for the linearized MHD equations, and numerical simulations of the full, nonlinear equations to study the propagation of a wavepacket through the domain.  We find that mode conversion between branches of the dispersion relation plays a dominant role in the resulting dynamics.  We identify the locations of mode conversion, and quantify both the amount of conversion surrounding the null point and the amount of dissipation near the null point.  The remainder of the paper is outlined as follows: in \S\ref{sec:initial} we describe the numerical code, LARE2D \citep{Arber:2001}, and discuss the initial atmosphere and magnetic equilibrium.    The same background atmosphere is used for both the ray tracing and numerical analyses.  \S\ref{sec:ray} describes the ray tracing procedure and important energy densities associated with the different types of waves.  \change{We briefly compare our ray tracing results to those of others, particularly McLaughlin and coauthors.}  \S\ref{sec:driver} describes the time--dependent boundary condition and \S\ref{sec:results} presents the results of the resistive MHD simulation.  We show how the wave pulse propagates through the atmosphere and quantify the amount of mode conversion between branches of the dispersion relation at topologically important locations.   We find that mode conversion strongly influences the propagation of wave energy density through the system.  We determine the amount of energy that escapes into the corona and note the accumulation and dissipation of current density at the null and along the separatrices.  In \S\ref{sec:driver-distribution} we return to the ray tracing analysis to model the wavepacket's propagation.  We estimate the amount of energy expected to arrive at the null and compare that to the numerical simulation results.  We discuss our results in a broader context in \S\ref{sec:discussion}, and finally summarize and conclude in \S\ref{sec:conclusion}.

\section{Numerical setup and initial conditions}\label{sec:initial}
\subsection{Background stratification}\label{sec:strat}
 We solve the resistive magnetohydrodynamic equations in 2.5 dimensions in Cartesian coordinates using the LARE2D code \citep{Arber:2001}.  The horizontal and vertical directions are $x$ and $y$\change{, and the out--of--plane direction is $z$}.  We notate spatial points by $\vect{x}=(x,y)$ and velocities by $\vect{v}=(v_x,v_y,v_z)$.  Individual components may also be referenced in the standard way by a subscript $i$ or $j$: $i= 1,2,3; j=1,2,3$.  In order, the equations describe mass, momentum, and energy conservation, and the magnetic induction:
  \begin{gather}
    \label{eq:continuity}\frac{D\rho}{Dt} = -\rho \del\cdot\vect{v}, \\
    \label{eq:momentum}\frac{D\vect{v}}{Dt} = -\frac{1}{\rho}\del P + \frac{1}{\rho}\vect{j}\times\vect{B} + \vect{g} + \frac{1}{\rho}\del\cdot S, \\
    \label{eq:energy}\frac{D\epsilon}{Dt} = -\frac{P}{\rho}\del\cdot\vect{v} + \zeta_{ij}S_{ij} + \eta j^2, \\
    \label{eq:induction}\frac{D\vect{B}}{Dt} = (\vect{B}\cdot\del)\vect{v} - \vect{B}(\del\cdot\vect{v})-\del\times(\eta\vect{j}).
  \end{gather}
  $D/Dt=\partial/\partial t + \vect{v}\cdot\del$ is the advective derivative.  \change{Owing to the translational invariance in $z$, all derivatives in the out--of--plane direction are zero.}  Our primitive variables are mass density, specific internal energy, plasma velocity, and magnetic field, $\rho, \epsilon, \vect{v}, \vect{B}$, respectively.  The current density is defined through $\vect{j}=\del\times\vect{B}/\mu_0$ where $\mu_0$ is the permeability of free space.  The stress tensor has components $S_{ij} = \nu(\zeta_{ij}-(1/3)\delta_{ij}\del\cdot\vect{v})$ with $\zeta_{ij} = (1/2)(\partial v_i/\partial x_j + \partial v_j/\partial x_i)$, and we use a uniform viscosity $\nu = 2.8\unit{kg}\unit{m}^{-1}\unit{s}^{-1}$.  We set the resistivity to $\eta = 116 \Omega\unit{m}.$  The viscosity and resistivity are such that the Reynolds number $R=10^5$ and the Lundquist number $S=10$, which we discuss in more depth below.

Temperature, internal energy, and pressure are related through the ideal gas law
  \begin{equation}
    \label{eq:ideal}P = \frac{\rho k_B T}{\mu_m} \quad \text{and} \quad \epsilon = \frac{k_B T}{\mu_m(\gamma-1)},
  \end{equation}
  with $k_B$ Boltzmann's constant and $\gamma=5/3$ the ratio of specific heats.  Gravitational acceleration is set to the solar value of $g=274\unit{m}\unit{s}^{-2}$, with $\vect{g} = -g\hat{y}$.  We use the fully ionized limit of MHD, but set the reduced mass to the average ion mass, $\mu_m=m_i = 1.25m_p$.  This choice more correctly reproduces coronal densities \citep[see discussion in][]{Leake:2013}.  The factor $1.25$ models the effect of heavy ions.

  We normalize the governing equations \eqref{eq:continuity}-\eqref{eq:induction} by writing each variable as a constant multiplying a normalized coordinate: $x = L_Nx^*,\ \rho=\rho_N\rho^*, $ and so on.  We set $L_N = 1.50\times10^{5}\unit{m},\ \rho_N = 3.03\times 10^{-4}\unit{kg}\unit{m}^{-3},$ and $B_N = 0.12\unit{T}$.  $\rho_N$ is the photospheric density, and $L_N$ approximately the scale height at the photosphere.  All other variables may be defined through these three.  The velocity normalization, for instance, is $V_N = B_N/\sqrt{\mu_0\rho_N} = 6177\unit{m}\unit{s}^{-1},$ the photospheric Alfv\'en speed for unit magnetic field.  Time normalization is $t_N = L_N/V_N = 24.28\unit{s}$.  The viscosity is normalized to $\nu_N=\rho_N V_N L_N$.  We define the Reynolds number $R=\rho_NV_NL_N/\nu$ using the photospheric Alfv\'en speed and scale height.  Using this, we set the normalized viscosity to $\nu^* = \nu/\nu_N=\frac{1}{R}=10^{-5},$ resulting in the value $\nu$ quoted above.  From here on we take all variables to be normalized unless explicitly stated, and suppress the star notation.

Our domain extends from the photosphere at the lower boundary up to the low corona.  The initial thermodynamic equilibrium is set to be invariant in $x$, so that $T(x,y,t=0) = T_0(y)$ (a subscript $0$ will be used to refer to initial state for all variables).  We model a low temperature photosphere, steep transition region, and isothermal corona as a hyperbolic tangent function:
  \begin{equation}
    \label{eq:tanh}
    T_0(y>0) = T_{ph} + \frac{T_{cor}-T_{ph}}{2}\Bigl(\fnc{tanh}\Bigl( \frac{y-y_{tr}}{w_{tr}}\Bigr) +1 \Bigr).
  \end{equation}
  The parameters are $T_0(0) = T_{ph} = 5778\unit{K}$, $T_{cor} = 150 T_{ph}$, $y_{tr} = 3.0\unit{Mm}$, and $w_{tr} = .75\unit{Mm}$.  The initial density profile is determined by numerical integration of the hydrostatic equation, $\del P_0 = -\rho_0 g \hat{\vect{y}}$, with $\rho_0(0) = 1.0\rho_N$ as a boundary condition.  \figref{fig:initial-atmo}(a) shows the initial hydrostatic equilibrium for temperature (left axis) and density (right axis) on a log scale.

We use a uniform staggered grid of $1024\times1024$ cells, with $\rho$ and $\epsilon$ defined at cell centers, $\vect{v}$ at cell vertices, and $\vect{B}$ at cell faces \citep[see][for details]{Arber:2001}.  The cell width is $\Delta_x=\Delta_y = \Delta = L_N/8,$ which sets the size of our domain to $x:(-9.6,9.6)\unit{Mm}$ and $y:(0,19.2)\unit{Mm}$.  

Finally, we estimate the magnitude of numerical diffusion caused by the finite difference scheme using the method of \citet{Arber:2007}.  We consider the 1D analog to \eqref{eq:induction}, $\partial_tB + C\partial_x B = 0$, Taylor expand the second order finite difference equation, and isolate the error term, whose coefficient is the effective numerical resistivity, $\eta_{num}$.  We find that $\eta_{num} = C \Delta^2/6L$, where $L$ is a typical length of the dynamic evolution and $C$ the fastest wave speed.  For a worst case scenario corresponding to a shock across three cells, $L=3\Delta$.  If $C\approx 20 V_N$ at locations where shocks form, we find that $\eta_{num}\approx \frac{\Delta}{L_N} = 0.15$ in the normalized units defined below.  We have run multiple simulations varying only the explicit resistivity and checking the solution in regions of strong gradients (there is little discernible effect outside these regions).  We found that explicit resistivity begins to dominate over the numerical resistivity at the expected value $\eta\approx 0.1$, confirming our approximate calculations.  We therefore use this value of resistivity for our simulations.  This ensures that resistive effects are due to the explicit term in equations \eqref{eq:energy} and \eqref{eq:induction}, while simultaneously keeping the resistive effects as small as possible, for the chosen grid resolution.  Resistivity is normalized to $\mu_0 L_N V_N$, so that $S=\mu_0L_Nv_N/\eta$ is the Lundquist number defined using the pressure scale height at the lower boundary.  The results are the values of resistivity and Lundquist number stated above.

  \subsection{Initial magnetic field}\label{sec:Bini}
  To the hydrostatic background we add an initial magnetic field derived from a flux function, $\vect{B} = \del\times\vect{A}=(\del A_z)\times \hat{z}$,
  \begin{equation}
    \label{eq:fluxfunc}
    A_z(x,y)=\sum_{p=-1}^1\frac{\psi_p}{\pi}\atan\frac{y-y_s}{x-x_p},
  \end{equation}
  while $A_x$ and $A_y$ are both zero.  As \figref{fig:initial-atmo}(b) shows, the flux function includes three sources of flux (per ignorable length) $\psi_p$, each located at depth $y_s = -0.825\unit{Mm}$, and having horizontal locations $x_0=0\unit{Mm},\ x_{\pm 1}=\pm3.75\unit{Mm}.$  The inner source $\psi_0$ has the opposite sign of flux of the outer two, $\psi_{\pm1}$.  The resulting magnetic field contains one null point, located at 
\begin{gather}
  y_\times = -y_s + \sqrt{\frac{x_1^2}{\abs{2\psi_1/\psi_0}-1}}.
\end{gather}
We choose $y_\times=3.75\unit{Mm},$ which sets $\abs{\psi_1/\psi_0} \approx 0.836,$ and $\psi_0 = -5 B_N L_N = -9\times 10^{10}\unit{Mx}/\unit{L_{ignorable}}$, creating a maximum vertical field at the lower boundary of $\approx 1\unit{kG}$ for each source.  The field strength along the lower boundary is shown in the bottom panel of \figref{fig:initial-atmo}(b).

The magnetic field generated by \eqref{eq:fluxfunc} is a potential field, whose $z$ component analytically satisfies Laplace's equation $\mu_0j_z(x,y) = (\del\times\vect{B})_z = -\del^2A_z = 0$, and $j_x=j_y=0$ as well.  However, the second order finite difference scheme used by LARE2D results in spurious currents, as may be found by substituting the Taylor series expansion of Eq~\eqref{eq:fluxfunc} into the finite difference scheme.  All odd order derivative terms cancel numerically, the second derivative term correctly reproduces the Laplacian, but even order derivative terms of order $4$ and higher do not cancel in general.  The initial condition is therefore slightly out of force balance.  We allow the system to come to equilibrium during an initial relaxation period for each simulation.  The resulting changes are minimal, though the largest changes are near the X--point.  The change in $B^2(x,y)$ is everywhere \change{$<1.1\%$} and less than $0.1\%$ for more than $99.5\%$ of the domain; the change in internal energy density is everywhere less than $0.1\%$.  For the remainder of this paper, we refer to $t=0$ as the end of this initial relaxation, and only discuss dynamics for $t>0$.  Subscripted variables $\rho_0, \vect{B}_0,\ldots$ therefore refer to this background initial condition.

\subsection{Boundary conditions}
\change{LARE2D requires two ghost cells surrounding the domain to implement the boundary conditions.  In the convection zone, the increasing sound speed will eventually cause waves to refract upwards at a depth that depends on the horizontal wavelengths of the waves.  We therefore use a reflective lower boundary, and add to it a time dependent driver, as described in \S\ref{sec:driver}.  The phase of the waves reflected from our lower boundary will not be accurate, but this does not affect our analysis in \S\ref{sec:results}, which focuses on the region near the null before any reflections reach it.}

\change{The side boundaries are line tied and the top boundary is zero gradient.  Linear damping regions are implemented for both, and these reduce the velocities in a cell based on distance to the boundary.  This has the effect of removing kinetic energy from the system.  The amount of reduction starts at $0$ at $y=102.4L_N\ (15.36\unit{Mm})$ or $\abs{x}=51.2L_N\ (7.68\unit{Mm})$ and increases towards the appropriate boundary (left, right, or top).  If $L_{damp}$ is the size of the damping region $(25.6L_N$ in $y$, $12.8\L_N$ in $x$) and $\Delta X$ is the distance from a cell within the damping region to the the start of the region, then at each time step the velocities in that cell are reduced by a factor $1 + 0.3\delta_t\frac{\Delta X}{L_{damp}}$, where $\delta_t$ is the numerical time step.  Our analysis will focus on dynamics near the null point and magnetic dome, and testing has shown that the damping regions prevent the majority of reflections, leading to a small contribution to the dynamics at locations of interest.}

\subsection{Background sound and Alfv\'en speed structure}

\begin{figure}[ht]
  \includegraphics[width=0.5\textwidth]{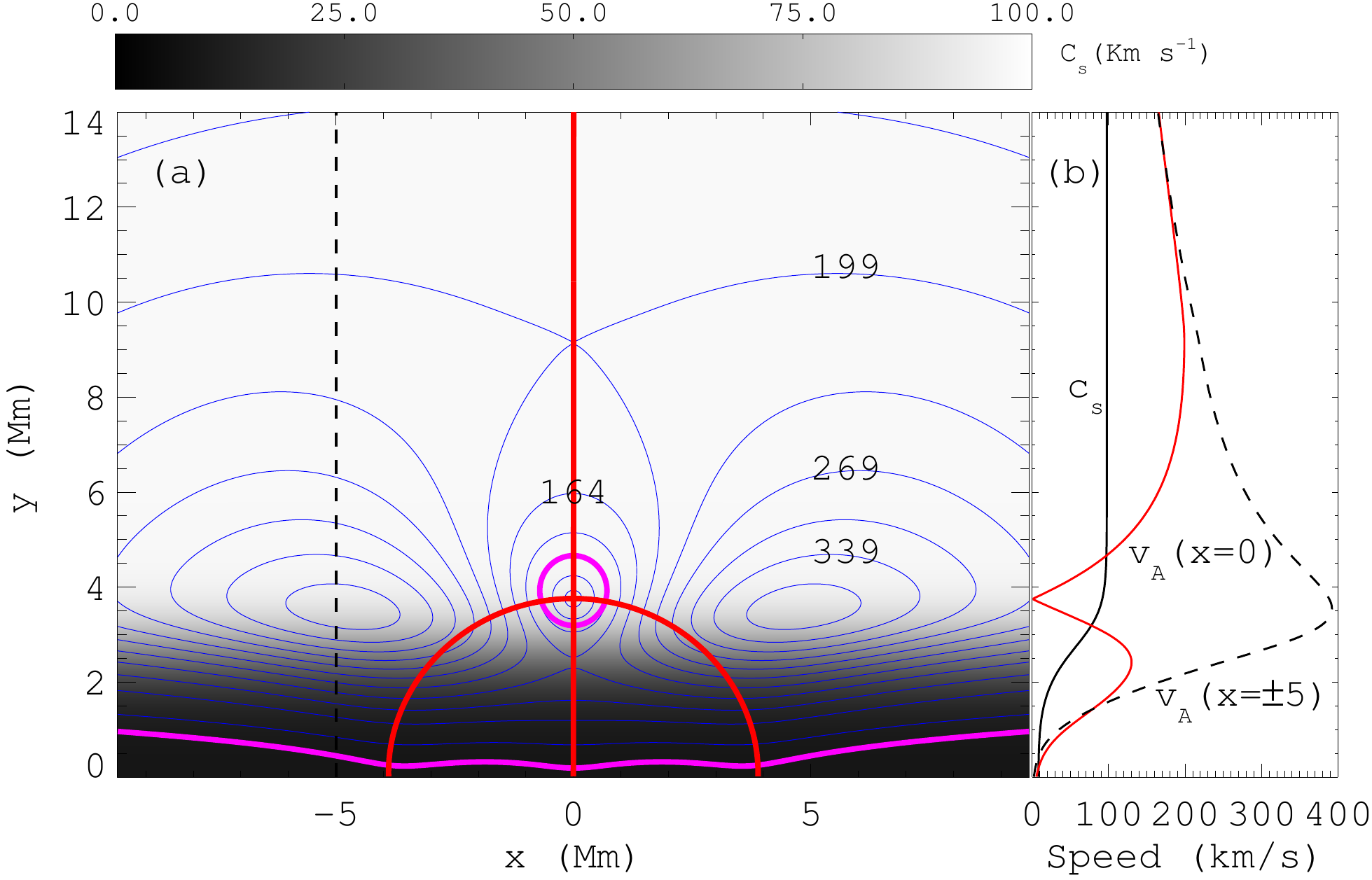}
  \caption{\label{fig:csva} (a) Variation of the sound speed (grayscale) and Alfv\'en speed (thin blue contour lines) within the domain.  The contours are in steps of $35\unit{km}\unit{s}^{-1}$ between $24$ and $374\unit{km}\unit{s}^{-1}$ (several indicated).  Thick red lines mark the locations of separatrices, and the thick magenta lines show where $c_s/v_A = 1.$  (b) The value of $c_s(y)$ throughout the domain (solid black), and $v_A(y)$ along two vertical slices: $x=\pm5\unit{Mm}$ (dashed) and $x=0\unit{Mm}$ (red).}
\end{figure}

An important aspect for wave propagation is the structure of the sound and Alfv\'en speed throughout the domain.  Hydrostatic stratification creates a sound speed that varies vertically only: $c_s(y) = \sqrt{\gamma k_b T_0(y)/\mu_m} = \sqrt{\gamma P_0(y)/\rho_0(y)}$.  This vertical sound speed stratification is shown as grayscale in \figref{fig:csva}(a) and as the solid black line in \figref{fig:csva}(b).  The sound speed ranges from $\approx 8-100\unit{km}\unit{s}^{-1}$ within the domain.

The Alfv\'en speed is structured horizontally as well as vertically: $v_A(x,y) = \frac{B_0(x,y)}{\sqrt{\mu_0\rho_0(y)}}$.  Thin solid blue lines in \figref{fig:csva}(a) show contours of the Alfv\'en speed, in steps of $35\unit{km}\unit{s^{-1}}$, with the value indicated at several levels.  The function has a minimum at the null point where $v_A(0,3.75\unit{Mm}) = 0$, and a maximum of $v_A \approx 400\unit{km}\unit{s}^{-1}$ in the two lobes on either side at $\vect{x}\approx(\pm 5,+3.75)\unit{Mm}$.  There are two saddle points along $x=0$, at $y\approx 2.4$ and $9.1\unit{Mm}$, though these are not on the same contour level.  \figref{fig:csva}(b) shows the Alfv\'en speed along two vertical slices, one passing through the null at $x=0\unit{Mm}$ (red), the other passing near the left maximum at $x=-5\unit{Mm}$ (dashed line).

\section{Linearized equations and ray tracing}\label{sec:ray}
\change{\subsection{Wave energy and dispersion relation}}
We will use LARE2D to solve the full, nonlinear set of equations \eqref{eq:continuity} to \eqref{eq:induction}.  However, we will also use linear theory to determine the form of a wave packet injected into the system, approximate its propagation through the (stationary) background, and help interpret the simulation output.  For this linear analysis, we set the viscous and resistive terms to zero, and add a perturbation to each background quantity: $\rho=\rho_0+\rho_1,\ P=P_0+P_1,\ \vect{B} = \vect{B}_0+\vect{b}$, and $\vect{v} = \vect{v}_1 (\hbox{with } \vect{v_0} = \vect{0})$.  The linearized ideal MHD equations are then
\begin{gather}
  \label{eq:lin-mass}\partial_t\rho_1 + \vect{v}_1\cdot\del\rho_0+ \rho_0\del\cdot\vect{v}_1=0,\\
  \rho_0\partial_t\vect{v}_1 = \rho_1\vect{g} - \del P_1 + \frac{1}{\change{\mu_0}}(\del\times\vect{b})\times\vect{B}_0,\\
  \partial_tP_1 = c_s^2\partial_t\rho_1+c_s^2\vect{v}_1\cdot\del\rho_0-\vect{v}_1\cdot\del P_0,\\
  \label{eq:lin-induction}\partial_t\vect{b} = \del\times(\vect{v}_1\times\vect{B}_0).
\end{gather}
We have assumed an adiabatic relation between the perturbed pressure and density, $\delta P_1 = c_s^2\delta\rho_1$.  These equations may be combined to yield a conservation relation for wave energy density and energy flux \citep[][ cf. Appendix \ref{sec:waveflux}]{Bray:1974}:
\begin{gather}
  \label{eq:wave-conservation}\partial_t\Biggl[\frac{1}{2}\rho_0v_1^2+\frac{P_1^2}{2\rho_0c_s^2}+\change{\frac{\abs{\vect{b}^2}}{2\mu_0}}-\Bigl(\frac{1}{2}\partial_y\rho_0g+\frac{\rho_0g^2}{2c_s^2}\Bigr)Y^2\Biggr] +\del\cdot\Bigl[P_1\vect{v}_1+\frac{1}{\change{\mu_0}}(\vect{B}_0\times\vect{v}_1)\times\vect{b}\Bigr]=0.
  \intertext{We can identify each term in \eqref{eq:wave-conservation} in order: the energy densities are kinetic ($E_K$), acoustic $(E_A)$, magnetic ($E_M$), and gravitational $(E_G)$, while the energy fluxes are acoustic $(\vect{F}_A)$ and magnetic (Poynting, $\vect{F}_M$):}
  \label{eq:wave-cons-short}\partial_tE = \partial_t(E_{K}+E_{A}+E_M+E_G) = -\del\cdot\vect{F}=-\del\cdot(\vect{F}_{A}+\vect{F}_{M}).
\end{gather}
We can further group the energy density terms into kinetic ($E_K$) and potential ($E_A, E_M, E_G$) parts.  In the gravitational term, $Y(t) = \int_0^tv_y(t^\prime)dt^\prime$ is the vertical displacement of a fluid element.  For frequencies greater than a few times the Br\"unt--V\"ais\"al\"a frequency ($\sim 4.5\unit{mHz}$), the gravitational term becomes increasingly less important than the other terms \citep{Hansen:2009}.  We will focus on frequencies of $\sim 40 \unit{mHz}$, and therefore drop this term from our analysis.

The above equation involves real quantities that may be directly calculated from the output of the simulation at each time $t$: $P_1(t)  = P(t)-P_0,\ \vect{b} = \vect{B}(t)-\vect{B}_0$, and so on.  This will be useful for determining the amount and type of wave energy at various locations in the simulation.  On the other hand, how the wave energy propagates through the system should be controlled by the properties of the background state.  To determine wave propagation, we first combine the linearized equations \eqref{eq:lin-mass}-\eqref{eq:lin-induction} into a single wave equation for the velocity perturbations by taking the time derivative of the momentum equation and substituting in the continuity, energy, and induction equations.  The result is
\begin{align}
  \label{eq:wave-full}\partial^2_{tt}\vect{v}_1 &= c_s^2\del(\del\cdot\vect{v}_1) + \frac{1}{\change{\mu_0}\rho_0}\Bigl\{\del\times\Bigl[\del\times(\vect{v}_1\times\vect{B}_0)\Bigr]\Bigr\}\times\vect{B}_0 + \vect{g}(\gamma-1)\del\cdot\vect{v}_1+\del(\vect{g}\cdot\vect{v}_1) \\
  & +(\vect{g}\cdot\vect{v}_1)\Bigl[\frac{\del c_s^2}{c_s^2}+\frac{c_s^2}{\rho_0}\del(\frac{\rho_0}{c_s^2})-\frac{\vect{g}}{c_s^2}\Bigr]\notag.
\end{align}
We assume each perturbed quantity varies in space and time only by a common phase term: $\vect{v}_1(\vect{x},t) = \vect{a}e^{i\Phi},\ \Phi = \vect{k}\cdot\vect{x} - \omega t$.  Next we apply the WKB approximation \citep{Weinberg:1962} that the phase function varies much more rapidly than any background quantity: $k=\abs{\del\Phi}\gg 1/h$, where $h$ represents any spatial scale of the background.   We again drop the explicit gravitational terms from consideration, though note that part of the stratification's effect is implicit through the spatial dependence of $c_s$ and $\rho_0$.  After applying these assumptions, we can write the wave equation in dyadic notation \citep[c.f.][who kept the explicit gravitational term]{Thomas:1982,Campos:1983}:
\begin{equation}\label{eq:dyad}
    \omega^2\vect{I}\cdot\vect{v}_1 = \Bigl[c_s^2\vect{k}\vect{k} - v_A^2(\hat{\vect{b}}\cdot\vect{k})\hat{\vect{b}}\vect{k}
      + v_A^2(\hat{\vect{b}}\cdot\vect{k})^2\vect{I} + v_A^2\vect{k}\vect{k}
      -v_A^2(\hat{\vect{b}}\cdot\vect{k})\vect{k}\hat{\vect{b}}\Bigr]\cdot\vect{v}_1,
\end{equation}
where $\vect{I}$ is the unit dyad.  Setting the determinant of this equation to zero, we find the dispersion relations for Alfv\'en ($\omega_A$) and fast-- and slow--magnetoacoustic waves ($\omega_\pm$):
\begin{equation}
  \label{eq:dispersion-relation}\mathcal{D}(\omega,\vect{k},\vect{x})=\Biggl(\omega_A^2- k^2v_A^2\cos^2\theta\Biggr)\Biggl(\omega_{\pm}^2 -k^2\Bigl[\frac{1}{2}(v_A^2+c_s^2)\pm\frac{1}{2}\sqrt{v_A^4+c_s^4 - 2v_A^2c_s^2\cos2\theta}\Bigr]\Biggr)=0.
\end{equation}
The angle between the propagation direction and the magnetic field (the attack angle or phase angle) is defined through $\hat{\vect{k}}\cdot\hat{\vect{b}}=\cos(\theta)$.  Equation \eqref{eq:dispersion-relation} has the same form as the standard relation for a homogeneous compressible plasma \citep{Kulsrud:2005}.  The WKB approximation we have used simply takes $v_A(\vect{x})$ and $c_s(y)$ to be spatially varying functions rather than uniform.   The determinant factors into $\mathcal{D}=\mathcal{D}_A\mathcal{D}_+\mathcal{D}_-$, which describe separate conditions on $\vect{k}$ for a given frequency $\omega$ for the Alfv\'en, fast, and slow mode, respectively (we will generically apply the subscripts $+,-,A$ to indicate solutions for each branch).  \change{Note that in each case the relation between frequency and wavenumber is linear, of the form $\omega\propto k$.  This is the relation for dispersionless waves, where each frequency wave propagates in the same way, just as for MHD waves in a homogeneous plasma.  Retaining the gravitational terms in \eqref{eq:wave-full} introduces dispersion at low frequencies near the Br\"unt--V\"ais\"al\"a frequency.  Our first order WKB approximation therefore includes the effect of refraction but excludes dispersion, which is unimportant for high frequency waves.}

Each mode will propagate through the system along a different path.  As explained in more detail in Appendix \ref{sec:wkb}, we can trace the path of a ray (say the fast ray) $\vect{x}(\tau)$, where $\tau$ parametrizes the distance along the ray, by picking an initial condition $\vect{k}(\tau=0),\ \vect{x}(\tau=0)$ and solving for $\vect{k}(\tau)$ and $\vect{x}(\tau)$ subject to the constraint that the correct condition is satisfied, i.e., $D_+=0$ for the fast ray).  The result is that the ray satisfies Hamilton's equations
\begin{gather}
  \frac{d\vect{k}}{dt} = -\frac{\partial \omega}{\partial\vect{x}}\Bigr\rvert_{\vect{k}}\\
  \label{eq:hamilx}\frac{d\vect{x}}{dt} = \frac{\partial \omega}{\partial\vect{k}}\Bigr\rvert_{\vect{x}},
\end{gather}
where $\omega$ comes from the condition $\mathcal{D}=0$.  We give explicit expressions for these equations for the fast and slow ray in Appendix \ref{sec:wkb}; the Alfv\'en waves simply follow field lines.  Note that $\partial_\vect{k}\omega = \vect{v}_g$, the wave's group velocity.  This is the velocity at which energy is transferred along the ray, and is equal to $\vect{F}/E$ from \eqref{eq:wave-cons-short}.  The energy relation is true up to the assumption that there is no dissipation and that the wave stays on a single branch of this dispersion relation, a point that we will return to repeatedly in the following.

\change{The fast and slow branches of the dispersion relation change their characters as the plasma shifts between pressure and magnetically dominated regions.  The fast mode is increasingly acoustic (potential energy dominated by $E_A$) for $\beta>1$ and magnetic (dominated by $E_M$) for $\beta<1$, while the opposite is true for the slow mode.  Mode coupling is allowed under certain conditions \citep{Tracy:2003}, where the essential requirement is that the gradient of the phase function $\Phi$ for each mode be similar in a region of space, so that, for example, $\vect{k}_+(\omega) \approx \vect{k}_-(\omega)$.  This can occur where the phase speeds ($v_\phi = \omega/k$) of the two modes are approximately equal, along the equipartition curves where $c_s = v_A$.}

\change{\citet{Cally:2007} and \citet{Hansen:2009} explain the typical terminology used in the helioseismic literature for mode conversion and transmission.  Physically this depends on which term dominates \eqref{eq:wave-cons-short}, $E_K,\ E_A,$ or $E_M$, and if that changes as a wave propagates.  Conversion refers to waves whose energy shifts from the magnetic to acoustic term (or vice versa) while staying on the same branch of the dispersion relation; transmission refers to a wave whose dominant potential energy term remains nearly the same as it propagates.  Thus, an acoustic fast wave originating from below the equipartition height may continue propagating as a magnetic fast wave above the equipartition height, and mode conversion is said to have taken place.  Because the ray theory is a solution along a single branch of the dispersion relation, it assumes perfect conversion.  These definitions depend only the properties of the wave, can be determined directly from the full MHD simulation presented in \S\ref{sec:results}, and are therefore to be preferred.}

\begin{figure}[ht]
  \begin{center}
    \includegraphics[width=0.48\textwidth]{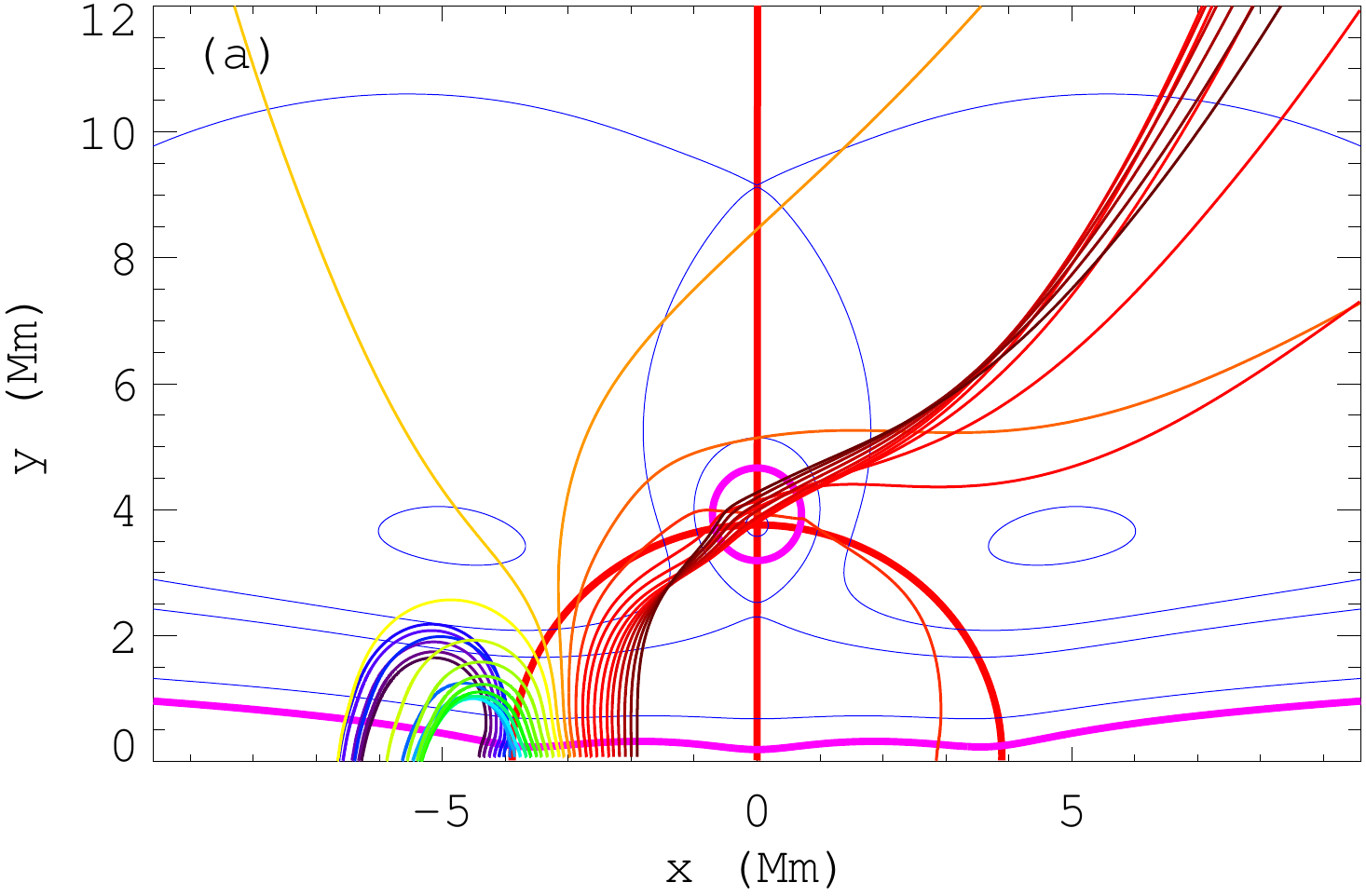}
    \includegraphics[width=0.48\textwidth]{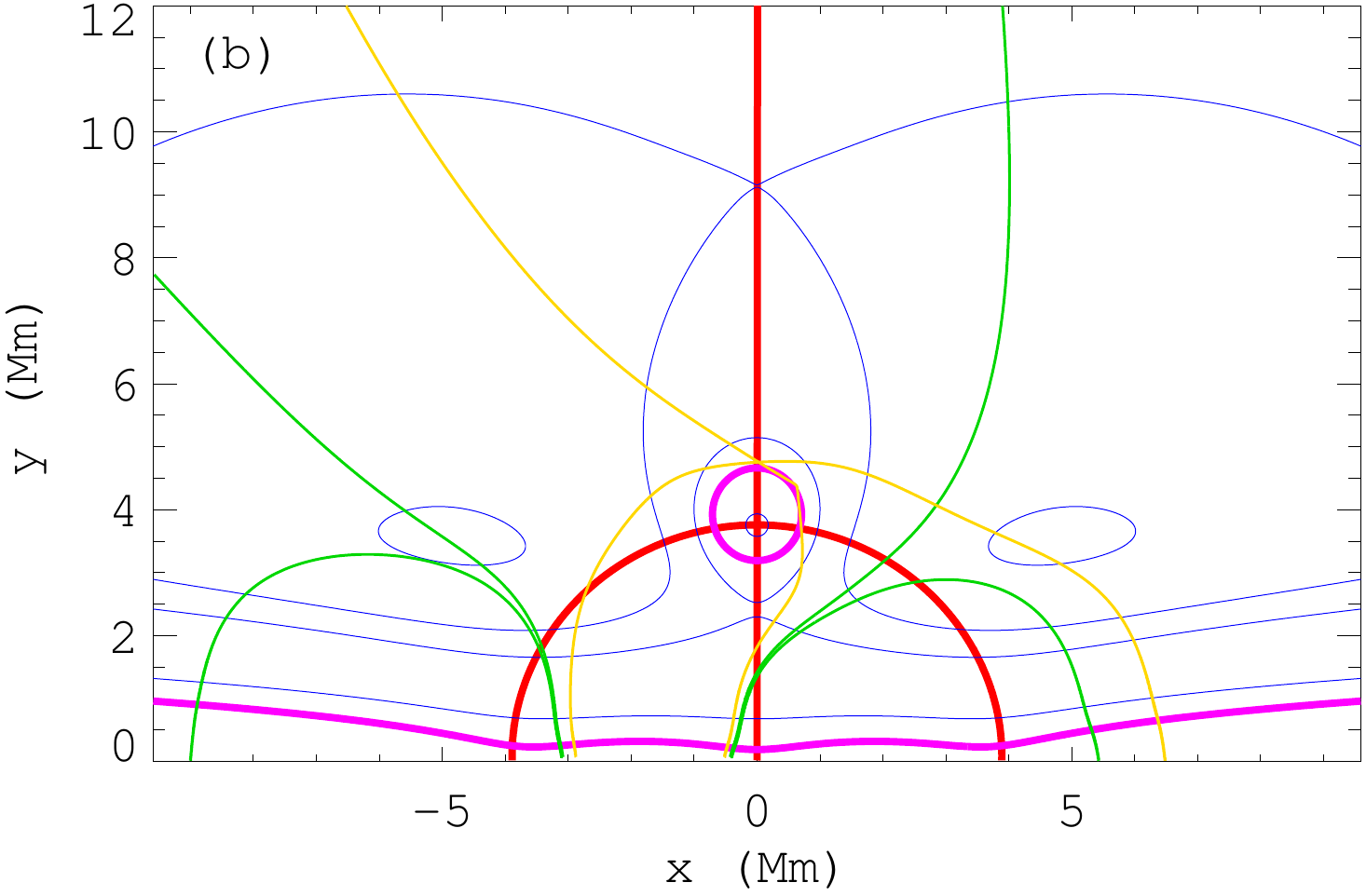} \caption{\label{fig:ray-paths} Fast ray paths through the domain. (a) A ray bundle, with each ray (rainbow colors) traced for an initially vertically propagating disturbance, initiated at intervals of 5 pixels along the lower boundary.  (b) 6 rays indicating boundaries in the domain for rays launched for $x<0 \unit{Mm}$.  Rays launch from the lower boundary between the two yellow paths refract in towards the null point, while rays launched outside of the outer two green paths refract back down to the photosphere.  Rays launched from the small areas between these curves reach the side or top boundaries without passing close to the null.  The pattern has reflection symmetry about $x=0\unit{Mm}$.}
  \end{center}
\end{figure}

\change{\subsection{WKB solutions}\label{sec:wkb-solutions}}
\figref{fig:ray-paths} shows the trajectories for bundles of fast magnetoacoustic rays\change{, which are solutions for the $\mathcal{D}_+$ branch of the dispersion relation}.  The slow waves \change{(solutions to $D_-$)} simply follow the field lines to a high degree of accuracy, and we do not show them here.  The thick red and magenta lines again show the separatrices and the two contours where $c_s=v_A$, while select contours of the Alfv\'en speed from \figref{fig:csva} are shown as thin blue lines, for easy comparison.  Over this background, panel (a) shows the paths for a bundle of fast rays initialized from the lower boundary as vertically propagating ($\hat{\vect{k}} = \hat{\vect{y}}$) fronts between $x\approx[-4.5,-2.0]$, with each ray depicted in a different color.  Clearly visible is the importance of refraction due to the inhomogeneous background: rays refract away from (towards) regions of high (low) phase speed in an amount that depends on the angle between their propagation direction and the local magnetic field direction.  From left to right in the figure, the rays refract back down towards the photosphere (purple--to--yellow rays), escape through the side or upper boundaries without passing near the $c_s=v_A$ region surrounding the null (yellow--to--orange), and pass near the conversion region surrounding the null point (orange--to--red).

An important feature of the ray solution is that some rays propagate directly through the $c_s=v_A$ region surrounding the null point.  This is because $\beta\neq 0$ and in the WKB approximation the fast ray describes a wave that transitions smoothly from magnetically dominated to acoustically dominated perturbations.  These rays cross each other multiple times, forming multiple sets of caustics.  \citet{Afanasyev:2012} have studied similar caustics both analytically and visually by determining ray trajectories near a linear 2D null point for a $\beta\neq 0$ plasma with uniform density and temperature.  In our case, tracing the paths of many rays which pass near the null point (not shown here) results in a very similar pattern of caustics as those in Figure 3 of \citet{Afanasyev:2012}.  The differences arise because of the stratification: near the null, our magnetic field is fairly linear in the horizontal direction, but not in the vertical direction.

In contrast, for a $\beta = 0$ plasma the slow mode solution vanishes from the dispersion relation, and the wave speed of the fast rays decreases to zero at the null.  In that case, the ray paths form logarithmic spirals focused on the null point \citep[][in particular, see Figure 7 of the former]{McLaughlin:2006a,Longcope:2012}, accumulating strong currents at the null which can then dissipate.  As has been noted before, the existence of a finite plasma pressure term thus makes focusing of wave energy on a null point more difficult than it otherwise would be.  We will explore this in more detail below by direct numerical simulation.

\figref{fig:ray-paths}(b) depicts a connectivity graph via a set of bounding rays for initially vertically propagating fast rays initialized in the left half of the domain, between $x\approx-3$ and $-0.5\unit{Mm}$.  The two leftmost green rays are traced from the centers of adjacent computational cells, and illustrate how quickly ray paths may diverge.  Rays traced further to the left of these, (initial location $x\lesssim-3\unit{Mm}$) all have turning points and close back down at the photosphere.  This is the typical behavior for low $\beta$ fast modes when the Alfv\'en speed increases with height.  Rays traced to the right of $x\approx -3\unit{Mm}$ display a much different behavior, and increasingly refract towards the null point.  The two yellow lines are bounding curves for rays that refract strongly toward the null point.  Rays launched between the rightmost green ray and the center of the domain (between $x=-0.5 \hbox{ and }0\unit{Mm}$) again exhibit a turning point and refract back downward to the photosphere.  The pattern repeats in mirror--image for rays launched from the positive $x$ side of the domain, due to the reflectional symmetry of the system.

\begin{figure}[ht]
  \begin{center}
    \includegraphics[width=0.48\textwidth]{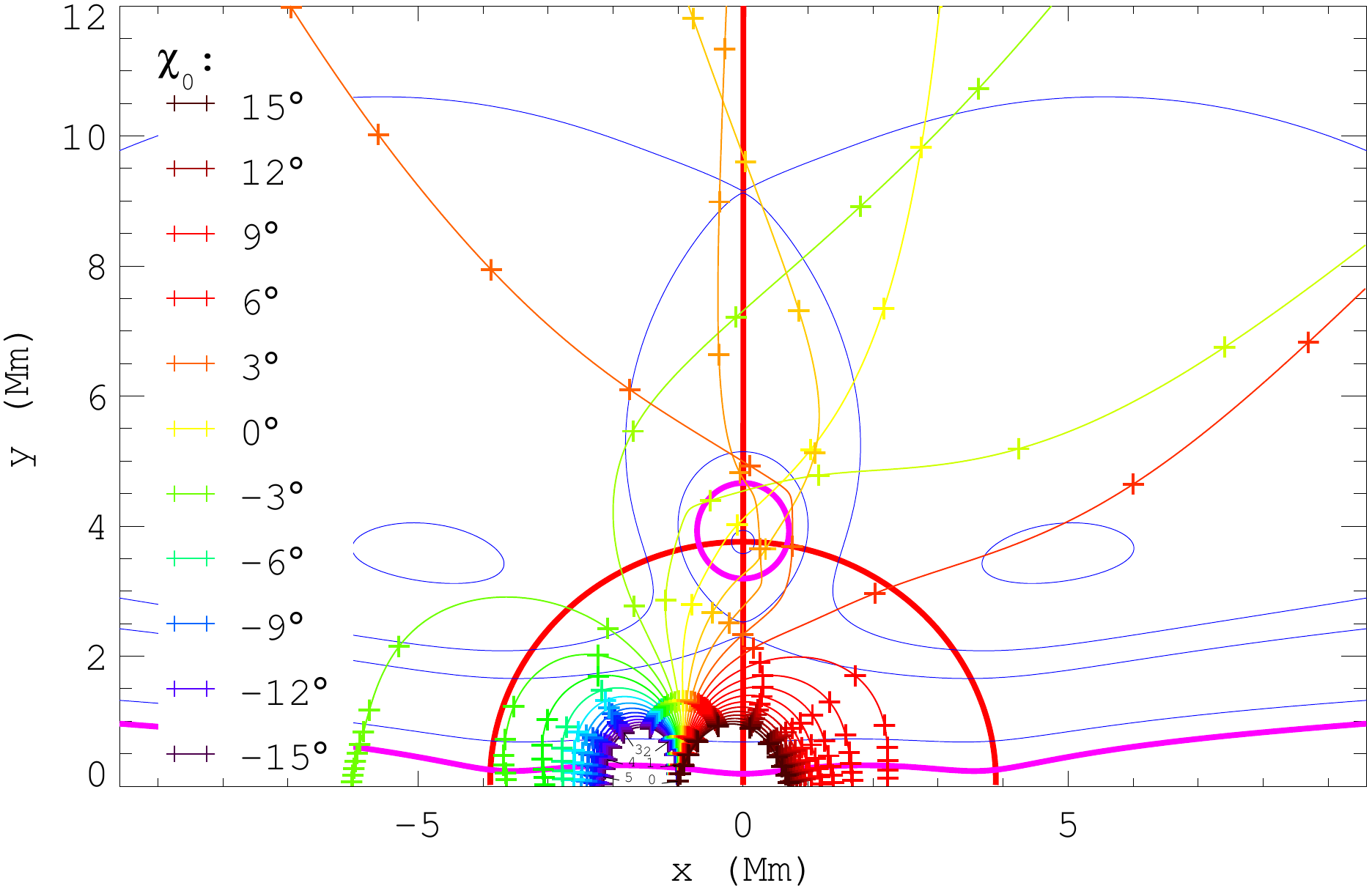}\caption{\label{fig:ray-angles} Fast ray paths for a single initial location $\vect{x}_0 = [-1,0.08]\unit{Mm}$ and a range of initial propagation directions $\chi_0$, up to $\pm 15^\circ$ to the vertical.  Plusses are spaced at travel times of $0.5t_N.$}
  \end{center}
\end{figure}
The path of a ray depends on its initial propagation direction, $\vect{\hat{k}}$, in addition to its initial location.  \figref{fig:ray-angles} shows the paths of fast rays launched from a single initial position near the lower boundary, $\vect{x}_0 = [-1.0,0.08]\unit{Mm}$, and initial propagation directions $\chi_0=\atan(k_x/k_y)$ up to $\pm 15^\circ$ to the vertical ($\chi=0^\circ$), in steps of $1^\circ$.  These values correspond to the center of the wavepacket we use to drive the numerical simulation, described in \S\ref{sec:driver}, and its initial range of propagation angles, as determined in \S\ref{sec:driver-distribution}.  The plus marks along each ray are equally spaced in units of travel time, at intervals of $0.5 t_N$, or, equivalently, through the phase distance $\tau$.  Each set of equal time points thus traces out a front of constant phase $\Phi(\vect{k(\tau)}\cdot\vect{x(\tau)} - \omega t(\tau))$.  The phase fronts are more closely spaced in the lower portion of the figure, where both the sound and Alfv\'en speeds are small, and more widely spaced higher up where the phase speeds are greater.  The set of all phase fronts fills in the phase function throughout the domain.  Rays passing close to the null equipartition region again display complex trajectories.  A steady prograde change in initial angle of propagation does not lead to a steady prograde change in the direction of the outgoing ray.  Instead, outgoing rays jump from prograde to retrograde change several times.  The rays cross, and so the phase function is, in general, multivalued throughout the domain.

\change{We can combine the ray tracing results to try to predict how an initial, localized disturbance will propagate through the system.}  A spatially localized wavepacket launched from the lower boundary will exhibit a combination of the behaviors illustrated by the ray paths of Figures \ref{fig:ray-paths} and \ref{fig:ray-angles}.  At each spatial location, the packet will be best described by a range of propagation directions.  In \S\ref{sec:driver-distribution} we will need to know the range of initial propagation angles that pass through the null point's equipartition layer.  To determine this, we initialized \change{fast} rays between $x=-5\unit{Mm}$ and $x=0.15\unit{Mm}$ with propagation directions $\chi=\pm15^\circ$ in steps of $0.1^\circ$.  We recorded each ray's closest approach, $d$, to the centroid of the equipartition layer, located at $\vect{x}_c=[0,3.9]\unit{Mm}$ (note that the equipartition centroid is shifted slightly above the null point, due to the stratification).  

\begin{figure*}[ht]
  \begin{center}
    \includegraphics[width = 0.5\textwidth]{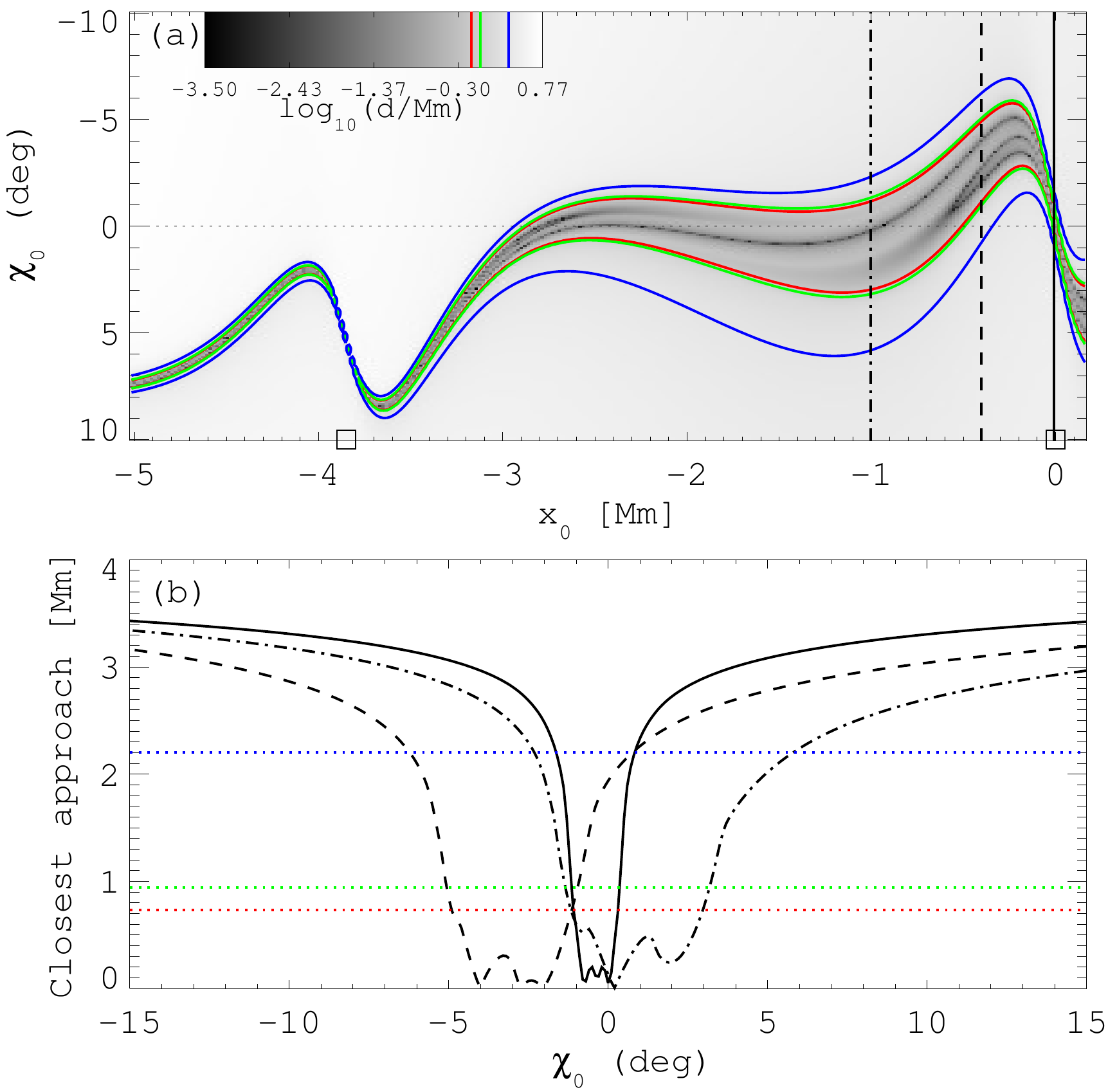}
    \caption{\label{fig:ray-ca} (a) Contour image showing the $\log$--scaled closest approach of each fast ray path to the centroid of the null equipartition boundary, located at $[0.0,3.9]\unit{Mm}$, as a function of initial location $(x_0)$ and propagation direction with respect to vertical ($\chi_0$).  Red, green, and blue lines are contours of $r_e\times[1,1.3, 3]=[0.735,\ 0.955,\ 2.21]\unit{Mm}$, and are also indicated on the colorbar.  The three vertical lines correspond to locations of the three profiles in (b), and the dotted horizontal line marks $\chi_0=0^\circ$.  Squares mark the intersection of separatrices with the lower--boundary.  (b) Slices through the top panel for rays initialized at $x_0=0$ (solid), $-0.4$ (dashed), and $-1\unit{Mm}$ (dash--dot).  Colored lines correspond to the contour levels from above.}
  \end{center}
\end{figure*}

The null's equipartition curve is approximately circular, with a radius $r_e\approx0.75\unit{Mm}$.  \figref{fig:ray-ca}(a) shows the result for angles between $\pm 10^\circ$, with the closest approach distance for each ray displayed in logarithmic scale as a function of the initial ray location (abscissa) and initial propagation angle (ordinate).  The distribution is nearly flat outside of this angle range, as seen in \figref{fig:ray-ca}(b).  Rays with positive (negative) angles initially propagate in the positive (negative) $x$ direction.  The red, green, and blue contours correspond to $1,\ 1.3,$ and $3$ times $r_e$.  We take $R_n = 1.3r_e=0.96\unit{Mm}$ to define a region of influence for the null.  This region is based on the equipartition scale height, $H_e^{-1} = \partial_\tau (c_s^2/v_A^2)\rvert_{c_s=v_A}$, so that $R_n=r_e+H_e$.  Therefore, rays initialized at the lower boundary with initial propagation angles falling between the green curves are strongly influenced by the null point.

The three vertical lines in \figref{fig:ray-ca}(a) show the location of slices through the contour plot, and are displayed as line plots in \figref{fig:ray-ca}(b).  The range of angles for which rays pass near the null reaches a maximum for rays initialized near $x=-1\unit{Mm}$ and a minimum around $x=-3.8$ and $0\unit{Mm}$, where the separatrices intersect the lower boundary (squares in top panel).  The actual range of angles undergoes significant variation, is typically not centered on $0^\circ$, and has multiple inflection points.  \change{Only two sets of vertically propagating rays, one between $x=-3.0$ and $-0.5\unit{Mm}$, and the other in a narrow range around $x_0=0$, pass close to the null.  The first of these corresponds to the region between the yellow rays of \figref{fig:ray-paths}(b).  Several dark bands run through the entire $x_0-\chi_0$ plot and indicate rays that pass very close to the center of the interaction region, almost directly through the null.  There are typically multiple distinct dark bands for a given $x_0$.  This is a manifestation of the prograde--retrograde--prograde behavior of outgoing rays discussed for \figref{fig:ray-angles}(b).  The angles at which the behavior changes for a given $x_0$ are located at the minima along vertical slices through the plot (see, e.g., the slices plotted in \figref{fig:ray-ca}(b)).}  Interestingly, the dark band that arises around $x_0=-1.5\unit{Mm}$ bifurcates at $x_0=-0.55\unit{Mm}$.  Moving further right, the three distinct bands at that point remain distinct across $x_0=0,$ which can be seen as the three small dips in the solid curve of \figref{fig:ray-ca}(b).

\change{\subsection{Conclusions from ray tracing}\label{sec:ray-conclusions}}
\change{We have found the ray tracing analysis to be a useful tool for understanding the properties of our model atmosphere, and, as will become apparent, for analyzing and interpreting the MHD simulation presented in the next two sections.  Here, we summarize our results from the ray theory and point out several important differences from previous ray tracing investigations.}

\change{We used a WKB method to estimate how a fast mode wavepacket crossing the lower boundary will propagate through the background field.  The WKB theory assumes that a ray stays on a single branch of the dispersion relation.  A wavepacket is a spatially localized disturbance with a distribution of wave vectors $\vect{k}$.  Thus, to see how a wavepacket will propagate, we traced rays from multiple initial positions and using multiple initial phase angles, with several examples show in Figures \ref{fig:ray-paths} and \ref{fig:ray-angles}.  We found that some rays will refract towards the null point, while most will refract back downwards towards the photosphere.  A small segment between these two regions appears able to propagate to the top of our domain (and hence to infinity), as shown by \figref{fig:ray-paths}(b) for initially vertically propagating rays.  For rays that refract back downward, the height of the turning point depends on the gradient of the phase velocity along each ray path (we have not attempted to analytically derive the turning point for our horizontally inhomogeneous background).  Rays reaching the interaction region surrounding the null exhibit complex behavior, with some eventually refracting back towards the lower boundary, some reaching the top and side boundaries, and many adjacent rays crossing paths, some at multiple locations.  The set of fast rays that pass near the null is a function of both the initial ray location and propagation direction.}

\change{We did not discuss solutions for the slow waves using $\mathcal{D}_-$ in any detail.  The properties of slow MHD waves are well known for homogeneous plasmas and can be found, for instance, in \S5.3.2 of \citet{Goedbloed:2004}.  In that case, the maximum departure of the group velocity from the magnetic field direction, the return angle $\theta_R$, is $\theta_R=-30^\circ$.  It occurs for $c_s/v_A=1$ and for a phase angle $\theta=0^\circ$.  For different phase angles and different values of the ratio $c_s/v_A$, the return angle is much smaller, typically $\lesssim5^\circ$.  To check that this carries over to the inhomogeneous case, we initialized slow rays at several hundred initial locations and wave vectors.  For a given ray, the maximum departure from the magnetic field direction is typically less than $5^\circ$, and less than $30^\circ$ for all rays, as in the homogeneous case.  For most of the distance along each ray we find departures of $<0.5^\circ$.  Rays that do make a substantial angle to the magnetic field typically do so only for small portions of their trajectories near the $c_s=v_A$ boundaries, after which they closely follow the field.}

\change{Our analysis differs from ray tracing used in studies of sunspots with slowly varying magnetic fields \citep{Cally:2007, Khomenko:2009a, Khomenko:2009b}.  They typically focus on the low frequency dispersion and the variation of the cutoff frequency due to the magnetic field, which we ignore.  These are very important effects for the interpretation of phase relations determined from observations \citep[see][and references therein]{Felipe:2012}.}

\change{Other authors have performed ray tracing through magnetic configurations containing nulls \citep{McLaughlin:2004, McLaughlin:2006a, McLaughlin:2008, Longcope:2012, Afanasyev:2012, McLaughlin:2016}.  Here, the focus is typically quite different compared to the sunspot studies, and addresses whether wave energy is reflected by the null or accumulated at the null, how to determine characteristic damping timescales, etc.  We will consider these studies more in the Discussion, \S\ref{sec:discussion}.}

\change{It is important to keep in mind how altering our initial condition would modify the ray behavior (the ease with which this is accomplished is one of the main advantages of the WKB method).  Suppose the magnetic field is everywhere reduced to zero, but the stratification is kept the same.  Then upward propagating fast rays, now degenerate to purely hydrodynamic sound waves, still refract off the increasing sound speed, and eventually reflect from a height where their frequency divided by wavenumber matches the local sound speed.  If instead we keep the magnetic field the same but set the density and temperature to uniform values, we find that a set of rays spiral in towards the null, as has been found many times before for both linear and more complicated nulls \citep{McLaughlin:2004, McLaughlin:2006a, Longcope:2012}.  Comparison with Figures \ref{fig:ray-paths} and \ref{fig:ray-angles} demonstrates that the vertical stratification breaks the radial symmetry close to the null and prevents the logarithmic inspiral.  These difference arise just from the linear (WKB) theory, and are in addition to any shock formation or mode conversion that may arise when solving the nonlinear MHD equations, as we will do in \S\ref{sec:results}.  The combination of stratified atmosphere, compressive waves, and nontrivial topology has not been well studied, and represents one significant advance of this present work.}

\change{Not included in \change{our WKB} analysis is the rate of mode conversion between fast and slow modes.  So, although many fast ray paths do pass through the interaction region around the null, as shown in \figref{fig:ray-ca}, it remains to determine how much energy stays on the $\mathcal{D}_+$ branch and how much converts to the $\mathcal{D}_-$ branch.  Slow waves thus generated will ultimately have different trajectories upon exiting the conversion region compared to the fast waves.  We now turn our attention to the numerical solution of Equations \eqref{eq:continuity} to \eqref{eq:induction} to answer this question.}

\section{Driver for nonlinear simulation}\label{sec:driver}

\floattable
\begin{deluxetable*}{CCRLRLCc}
\tablecaption{Wavepacket driver parameters \label{tab:driver}}
\tablecolumns{8}
\tablewidth{0pt}
\tablehead{
\colhead{Parameter} &
\colhead{Definition} &
\twocolhead{Value (normalized)} &
\twocolhead{Value (MKS)} &
\colhead{(Inverse variable)} & 
\colhead{Description}
}
\startdata
f_d & \omega_d/2\pi         & 6.0/2\pi & t_N^{-1}   & 40     & \unit{mHz}           & (25.42\unit{s})          & Central frequency\\
k_y      & \omega_d/c_s   & 4.533 & L_N^{-1}  & 30    & \unit{Mm}^{-1}        & (\lambda = 0.2\unit{Mm}) & Central wavenumber\\
w_x      & \ldots         & 1.977 & L_N      & 0.297  & \unit{Mm}            & \ldots                   & Width (X) \\
w_y      & \ldots         & 0.4567 & L_N     & 0.069  & \unit{Mm}            & \ldots                   & Width (Y) \\
T_d      & \omega_dw_y/c_s & 2.07 & t_N      & 50.25  & \unit{s}             & \ldots                    & Driver duration\\
v_d      & \ldots         & 0.1 & V_N     & 0.6177  & \unit{km}\unit{s}^{-1} & \ldots                    & Driver amplitude \\
\enddata
\end{deluxetable*}

To study energy propagation and dissipation through our system, we use a time dependent lower boundary to introduce acoustic wave packets and mimic a wave generation mechanism in the solar atmosphere.  The driver properties are summarized in Table \ref{tab:driver}.  The perturbations are defined through a spatially dependent vertical velocity, $\vect{v}_1 = (0,v_y,0)$, with amplitude $v_d$:
\begin{equation}
  \label{eq:vdrive}v_y(x,y,t) = v_d\exp\Bigl[-\frac{(x-x_d)^2}{2w_x^2}-\frac{(y-y_d-c_st)^2}{2w_y^2}\Bigr]\sin\Bigl[k_y(y - y_d)-\omega_d t\Bigr].
\end{equation}
\change{For a homogeneous plasma,} the density and energy perturbations consistent with the velocity perturbation are
\begin{equation}
  \label{eq:de-drive}\rho_1 = \rho_0v_y  k_y/\omega_d\qquad\text{and}\qquad \epsilon_1 = \rho_1(\gamma-1)\frac{\epsilon_0}{\rho_0}.
\end{equation}
\change{The form of the above variation ignores components in the $x$ direction, but nevertheless primarily excites a fast acoustic wavepacket at the lower boundary, as discussed below.  Similar functional forms have been used to model, for instance, p--mode excitation in sunspot umbra \citep{Moradi:2015}.  We do not introduce any perturbations to the out--of--plane variables.  Because our initial condition has $v_z=B_z =0$, and because all out--of--plane derivatives are zero in 2.5D, it is clear from the momentum and induction equation that no out--of--plane component can later be generated.  This decouples the Alfv\'en mode from our simulation (both numerically and analytically), and hereafter we focus only on the fast and slow magnetoacoustic modes.  LARE2D does update $v_z$ and $B_z$, and we have verified that they remain zero throughout the computation, as expected.}

Equation \eqref{eq:vdrive} describes a 2D Gaussian wavepacket advected upwards at the local sound speed $c_s$.  We set the angular driving frequency $\omega_d = 6 t_N^{-1}$ (physical frequency $f_d=\omega_d/2\pi\approx 40\unit{mHz}$).  The vertical wavenumber is set using the driving frequency and the sound speed at the lower boundary, $k_y = \omega_d/c_s(0)\approx 30\unit{Mm}^{-1}$.  The packet has horizontal and vertical widths $(w_x, w_y) = (0.297, 0.069)\unit{Mm}$, and an initial centroid location $(x_d,y_d) = (-1.0, -0.2)\unit{Mm}$.  The widths were set so that the wavepacket amplitude falls to $0.01\%$ of its maximum value within $\pm1\unit{Mm}$ horizontally and 1 wavelength vertically.  As might be ascertained by considering \figref{fig:ray-ca}(a), and as we describe in more detail in \S\ref{sec:driver-distribution}, choosing an initial position for the wavepacket of $x_d=-1\unit{Mm}$ maximizes its interaction with the null.  We have chosen the packet's parameters to make this the case, and to break the symmetry of the system by localizing the packet to one side of the central separatrix field line.

We drive our simulations by adding the above velocity, density, and energy perturbations into the lower boundary ghost cells.  \change{We do not add a perturbation to the magnetic field in the ghost cells so that, in the ghost cells, the perturbation is purely acoustic.  The acoustic wave thus introduced immediately couples to a magnetoacoustic fast wave and generates perturbations (of low amplitude) to the magnetic field in the domain.}  This is done for simplicity in implementing the boundary conditions.  Our topology does not permit us to write down the global normal modes of the system, but the adiabatic condition does enforce the self--consistent relation between the perturbed velocity, density, and internal energy described by \eqref{eq:de-drive}\change{, at least up to the approximation that there is a single vertical wavenumber $k_y$.}  Near the lower boundary, the acoustic energy and flux are the dominant terms of Equation \eqref{eq:wave-conservation} and the acoustic wavepacket approximation is good.

Advection of the packet upward at speed $c_s$ means the vertical width of the wavepacket, $w_y$, sets the duration of driving, $T_d$.  The effective temporal envelope produces a range of frequencies peaked about the driving frequency.  The vertical wavenumber is set explicitly through $k_y$, but horizontal modes $k_x$ will also be excited due to the horizontal gradients in our wavepacket amplitude and the background field.  Their existence leads to a range of initial propagation angles, $\chi = \atan(k_x/k_y)$.  We will estimate the distribution of $k_x$ and its effect on the dynamics in detail in \S\ref{sec:discussion}, but for now we simply note that the final result of the boundary driving is a distribution of waves in propagation direction, frequency, and total wavenumber $k = \sqrt{k_x^2+k_y^2}$, introduced at the lower boundary.  Properties of the driver are summarized in Table \ref{tab:driver}.

\section{Simulation of pulse propagation and conversion}\label{sec:results}
\subsection{General description of simulation results}\label{sec:sim-gen}
  \begin{center}
    \begin{figure*}[hp]
      \includegraphics[width=0.95\textwidth]{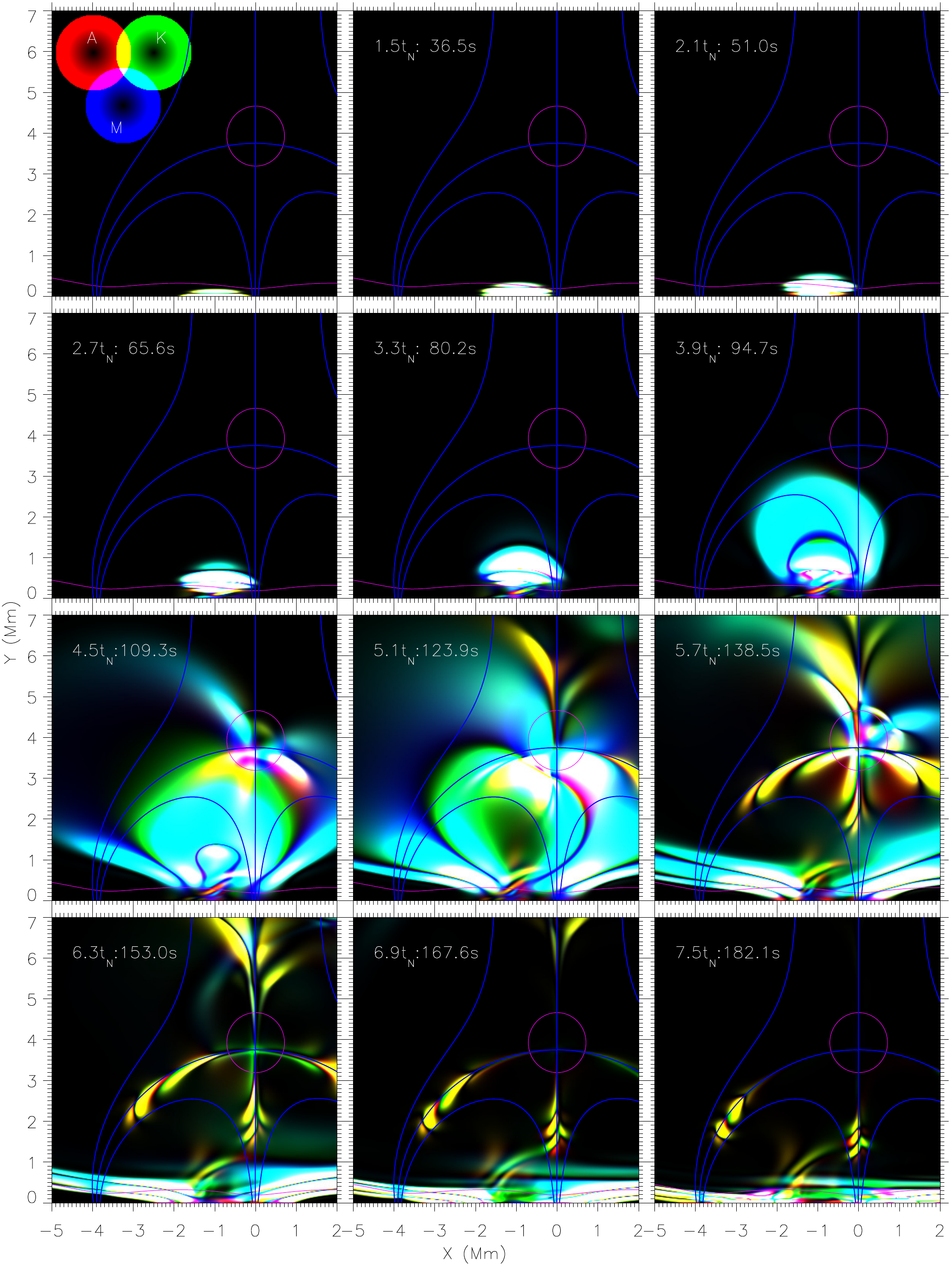}
      \caption{\label{fig:en-time} Time sequence of energy densities for acoustic (red), kinetic (green), and magnetic (blue) energy density terms from Equation \eqref{eq:wave-conservation}.  Blue lines are select magnetic field lines of the background state, and the two $c_s=v_A$ curves are shown in magenta.  The color wheels in the upper left show the overlap of different energy densities, and are on a linear scale from $0$ at each center to $1\times 10^{-2}\unit{J}\unit{m}^{-2}$ at the radii where the circles intersect, and then constant beyond that.  Horizontal distance is indicated in the top left panel, vertical distance in the center left.  Animation available in the online material.}
    \end{figure*}
  \end{center}
  
  Our analysis of the simulation will focus on the energy density and flux terms of Equation \eqref{eq:wave-conservation}.  \figref{fig:en-time} contains twelve frames from a movie showing the interaction of the introduced perturbation with the null point.  The images consist of three color channels---red, green, and blue---whose intensities correspond to the acoustic, kinetic, and magnetic energy density terms from Equation \eqref{eq:wave-conservation}.  The color channels are additive, e.g., green+red = yellow, as shown by the color wheels in the upper left panel.   Each color wheel is scaled linearly between zero (black) at the center and $1\times10^{-2}\unit{J}\unit{m}^{-2}$ at the radii where the disks overlap.  The outer $\sim 30\%$ of each disk is of uniform intensity to most effectively show the overlap of each energy term with the others.  Overlap of all three terms shows up as white in the figure.

  The energy density terms correspond to the energy carried by the waves, and together with the fluxes $(\vect{F}_{A},\ \vect{F}_M)$ they satisfy a conservation relation for wave energy, independent of the system's total energy.  \change{Because the wavepacket greatly expands and breaks into many individual pieces, no level of saturation for depicting the energy densities is entirely satisfactory.  The saturation level of $1\times10^{-2}\unit{J}\unit{m}^{-2}$ in \figref{fig:en-time} was chosen to fairly accurately depict the energy densities after the wavepacket has rapidly expanded.}  Early in the simulation, say at $t=2.1t_N$, the peak energy density of the wavepacket is around \change{$42\unit{J}\unit{m}^{-2}$}, and the figure is highly saturated.  \change{The energy density at the peak of the wavepacket ($\vect{x}=(-1.,0.18)\unit{Mm}$) is proportioned $32\%$, $50\%$, and $18\%$ between acoustic, kinetic, and magnetic terms, and the kinetic and potential terms are in equipartition.  Later on, the pulse expands into the low $\beta$ regions and breaks into multiple packets.  The peak energy densities of the yellow pulses at $6.3t_N$, near $\vect{x}=(-2.8,2.4)\unit{Mm}$, are $\sim 0.22\unit{J}\unit{m}^{-2}$, and here the energy density is divided $54\%,45\%$, and $1\%$ between the acoustic, kinetic, and magnetic terms.  Thus, the saturation level does a fairly accurate job at depicting the ratios the energy terms at later times, after around $3.5t_N$.}

Cyan portions of the figure correspond to regions with roughly equal parts kinetic and magnetic energy densities (blue plus green), the partition of energy associated with fast magnetoacoustic waves in a low $\beta$ plasma and slow waves in a high $\beta$ plasma.  In the same way, yellow regions have equal parts kinetic and acoustic energy densities, and correspond to slow magnetoacoustic waves in the low $\beta$ case and fast waves in the high $\beta$ case.  In the following, we will use ``magnetic'' or ``acoustic'' waves to indicate which of these energy density terms from \eqref{eq:wave-conservation} dominates the other.  In many circumstances, each type of wave for an ideal magnetohydrodynamic fluid exhibits equipartition of energy between its kinetic and potential terms \citep{Zweibel:1995}.  Depending on the properties of the wave (standing or traveling, short or long wavelength) and the background medium (static, slowly varying, or turbulent fluctuations about an average equilibrium), equipartition may hold only weakly (in a spatially and/or temporally averaged sense), strongly (at each space--time point), or not at all.  Our simulation uses a high frequency wave, and the Ohmic dissipation term is only important in regions of strong gradients.  As a result, for most locations in our simulation, we should find that the kinetic energy is in equipartition with the sum of acoustic and magnetic terms.  \change{We can think of allotting fractions }of the kinetic term between the fast and slow wave in proportion to the magnetic and acoustic terms to get the total energy for each type of wave.  This approach is only an approximation, and will work best at locations where the two waves are strongly distinct (very high or low $\beta$ regions).  However, it will still prove useful in understanding the propagation of energy through the system.  \change{\citep[For a discussion of important cases where strong equipartition does \emph{not} hold, and implications for the interpretation of observations, see][]{Goossens:2013}.}

Below the lower equipartition height, the wavepacket is predominantly acoustic \change{(recall the color saturation, discussed above)}.  The first several panels of \figref{fig:en-time}, up to $\sim 3.3t_N$, show the slow upward propagation of the wavepacket.  As the pulse crosses the lower equipartition layer, the wave energy mostly remains on the fast branch of the dispersion relation: it converts from an acoustically dominated (yellow) high $\beta$ fast wave to a magnetically dominated (cyan) low $\beta$ fast wave.  \change{The ongoing conversion is apparent in the division of energy at time $2.1t_N$ quoted above, resulting in $20\%$ of the energy in magnetic perturbations.} 

The pulse rapidly expands in size as it reaches the higher phase speed portions of the domain, starting around $3.3t_N$.  Between $3.3$ and $5.0t_N$, the outer portions of the wavefront turn over and become downward propagating.  The rapid expansion and the turnover are in agreement with the properties of the constant phase fronts from \figref{fig:ray-angles}.   In the region surrounding the null, but still outside the magenta contour (e.g., still low $\beta$), a portion of the wavepacket refracts towards the null, and a portion refracts back downwards towards the photosphere.  Some of the initial wavefront passes around the null to reach the upper boundary, and therefore escapes the system. These results again agree qualitatively with the results of the ray--tracing analysis of \S\ref{sec:ray}.  \change{Note that this is very different from what would happen if the magnetic field were uniform.  Then, the phase speed always increases with height, and the fast wave will refract back towards the photosphere.  The strong inhomogeneity in the field due to the presence of the null is responsible for the radically different wave behavior.}

Around the null we see a continuous transfer of energy from the fast to the slow branch of the dispersion relation, and from magnetic to acoustic energy densities, starting when the wave approaches the $c_s=v_A$ region surrounding the null, around $4.5t_N$.  This is visible in the animation of \figref{fig:en-time} as the appearance of the yellow colored pulses propagating away from the null.  They are first visible at $t=4.1t_N$, in regions outside the equipartition layer where the plasma $\beta$ is still less than, but close to, the equipartition value of $\beta=1.2$.  Energy transfer continues in the region spanning the equipartition contour.  At least part of this behavior does not neatly fit into either of the currently used categories of transmission or mode conversion, as it involves, purely in the low $\beta$ region, a jump from one branch of the dispersion relation to the other and a change from magnetically to acoustically dominated energy density at the same time.

While the fast mode may propagate in any direction relative to the magnetic field lines, the slow mode is heavily guided along the field.  In \figref{fig:en-time}, this shows up in the low $\beta$ regions as the yellow portions of wave energy density guided along the four separatrix field lines leaving the null.  If one had a detector sitting on the separatrix at, say, $\vect{x}=[-2,3.5]\unit{Mm}$, one would see a fast magnetic wave headed towards the null, and a time later a slow acoustic wave headed away from the null.  Because the magnetic field lines concentrate moving away from the null and back toward the photosphere, the energy leaving in the slow mode becomes increasing concentrated around the separatrix field lines as it propagates downwards towards the magnetic foot points.

In summary, the simulation shows that the initial wavepacket splits up into numerous sub--packets as it propagates.  A small portion remains acoustic (transmits across the lower $c_s=v_A$ boundary as a slow wave) and is confined to low lying field lines in the low $\beta$ region.  \change{This accounts for $\approx10\%$ of the injected energy.}  Most of the initial packet remains a fast wave when crossing the lower equipartition height and subsequently refracts either left or right and returns to the lower boundary.  However, a part of the upward propagating fast wave is further refracted in towards the null point.  This portion appears to mode convert near the null, and largely leaves the null region as slow acoustic waves confined to field lines near each of the null's four separatrices.  The next two subsections will cover the mode conversion process near the null in more detail.

\subsection{Energy density time--distance diagrams}\label{sec:time-distance}
\begin{figure}[Ht]
  \begin{center}
    \includegraphics[width=0.4\textwidth]{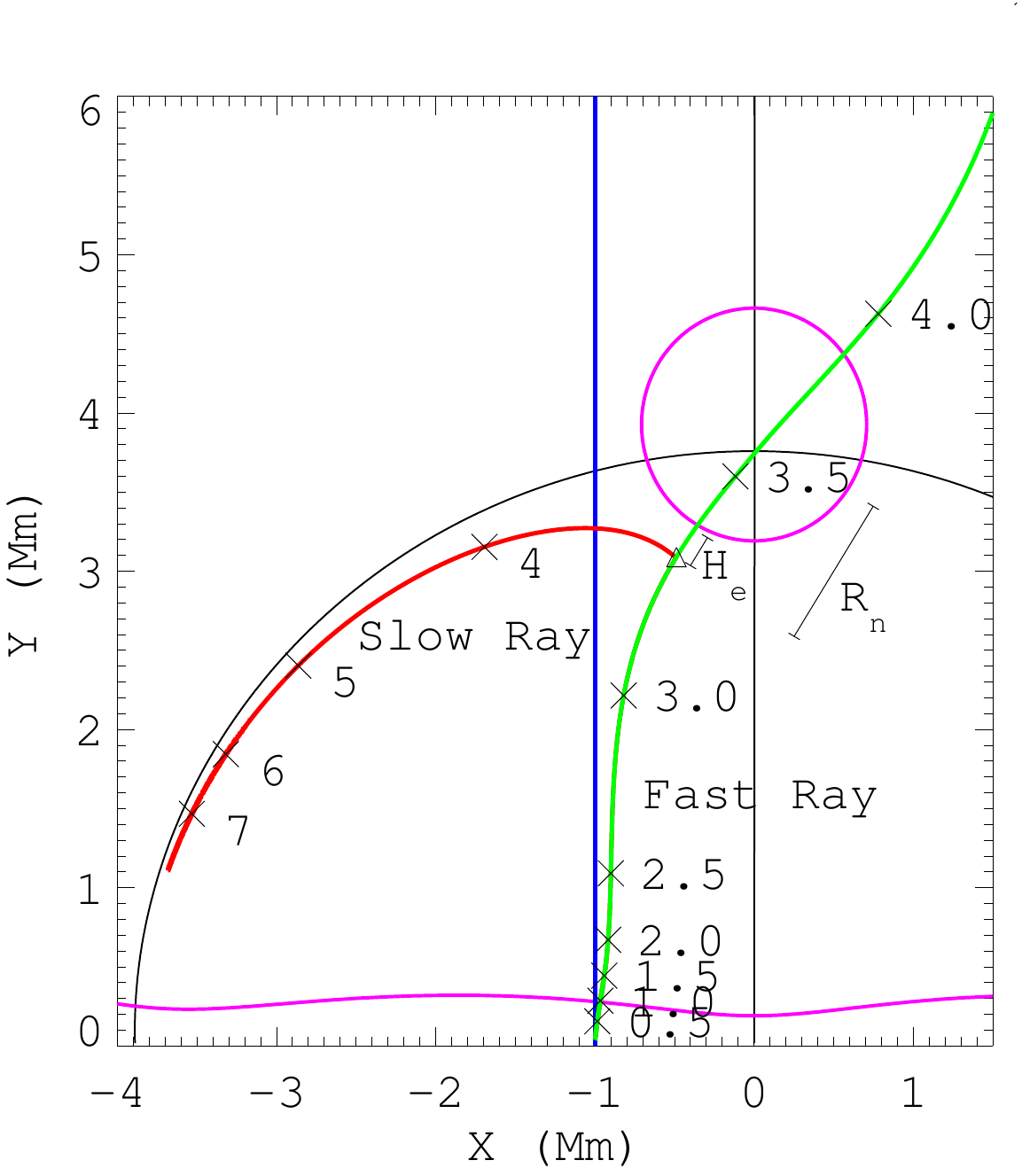}
    \caption{\label{fig:phase} Phase distance along two curves near the null point: a fast ray (green), a slow ray (red).  Also indicated are the reference slice at $x=-1\unit{Mm}$ (blue), the point at which the two rays are joined (triangle, see text), and the lengths of the equipartition scale height $H_e$ and the null's region of influence $R_n$.}
  \end{center}
\end{figure}

\change{According to \citet{Schunker:2006}, the amount of wave mode conversion depends on the angle of propagation relative to the magnetic field and the equipartition layer scale height $H_e$ (see Equation \eqref{eq:transmission}).  In \S\ref{sec:wkb-solutions} we found that $H_e\approx 0.23\unit{Mm}$ for the equipartition layer surrounding the null.  Because the equipartition curve is approximately circular with radius $r_e=0.73\unit{Mm}$, we defined a radius of influence for the null as $R_n = r_e+H_e \approx 1.3r_e = 0.96\unit{Mm}$.  These distances are indicated in \figref{fig:phase}.  Based on Equation \eqref{eq:transmission}, we expect mode conversion to become important as waves reach this distance from the null.}

The general description of mode conversion \change{reported in the previous section} is supported by combining the ray tracing and numerical approaches.  \figref{fig:phase} shows several curves in the vicinity of the null, with the separatrices shown in black.  The blue vertical line located at $x=-1\unit{Mm}$ passes through the initial center of the introduced wavepacket.  The green curve is the path of a fast ray initialized at the lower boundary, with crosses showing equal phase distances $(\tau)$ or, equivalently, equal time differences in units of $t_N$, as the position of a phase point moves at the group velocity along the curve: $\vect{x}(\tau(t)) = \int_0^t \vect{v}_g(\vect{x}(t^\prime),t^\prime)dt^\prime$, \change{where $\vect{v}_g$ is defined below Equation \eqref{eq:hamilx}}.  The red curve is the path of a slow ray, with crosses indicating a phase point moving at the slow speed; it closely follows a magnetic field line.  The slow ray was initialized from a location along the fast ray path, indicated by the triangle at $\tau\approx 3.25t_N$ ($\vect{x}(\tau) \approx [-0.49,3.11]\unit{Mm})$.  This is where the fast ray reached one conversion--scale--height $H_e$ away from the equipartition contour.  The lengths of $H_e$ and $R_n$ are also indicated in the figure.

\begin{figure}[ht]
  \begin{center}
    \includegraphics[width=0.41\textwidth]{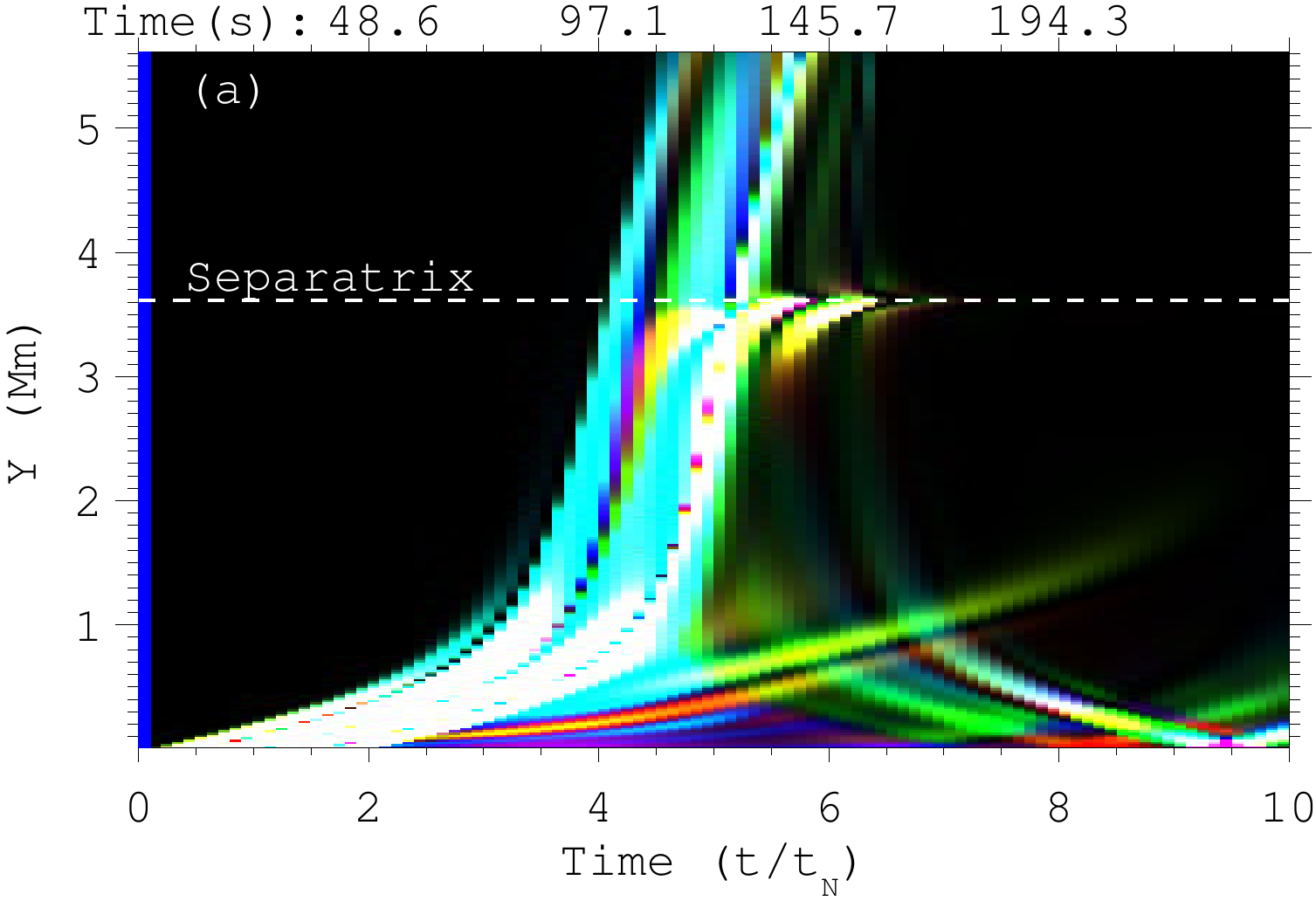}\hspace{2eM}

    \includegraphics[width=0.48\textwidth]{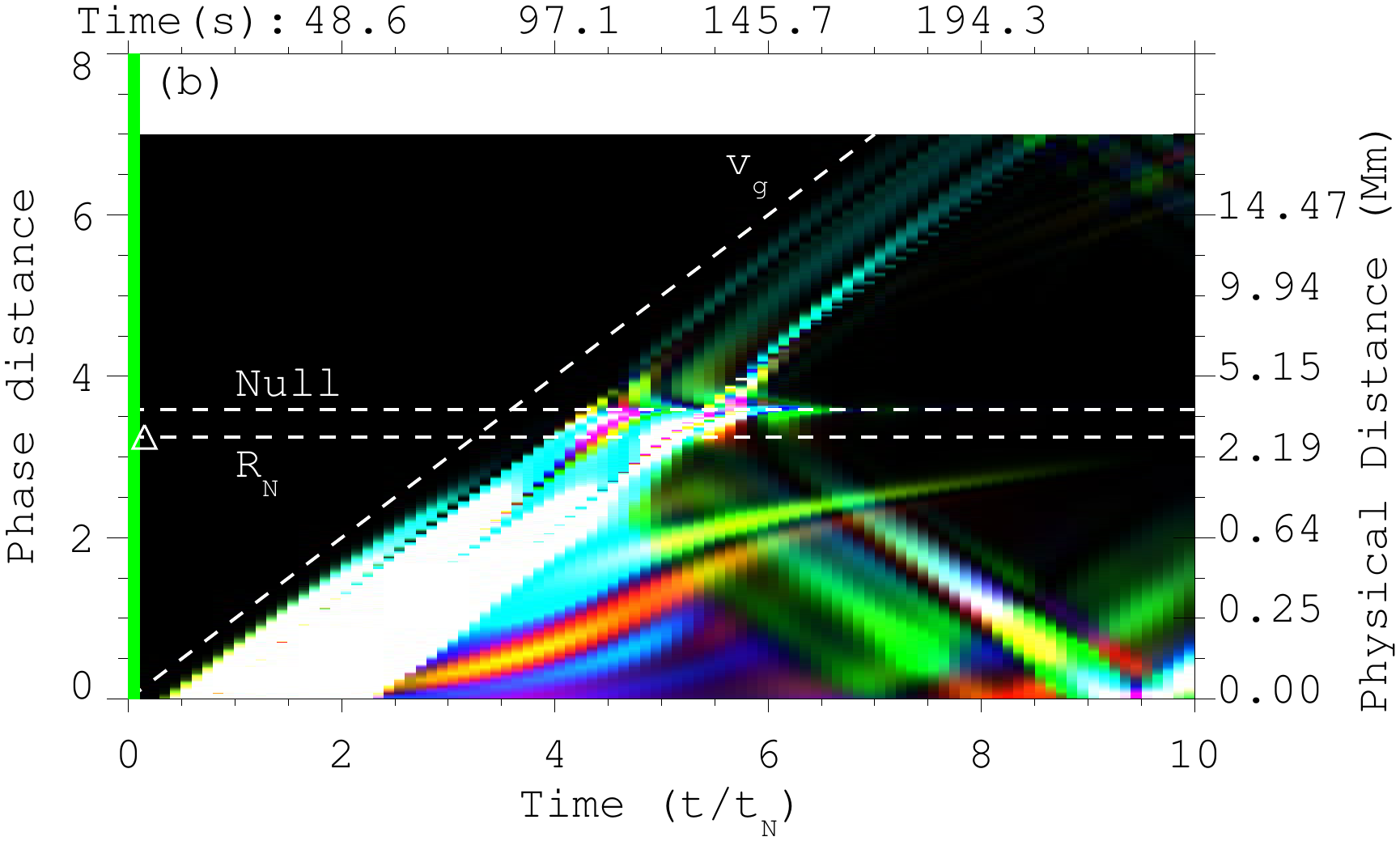}

    \includegraphics[width=0.48\textwidth]{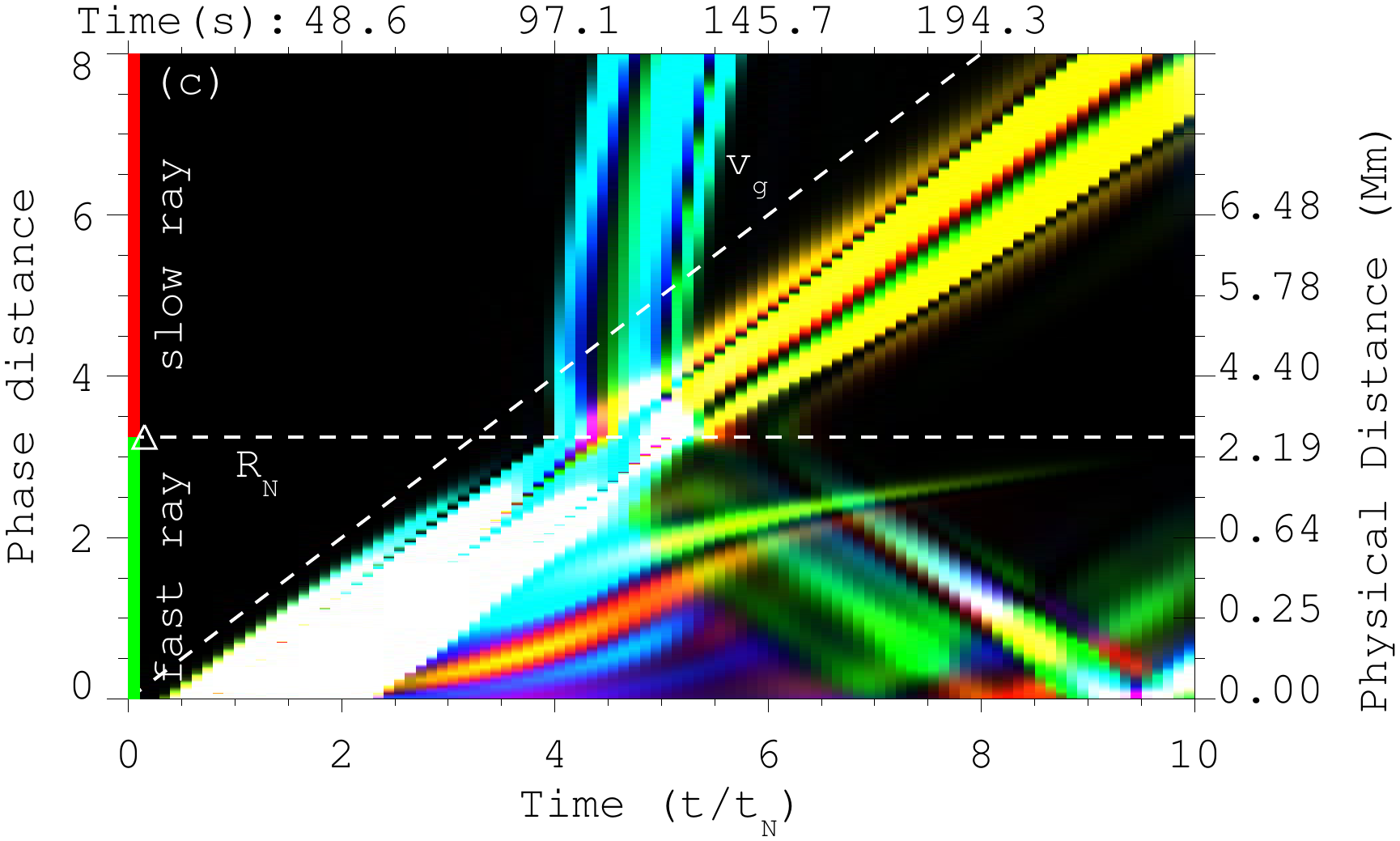}
    \caption{\label{fig:enrays}Energy densities along the 3 curves from \figref{fig:phase}:  (a) Vertical slice along $x=-1\unit{Mm}$ (blue curve, \figref{fig:phase}).  (b) Fast ray (green curve, \figref{fig:phase}) passing near the null point.  (c) Fast--to--slow conversion ray path, each ray traced using WKB approximation, then joined together (green then red curves, \figref{fig:phase}).  The color scheme and scale are the same as in \figref{fig:en-time}.  The right axis of (b) and (c) show distance in physical units along each ray, while the left axes show the phase distance.}
  \end{center}
\end{figure}
We extract the energy densities displayed in \figref{fig:en-time} along the blue, green, and red curves shown in \figref{fig:phase} at each simulation output time to generate the time--distance plots shown in \figref{fig:enrays}.  The color scheme and scaling is the same as in \figref{fig:en-time}.  In each of the panels, the abscissa measures time in units of $t_N$, and the ordinate measures distance along the curve.  For panel (a), distance is measured in $\unit{Mm}$, but for (b) and (c), distance is measured in phase coordinates.  For reference, the thick colored line at $t=0$ in each panel indicates the line--color of the line in \figref{fig:phase} along which the energy densities were extracted.

\figref{fig:enrays} (a) is the time--distance diagram of energy density along the blue vertical line at $x=-1\unit{Mm}$ of \figref{fig:phase}.  The initial shallow sloped portion at the bottom of the panel shows the upward propagating initial pulse (magnetoacoustic fast wave), moving at the relatively slow sound speed below the equipartition height at $y\sim 0.5\unit{Mm}$.  Some of the energy remains acoustically dominated, switching from the fast to the slow branch of the dispersion relation.  This behavior is visible as the shallow sloped green/red/yellow ridges beyond $t=5t_N$, which shows that a portion of the wave refracts and becomes downward propagating around $y=1.5\unit{Mm}$.  However, most of the energy remains on the fast branch and converts to a magnetically dominated disturbance.  This energy streams upward at a much faster speed starting around $t=3t_N$, due to the increasing Alfv\'en speed.  As the upward propagating pulse approaches the separatrix layer, indicated by the horizontal dashed line, we again see two distinct portions.  The cyan (magnetic, low--$\beta$ fast wave) portion passes across this topological barrier, while the yellow (acoustic, low--$\beta$ slow wave) portion is confined to field lines close to but underneath the separatrix.

Note that little of the behavior seen in panel (a) directly represents the physical energetics of the plasma.  The velocities are mostly phase velocities caused by different portions of the wavefront refracting across the vertical slice at different times.  In particular, the relation between the apparently upward propagating magnetic energy and the acoustic energy confined near the separatrix is unclear.  \change{On the other hand, vertical integration of the different energy channels of panel (a) would most closely approximate line--of--sight observations.  This illustrates the care required to interpret observations when one cannot assume the field is nearly homogeneous.}

In contrast, \figref{fig:enrays} (b) shows the energy densities as a function of time along the curved, fast--ray path shown in green in \figref{fig:phase}.  Distance is now measured as a phase through the time--integral of the group velocity along the curve, the $\vect{x}(\tau(t))$ defined above.  The white region at the top of the panel is due to this fast ray exiting the computational domain, and is included to keep a consistent scaling between panels (b) and (c).  The location of the start of the conversion region, $R_N$ (the triangle here and in \figref{fig:phase}), and the null point are each indicated by horizontal dashed lines.  There are two immediately obvious and important differences compared to panel (a).  i) Almost all the energy is confined to straight line paths with a slope equal to the local fast--wave group velocity (slope of 1 in the figure's units, diagonal dashed line).  ii) Almost none of the energy makes it beyond the null point.  The second point is true even though energy propagation past the null is allowed due to the finite plasma $\beta$, because, unlike for a cold plasma, the group velocity of the fast ray does not drop to zero.  Even so, it is evident that little energy actually follows this trajectory beyond the null.

\figref{fig:enrays} (c) shows the energy density time--distance plot for a hybrid curve of \figref{fig:phase}, where the slow ray (red) has been stitched on to the fast ray (green) at the phase position where the fast ray crosses the interaction region surrounding the null (triangle).  The lower portion of the plot, up to $t\approx 3.75t_N$ (horizontal dashed line), is the same for panels (b) and (c), with distance measured as phase position along the fast ray.  Above this, distance is measured in phase position along the slow ray.  Each portion is indicated by the red and green lines on the left of the plot.  The magnetically dominated energy propagating along the fast ray transitions smoothly to acoustically dominated energy propagating along the slow ray, starting around $t=4t_N$.  In both cases the slope is 1 (compare again to the diagonal dashed line), indicating that the energy propagates at the fast group velocity in the bottom portion and at the slow group velocity in the top portion of this plot.  Near the $v_g$ label is a steep sloped region of magnetically dominated energy density, also starting around $t=4t_N$.  It is fully contained in the slow ray portion of the diagram and is due to a separate portion of the fast wave front that refracts across the slow ray path.  It has a high phase speed as it crosses the red path because the phase front is expanding and crosses the slow ray line at a high angle (compare with the animation of \figref{fig:en-time}).  This is the outer portion of the fast--mode pulse that refracts back downwards towards the photosphere.

The panels of \figref{fig:enrays} only show the energy density along a select number of paths through the domain.  However, the fact that the rate of propagation all along the fast--joined--to--slow ray is precisely the local group velocity is a strong indication that at least some energy follows this approximate path.  This is, again, in sharp contrast to the first panel, which shows differences in propagation between the acoustically and magnetically dominated regions, and to the middle panel, which shows energy propagating along the fast ray originally, then abruptly vanishing at the null point.  

Energy densities extracted along other ray paths (not shown) all show similar results.  For instance, energy densities extracted for the 31 rays traced in \figref{fig:ray-angles} each show propagation of energy along the entire path, except those that pass within $R_n$ of the center of the null equipartition region.  Instead, the time--distance plots along those latter rays look similar to \figref{fig:enrays}(b).  This suggests that wave energy enters the region around the null as a fast mode, but it does not exit the region as a fast mode.  The energy must either convert to the slow mode or dissipate locally.  We will now attempt to determine the fate of this energy.

\subsection{Energy densities and fluxes near the null and in the corona}\label{sec:en-den-flux}

\begin{figure}[ht]
  \begin{center}
    \includegraphics[width=0.48\textwidth]{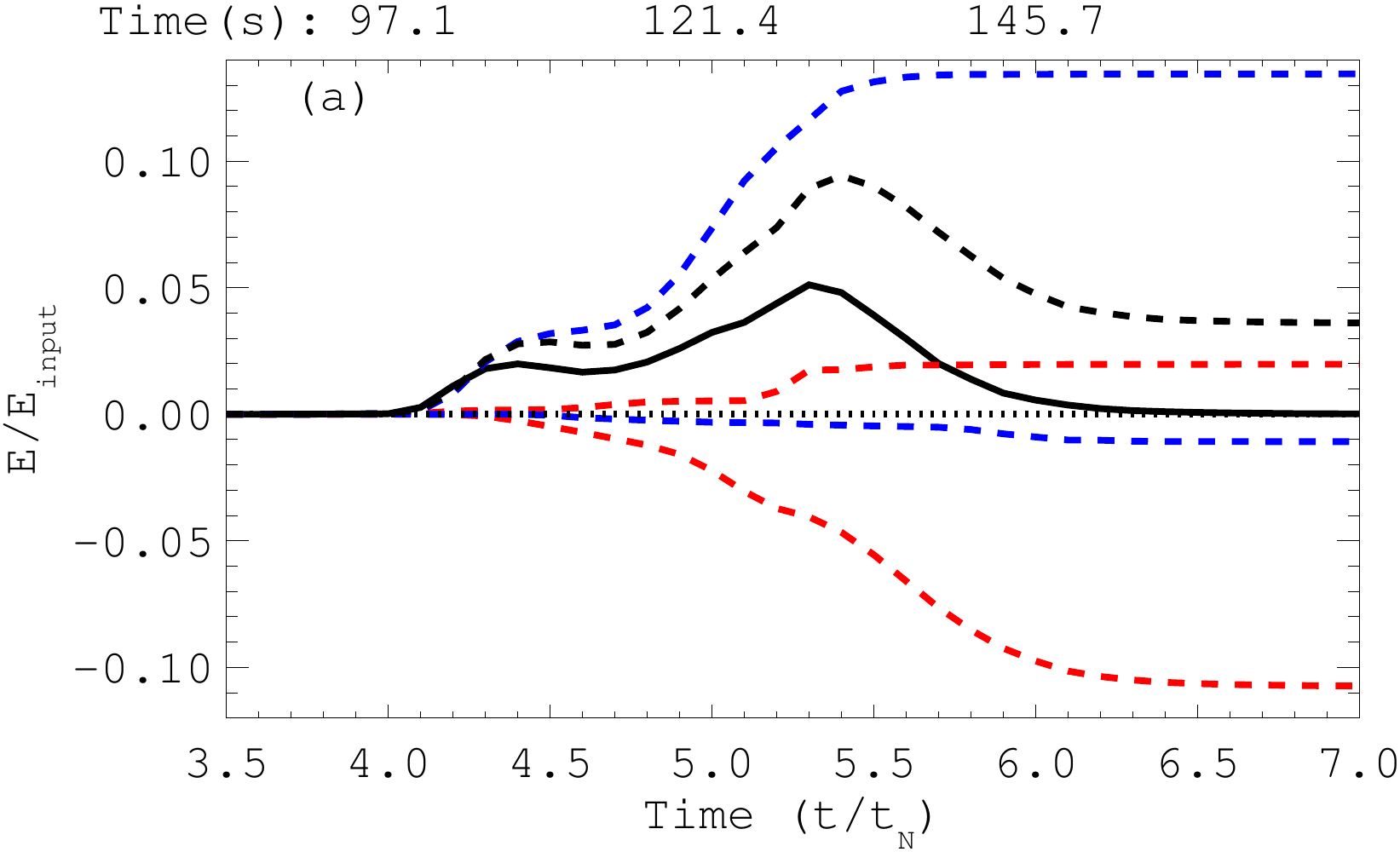}
    \includegraphics[width=0.48\textwidth]{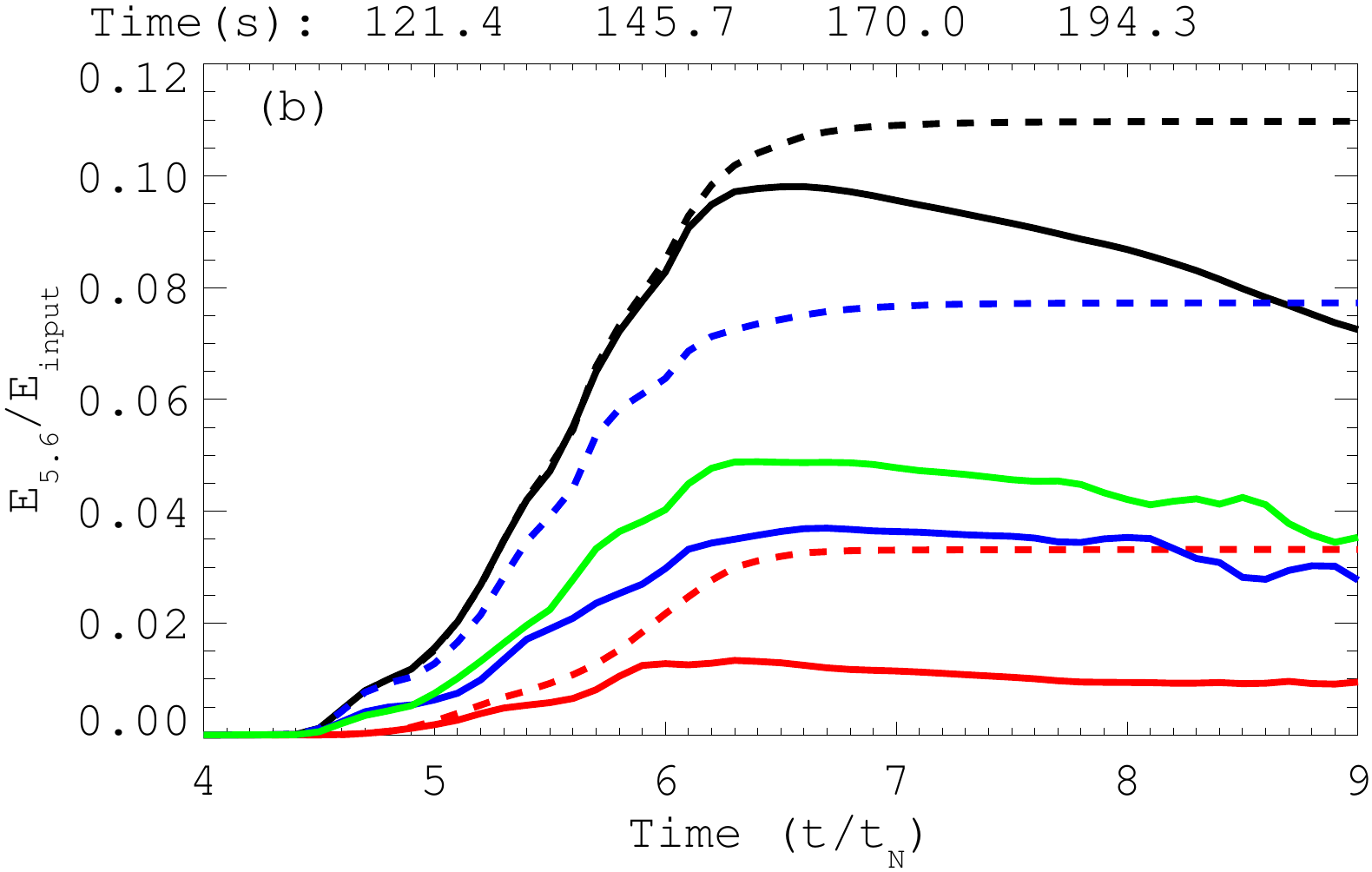}
    \caption{\label{fig:flux-null} Spatially integrated wave energy densities (solid curves), and temporally integrated fluxes (dashed curves), each as a function of time and normalized to the total energy of the initial wavepacket.  Red, blue, green, and black show acoustic, magnetic, kinetic, and total integrated quantities.  (a)  Integrated energy fluxes across and energy densities within the equipartition curve surrounding the null.  Dashed positive curves show inward flux (e.g., inward acoustic flux is positive dashed red), and negative dashed curves show outward flux.  The net flux of acoustic and magnetic terms through boundary is shown in black, and is positive.  (b) Integrated energy fluxes across and energy densities above the height $y\approx 5.6\unit{Mm}$.  The dashed curves are net fluxes for each term, including both upward and downward.  The downward fluxes are $<1\%$ of the upward fluxes at all times.  Note the difference in scale between (a) and (b).  See text for details.}
  \end{center}
\end{figure}

We can estimate the total amount of mode conversion in the region surrounding the null by integrating each term of \eqref{eq:wave-conservation}.  The energy inside a region $\mathcal{A}$ is given by the integral of the energy density terms at each time,
\begin{gather}
  W(t) = \int_\mathcal{A}\Bigl(E_{K}(t)+E_A(t)+E_M(t)\Bigr)d\mathcal{A},
  \intertext{while total flow of energy into or out of the region is given by the integral of the acoustic and magnetic energy fluxes across the boundary $\partial\mathcal{A}$}
  W_{flux}(t) = \int_0^t \int_{\partial \mathcal{A}}\Bigl(\vect{F}_{A}(t^\prime)+\vect{F}_{B}(t^\prime)\Bigr)\cdot \hat{\vect{n}}dl dt^\prime.
\end{gather}

Plots of $W(t)$ and $W_{flux}(t)$ for two different areas are shown in \figref{fig:flux-null}.  In panel (a), $\mathcal{A}$ is taken to be the circle of radius $R_n$ centered on the centroid of the $c_s=v_A$ region, resulting in $W(t)$ displayed as the solid black line in \figref{fig:flux-null}(a), normalized to the total wave energy introduced through the lower boundary, $E_{input}$.  A fiducial at $E/E_{input}=0$ is marked by the dotted line.  We decompose the net flux into inward and outward terms, with inward magnetic (acoustic) shown as positive dashed blue (red) and outward shown as negative.  About $15.5\%$ of the injected wavepacket's total energy crosses the null's equipartition boundary (blue and red positive dashed curves).  As expected, most of the energy arrives as a Poynting flux, consistent with the wave being predominantly a fast--mode wave in the $\beta<1$ region.  However, most of the energy leaving the null area is acoustic in nature, and slightly lags the inward directed Poynting flux, showing that mode conversion has taken place.  The conversion is fairly efficient, with a ratio of $0.7$ between the net acoustic flux and net Poynting flux across the boundary.  Of the energy that passes into the conversion region, $7\%$ appears to exit in a form consistent with how it entered.  The remaining $23\%$ is energy that enters the region but never exits: this energy heats the plasma through Ohmic and shock dissipation, or is lost due to uncaptured numerical diffusion.  We briefly describe the current accumulation at the end of this section.

We can also calculate the energy that continues propagating upward into the corona by considering a region $\mathcal{A}$ in the upper portion of our domain.  In \figref{fig:flux-null}(b), we show the results of computing the net flux across and energy above $y = 5.6\unit{Mm}$, which is above the transition region but below the upper damping region.  Here, solid lines show energy density terms integrated over the entire domain above $y= 5.6\unit{Mm}$ at each time, and dashed lines show the temporally integrated net fluxes across that boundary, up to each time.  The downward fluxes are $\lesssim 1\%$ of the upward fluxes, which is why we do not independently display them here, as we did in panel (a).  Spatially integrated acoustic, kinetic, and magnetic energy densities are shown in solid red, green, and blue, respectively, while the Poynting and acoustic fluxes are shown in dashed blue and red, respectively.  

As briefly discussed in \S\ref{sec:sim-gen}, we do indeed find that the kinetic energy is typically in equipartition with the sum of the acoustic and magnetic energies, i.e., the solid green curve is roughly the sum of the solid red and blue curves.  The exception is when the wavepackets begin to interact with the boundaries, causing the matched oscillations seen around times $t=8-9t_N$ in \figref{fig:flux-null}(b).  The kinetic energy also appears proportionally allotted between the acoustic and magnetically dominated waves.  In other words, each type of wave appears to be in equipartition, independently.  For instance, the Poynting flux (dashed blue) accounts for both the magnetic energy density and its associated kinetic energy, so that the dashed blue line is essentially double the solid blue line; the same is true for the acoustic terms in red.  Then, the second half of the dashed blue and red lines come from the allotted proportions of the solid green line.  \change{Because these are integrated fluxes, this plot only demonstrates equipartition between kinetic and potential energies in a spatially averaged sense.}

The integrated flux across the boundary and energy above the boundary are two independent measures of the total wave energy above $y=5.6\unit{Mm}$.  These are shown in black, and match very well until they begin to diverge around $t=6t_N$.  This is expected, as at that point the waves reach the damping regions at the edges of our computational domain and eventually begin to exit the domain through the upper boundary.  Because the Alfv\'en and sound speeds are fairly uniform above this height, the waves undergo little additional refraction.  The total integrated flux across the height $y=5.6\unit{Mm}$ is thus the total proportion of energy from the initial wavepacket that we estimate continues propagating upward into the model corona.  This accounts for $11\%$ of the initial energy of the wavepacket, with $3.3\%$ associated with a slow magnetoacoustic wave and $7.7\%$ with a fast magnetoacoustic wave.

Essentially all of the acoustic flux across $y=5.6\unit{Mm}$ originated from mode conversion at the null point.  \figref{fig:flux-null}(a) shows that $\approx 11\%$ of the total energy left the null point region as an acoustic wave.  It appears that, at least for this simulation, the converted energy that leaves the null is roughly equally distributed along each of the null's four separatrices, with perhaps a slight bias towards the upward leg.  The animation of \figref{fig:en-time} appears to qualitatively support this conclusion.

On the other hand, the Poynting flux leaving the null accounts for just $\approx 1\%$ of the total energy, while $7.7\%$ was found to cross $y=5.6\unit{Mm}$.  At least $6.7\%$ must therefore be due to a portion of the fast mode wave that refracts around the null but does not enter the null's mode conversion region.  This would be the portion of the wave front discussed in \S\ref{sec:ray}, such as rays initiated between the green and yellow curves of \figref{fig:ray-paths}(b).  That figure only shows rays initiated with $k_x=0$.  An analysis similar to one resulting in \figref{fig:ray-ca} could be made to determine the distribution of rays initiated at the lower boundary that exit the top of the system.  We will not perform that analysis here, but simply note that a structured magnetic field can allow some fast wave energy to propagate into the corona\change{ whereas for a uniform field it would eventually refract back to the photosphere.}

\subsection{Current accumulation near the null}\label{sec:jz}
\begin{figure*}[ht]
  \begin{center}
    \includegraphics[width=\textwidth]{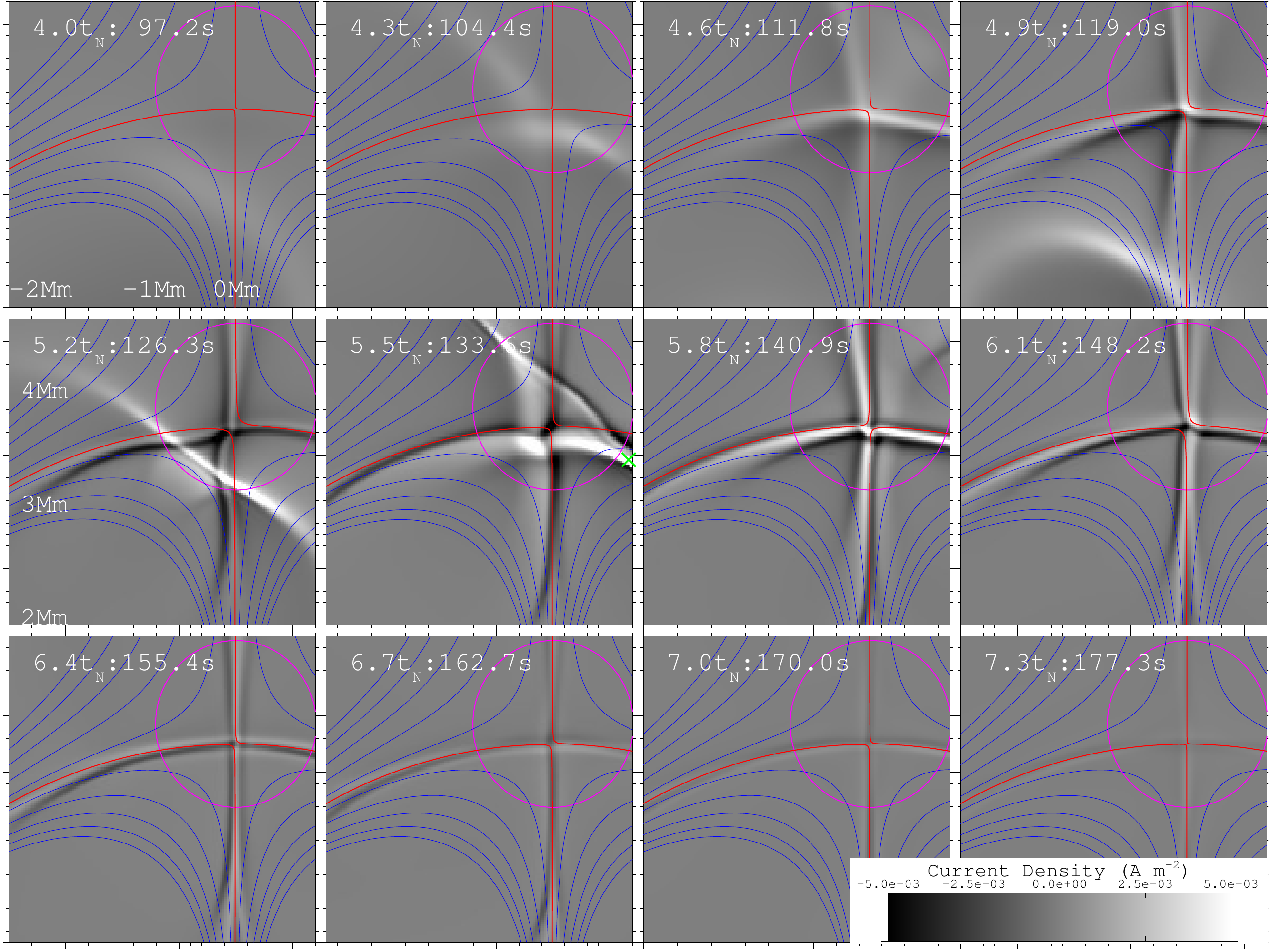}
    \caption{\label{fig:current-time} Time sequence of current density (grayscale) and field lines (blue) for a region near the null.  Magenta shows the $c_s=v_A$ contour, and dashed--red lines are the separatrices.  The top left and middle left panels shows the horizontal and vertical spatial scales, respectively.  Time is marked in both normalized and physical units.}
  \end{center}
\end{figure*}

In the previous section we found that $23\%$ of the energy that enters the null point region is never measured to exit it.  This is likely due to the localization of currents to the null point and separatrices and the subsequent dissipation of those currents.  \figref{fig:current-time} shows a time sequence of the out--of--plane current density (grayscale) around the null.  White shows positive (out of page) and black negative (into page) currents.  Blue lines are magnetic field lines, red lines are the approximate locations of the separatrix field lines, and the magenta curve again marks the $c_s=v_A$ equipartition contour.  The grayscale is saturated at $\pm5\times 10^{-3}\unit{A}\unit{m}^{-2}$.  The maximum current density is $\abs{j_z}\approx0.016\unit{A}\unit{m}^{-2}$ at $t=5.5t_N$, marked in the figure by the green cross (this is the greatest current density for a larger field of view than shown in the figure, as well).  

The first pulse of the wavepacket passes through the null point region around $4t_N$.  This stresses the field and leads to regions of current density localized along each separatrix, forming a set of current ribbons by $4.6t_N$.  The second pulse of the wavepacket steepens to form a shock as it approaches the null, around $5.2t_N$.  \change{This appears to be the fast oblique magnetic shock reported by \citet{McLaughlin:2009}, and is accompanied by a strong spike in the current density, as found by \citet{Gruszecki:2011} and \citet{Afanasyev:2012} and references in those papers.  We will defer details of shock formation and the development of the current sheet to a later paper.}

The null point collapses to form a current sheet originally oriented at $\sim45^\circ$ to the separatrices.  Next, a further accumulation of current density is evident in the ribbons which form along each of the separatrices.  The transverse (cross--field) length scale of each ribbon decreases until it reaches the diffusion scale, $\ell = L_N/S$, at which point the ribbons Ohmically dissipate.  This is most easily seen for $t>5.8t_N$.  The reduction in current density is not associated with plasma flows or energy fluxes.  This dissipation accounts for the $23\%$ of the energy that enters the equipartition region but does not leave, quoted above.

The current sheet later collapses again at $-45^\circ$ to the separatrices, rotated by $90^\circ$ to the original collapse direction.  The process repeats itself, with apparent oscillatory reconnection at the null.  This type of behavior has been observed in other studies \citep{McLaughlin:2009, Murray:2009, McLaughlin:2012a, McLaughlin:2012b}.  A more detailed discussion of the formation of the current sheet, its dissipation, the oscillatory reconnection, and how each depends on the initial properties of the wavepacket, will be presented in a future work.

\section{Comparison of the WKB and numerical simulation results}\label{sec:driver-distribution}
As we have noted above, the total energy introduced to the system by our wavepacket is distributed over a range of initial propagation angles at each initial spatial location.  We would like to estimate the total power carried by the wavepacket in each of these directions, and see how the WKB estimate of where the energy ends up, determined by following sets of rays, compares to the results of the numerical simulation.  This approach has recently gained traction for understanding propagation through model sunspots \citep[see][and references therein]{Felipe:2012} when the field gradients are small enough that the analytic predictions may approximately apply.  However, it has also been used to interpret observations from regions with more complex topologies where the homogeneous theoretical predictions may apply less well \citep{Stangalini:2011, Kontogiannis:2014}.  Our work allows a critical comparison to be made.

The nominal wavevector of the wavepacket described by Equation \eqref{eq:vdrive} is $\vect{k} = k_y\hat{\vect{y}}$.  However, any horizontal gradients will produce nonzero horizontal modes with finite values of $k_x$.  We can approximate the distribution by Fourier transforming \eqref{eq:vdrive} in the $x$ direction (this approach ignores the effect of horizontal variations in the background magnetic field).  The resulting spectrum is simply a Gaussian in $k_x$,
\begin{gather}
  \tilde{v}(k_x) = \frac{1}{2\pi}\int_{-\infty}^\infty v_d(y,t)  e^{-\frac{(x-x_d)^2}{2w_x^2}}e^{ik_xx}dx \\
  \label{eq:power-kx}=v_d(y,t)e^{ik_xx_d}\frac{w_x}{\sqrt{2\pi}}e^{-\frac{1}{2}k_x^2w_x^2},
\end{gather}
where we have combined the amplitude and temporal/vertical behavior into $v_d(y,t)$.  As always, a more localized pulse (smaller $w_x$) requires larger horizontal wavenumbers.  Parseval's theorem relates the signal power to the spectral power, $\int_{-\infty}^\infty \abs{v(x)}^2dx = \int_{-\infty}^\infty \abs{\tilde{v}(k_x)}^2dk_x=\int_{-\infty}^{\infty}P(k_x)dk_x$.  $P(k_x) = \tilde{v}\tilde{v}^*$ is the power spectral density and $^*$ denotes complex conjugation.  Inserting Equation \eqref{eq:power-kx} and normalizing so that $\int P(k_x)dk_x = 1$, we find that the power spectral density in each horizontal mode is $P(k_x) = \frac{w_x}{\sqrt{\pi}}e^{-k_x^2w_x^2}$.  This is our estimate of the distribution of $k_x$ due to the finite horizontal width of our wavepacket.  In order to relate this to an initial range of ray propagation directions, we change variables to $\chi$ using the relation $k_x = k_y\tan\chi,$ with $k_y$ fixed.  Doing so, we find that the power into $d\chi$ at angle $\chi$ is
\begin{equation}
  \label{eq:p-dchi} P(\chi) = \frac{w_xk_y}{\sqrt{\pi}}e^{-w_x^2k_y^2\tan^2\chi}(1+\tan^2\chi).
\end{equation}
\change{This distribution differs from a Gaussian peaked at $\chi=0$ by the factor $(1+\tan^2\chi)$, and has the effect of shifting power to angles slightly away from the vertical, relative to a Gaussian distribution.  For our parameters, the difference is extremely small: approximately $0.6\%$ of the total power is redistributed to greater angles.  Substituting }the values of our wavepacket, $w_x=0.296\unit{Mm}$ and $k_y = 30\unit{Mm}^{-1}$, shows that the $1\sigma$ level of the distribution is $\chi\approx \pm12.5^\circ$ about the vertical.

Next, we use these distributions to model a wavepacket as a bundle of rays and estimate the proportion of the wavepacket's initial energy that reaches the null point region.  The estimation depends on three factors: i) the distribution of the wavepacket's power in initial location $x$; ii) the distribution of the wavepacket's power in initial propagation angle $\chi$; and iii) the range of angles at a given initial location $x$ for which rays pass within $R_n$ of the null.  For i) and ii), we assume the wavepacket's power is separable in $x$ and $\chi$, i.e., $P(x,\chi)=P(x)P(\chi)$, so that it has the same distribution of $\chi$ for each $x$, and only the relative amplitude varies in $x$.  $P(\chi)$ is given by Equation \eqref{eq:p-dchi}, and $P(x) = \frac{1}{w_x\sqrt{\pi}}e^{-(x-x_d)^2/w_x^2}$.  Again, each power density function is normalized so that $\int P(x,\chi) = \int P(x)\int P(\chi) = 1$.  For iii) we use the green contours of \figref{fig:ray-ca} to define the position--dependent limits $\chi_1(x)$ and $\chi_2(x)$.  The total power directed towards the null is then
\begin{gather}
  P_{null} = \int_{-\infty}^\infty dx \int_{\chi_1(x)}^{\chi_2(x)} d\chi P(x) P(\chi)
  \intertext{which, after taking the $\chi$ integral, is}
    \label{eq:px}P_{null}= \int_{-\infty}^{\infty}\frac{1}{\sqrt{\pi}w_x}e^{-\frac{(x-x_d)^2}{w_x^2}}\frac{1}{2}\bigl[\erf(k_yw_x\tan\chi_2(x))-\erf(k_yw_x\tan\chi_1(x))\bigr]dx,
\end{gather}
where $\erf(x)$ is the error function.  Note that $P_{null}$ is, implicitly, a function of the wavepacket parameters $x_d$, $w_x$, and $k_y$.

\begin{figure}[ht]
  \begin{center}
    \includegraphics[width=0.48\textwidth]{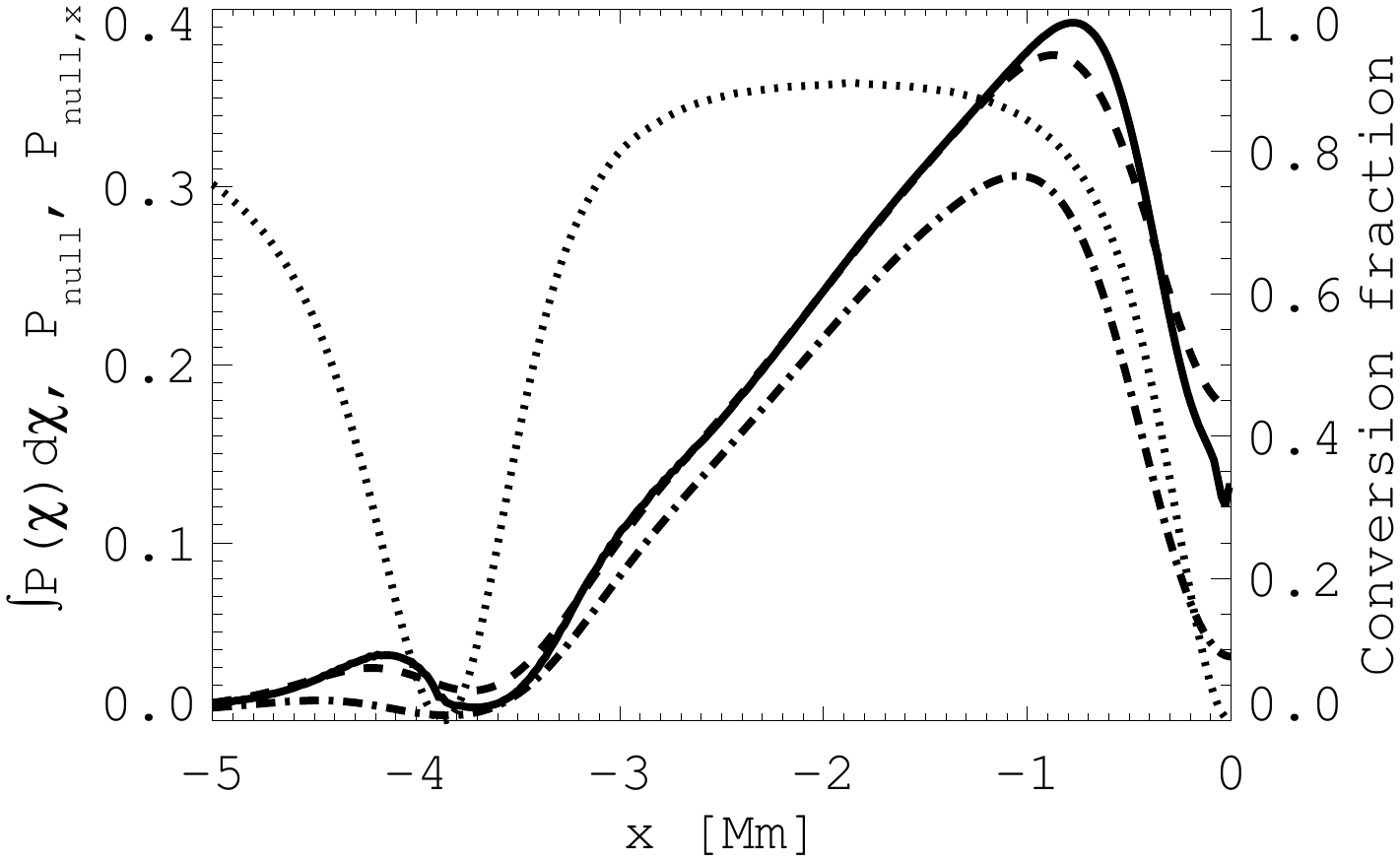}
    \caption{\label{fig:power-th} Solid curve: proportion of wave energy at each $x$ that passes within one interaction distance $R_n$ of the null (see text).  Dashed curve: estimate of the wave packet energy that approaches the null, as a function initial injection location, assuming no conversion.  Dash--dot: estimate of wave packet energy that approaches the null as a function of injection location, assuming conversion at the lower $c_s=v_A$ location.  Dotted: Conversion fraction (right axis) estimated using Eq.~(26) of \citet{Schunker:2006}.}
  \end{center}
\end{figure}

The solid curve of \figref{fig:power-th} plots $\int_{\chi_1(x)}^{\chi_2(x)}P(\chi) d\chi$ as a function of $x$ for the wavepacket parameters ($w_x,\ k_y$) described in \S\ref{sec:driver}.  It shows, for each $x$, the total proportion of the energy in the range of propagation angles that pass within one $R_n$ of the null.  The function is symmetric about $x=0$, and we only plot $x<0$ here.  By multiplying this function by the spatial power distribution of the wavepacket and integrating over the lower boundary, we can estimate the proportion of the total injected energy that makes it to the interaction region surrounding the null, $P_{null}$.  Performing the calculation when $x_d = -1\unit{Mm}$, we find $P_{null} = 0.37$, which is the value of the dashed line in \figref{fig:power-th} at $x=-1\unit{Mm}$.  The entire dashed line is the result of repeating the integration for different values of the initial wavepacket centroid location $x_d$ to create the function $P_{null}(x_d)$.  It is therefore a convolution of the solid curve with the Gaussian describing our wavepacket, and it shows the fraction a wavepacket's energy that will reach the null point as a function of the packet's initial centroid location.

In \S\ref{sec:en-den-flux} we found that $\sim15.5\%$ of the initial wavepacket's energy reached the region surrounding the null, less than half of the ray theory estimate.  However, the WKB estimate that $37\%$ of the injected energy should reach the region around the null represents an upper limit for two reasons: it assumes that (i) all of the energy remains on the fast--mode branch of the dispersion relation, and (ii) that only the impact parameter $d$ of the ray to the null matters, not the angle of approach.  Relaxing each assumption potentially reduces our estimate for the amount of energy that can either reach the null or mode convert in its vicinity.  We begin by accounting for our imperfect boundary driving.  \change{As mentioned in \S\ref{sec:driver}, part of the wavepacket driver immediately couples to the high $\beta$ slow mode.  Actually determining how much is complicated by the fact that the magnetic energy tends to ``pile up'' relative to the acoustic energy, due to the difference in the fast and slow group velocities.  This makes a simple ratio of the two quantities misleading at the driving location.  By studying the energy densities and fluxes near the injection site, we estimated that $\gtrsim90\%$ of the energy goes into the fast mode.  To be conservative, we will therefore reduce our estimates by $10\%$ at the end of this calculation, to account for this.}

We use the transmission formula Equation \eqref{eq:transmission} to estimate the amount of conversion from the fast to the slow mode at the lower equipartition region.  Applied here, $H_e$ refers to the scale height of the lower equipartition layer, measured along each ray path.  This formula has recently enjoyed broad application in interpreting observations of photospheric and chromospheric oscillations and their relation to the local magnetic field \citep{Stangalini:2011}.  The formula is for the transmission of a high--$\beta$ fast acoustic wave to a low--$\beta$ slow acoustic wave.  Its complement, $C=1-T$, gives the conversion coefficient for a high--$\beta$ fast acoustic wave to a low--$\beta$ fast magnetic wave, under the assumption that there are no reflections.  We take $k=k_y$, $\theta=\theta(x)$ the inclination of the magnetic field along the lower boundary, and $H_e=H_e(x)$ the equipartition scale height measured separately for each vertical column.  The conversion coefficient calculated for each $x$ is shown as the dotted line in \figref{fig:power-th}.  Performing the integration for our standard wavepacket parameters, we find that $P_{null,c}=\int\int P(\chi)P(x)C(x)d\chi dx = 0.31$, so that $6\%$ less of the initial wavepacket's energy is expected to reach the null, compared to the case of full conversion.  \change{Including the initial $10\%$ loss to the slow mode due to imperfect driving, we find $P_{null,c}=0.28$.  The new estimate for the ray theory is about $80\%$ greater than the amount of energy to reach the null found from direct numerical simulation.}

Finally, we repeat the above calculation to account for conversion for wavepackets introduced at any location $x_d$.  The result shown as the dash--dotted line of \figref{fig:power-th}.  For example, by this estimate, a packet initiated directly underneath the null point ($x_d=0$) will have $P_{null}=0.036$.  In that case, the field at the lower equipartition region is nearly vertical, and from \citet{Schunker:2006} Equation (26) we expect a high degree of acoustic transmission.  The transmitted rays are slow rays in the low $\beta$ portion of the domain and are strongly constrained to follow sets of field lines that diverge from the null point.  We therefore expect little energy to reach the region around the null.

Conversion at the lower equipartition layer therefore has a significant effect on the amount of energy that reaches the null, but even when conditions are most unfavorable we still expect some of the injected energy ($3.6\%$ in the case just discussed) to reach the upper conversion region.  We repeated the numerical simulation described in \S\ref{sec:results} using $x_d=0\unit{Mm}$ for the centroid of the injected wavepacket and found that the numerical results qualitatively support the conclusions from the ray theory.  Furthermore, a set of simulations in which we varied $x_d$ reproduced the general trends of \figref{fig:power-th}.  However, a detailed comparison between the full set of simulations is outside the scope of the present work.

\section{Discussion}\label{sec:discussion}
\change{We have combined a numerical simulation with a WKB method to study wave propagation in a stratified atmosphere containing a magnetic null.  The presence of the null creates strong gradients in the field that substantially modifies wave behavior compared to more slowly varying fields \citep{Khomenko:2009a, Rijs:2015}.  On the other hand, our stratified atmosphere, finite $\beta$, and boundary driving distinguishes our simulations from most studies of MHD waves near null points \citep[see review by][]{McLaughlin:2011}.  We therefore studied how two previously known effects, refraction and mode conversion, combine to modify the wave behavior in the low solar atmosphere.  Crucially, we were able to quantify each effect in terms of the energy of a wave packet.}

\change{Our quantification of mode conversion at the null in terms of wave energy is a novel result.  We found that conversion at the null between incident magnetic fast waves and exiting acoustic slow waves was very efficient, at about $70\%$.  This result leads to important differences with wave--null interaction studies based on the linear theory.  Those studies rely on reflection and refraction of the fast wave back towards the null point to dissipate energy at the null in logarithmic time \citep{Craig:1991, Hassam:1992, Longcope:2007, Longcope:2012}.  Based on our result, we expect that extending the linear analysis of \citet{Longcope:2012} to include pressure forces and mode conversion will show that reflected waves are dominated by slow waves of azimuthal mode $m=4$, concentrated to the four separatrices.  This may substantially reduce the amount of wave energy dissipated at the null, depending on the size of the equipartition region.}

Several details of the present and related simulations are beyond the scope of the current paper, but are worth summarizing here.  We have run similar simulations with different numerical resolutions, transition region heights, and wavepacket properties ($x_d, w_x, \omega_d$), that collectively support the conclusions reached in this paper.  Perhaps the most striking is that when the packet's injection location is varied, the energy that reaches the null follows the pattern from the ray theory shown in \figref{fig:power-th}.  The combination of mode conversion at the lower $c_s=v_A$ height, refraction towards the null due to the varying fast mode speed, and subsequent conversion near the null is responsible for this effect.  The wavepacket we studied in this investigation was situated to maximally refract towards the null, as determined by the WKB method.  Therefore, the direct dissipation at the null is likely an upper bound, at least for the high frequency ($\sim40\unit{mHz}$) waves treated here.  A fuller accounting of the shock formation, current accumulation and dissipation, and how the energetics vary with the parameters of our system will be presented in a follow up paper (Tarr et. al. 2017, in preparation).

\subsection{Presence of nulls on the Sun}\label{sec:null-presence}
\change{A natural question is how often one expects a low lying null, so critical to our investigation, to arise on the Sun.  }Current estimates for the number of nulls, and the strong gradients in the Alfv\'en speed that come with them, give roughly one per supergranular cell above $1.5\unit{Mm}$ \citep{Close:2004,Regnier:2008,Longcope:2009}, and likely more below that height.  \citet{Longcope:2009}, for instance, used a spectral method to determine that about 19000 nulls with a height above $1.5\unit{Mm}$ are present due to the quiet sun field at any given time, about 1 null per $300\unit{Mm}^2$.  Roughly half of these occur between $1.5$ and $4.5\unit{Mm}$, and the statistics vary only by some $10\%$ over the two solar minima and 550 magnetograms considered.  Since high moments drop off more rapidly with height, they expect more nulls below $1.5\unit{Mm}$, but the noise threshold of MDI magnetograms, the basis of their analysis, prevents estimates below that height.  \citet{Freed:2015} provided an observational study of the corona using AIA data to determine the distribution of nulls with height, and found rough agreement with the work cited above, for the heights they were able to observe.

The presence of many nulls between the photosphere and the transition region makes it likely that the physical processes we describe in this paper are rather common, namely, that convectively driven waves pass through multiple conversion sites as they propagate.  This idea is supported by other simulations.  \citet{Nutto:2012} simulated a convectively unstable network region that did not contain, \emph{a priori}, a null point.  They found that multiple overlaying equipartition regions form self consistently.  Some even have the roughly circular shape that may indicate a null, though the authors did not specifically check for the presence of one (see their Figure 3, panel $t=200\unit{s}$).  They found evidence for multiple conversions between fast and slow, acoustic and magnetic pulses at each of the equipartition regions, and found that a slow mode acoustic wave is able to propagate outward beyond the chromosphere, as did we.

\change{We therefore expect the processes we describe in this paper to be applicable to regions of plage, network, and quiet sun.  We anticipate that the relative importance of refraction and mode conversion will vary based on the relevant length scales: the height of the null, distance between magnetic concentrations, and the size of the null's equipartition region.  Future studies are required to determine which effects dominate in different atmospheric regions.  The present simulation corresponds most nearly to a 2D cut through a small ephemeral region or strongly enhanced network flux.}

\change{Active regions present a different challenge.  The wave fields in quiescent active regions with highly symmetric umbra and penumbra are likely well described by slowly varying sunspot models, as evidenced by the good agreement between observations and simulations of acoustic halos \citep{Rajaguru:2013,Rijs:2016}.  Strong gradients in the magnetic field in general, and nulls points in particular, probably play a more important role in active regions with more complicated structure.  Nulls are found above the photosphere in some active regions, but not the majority; at the same time, active regions with nulls are much more likely to flare than those without \citep{Barnes:2007}.  Coronal nulls have separators connected to them, and photospheric driving will localize currents to these separators \citep{Longcope:2001,Parnell:2008}.  Focusing of wave energy at these locations may provide a way to destabilize these current sheets, accounting for the increased rate of flares, or even sympathetic eruptions \citep{Gruszecki:2011}.  Energy leaving the reconnection site as a slow mode will be guided and focused along the separatrices to the photospheric foot points, the flare ribbons, which we will briefly consider below.}

\subsection{Comparison with studies of nulls}
\change{Our analysis is most closely related to other studies of MHD waves near nulls, and here we review those results related to our own.  The key properties for comparison are that our magnetic field is nonlinear, our plasma has finite $\beta$, and our density is nonuniform.  The most well--studied situation is a linear null for which $\beta=0$ and the density is uniform.  In that case, wave energy accumulates at the null because of refraction, as seen, for instance, by the inward spiral of rays \citep{McLaughlin:2004}.  Focusing of the wave causes an enhancement in current density at the null, which then dissipates to locally heat the plasma.  By repeated reflections between the null and an external conducting boundary, all of a wave's initial energy is dissipated at the linear null \citep{Craig:1991,Hassam:1992}.}

\change{When the field around the null is nonlinear, saddle points and extrema in the phase speed will cause the phase front of an incoming plane wave to split, as reported by \citet{McLaughlin:2006a} and seen in our work for the rays (\figref{fig:ray-paths}) and simulation output (\figref{fig:en-time}).  \citet{Longcope:2012} determined the energetic consequences of this for a $\beta=0$, uniform density, quadrupolar field, and found that only $\sim40\%$ of the energy dissipated at the null, while the rest propagated away to infinity as a fast wave.  Note that those authors studied a wave initiated at the null, rather than one initially propagating towards the null; later reflections carried only part of the wave back to the null.  The difference between the quadrupole and linear null is the introduction of a new length scale, the distance over which the linear term in the Taylor series expansion of the magnetic field about the quadrupole null is valid.  An incoming wave that is planar on that scale will refract towards the null, and the rays describing it will form the usual logarithmic inspirals.  Waves on larger scales will split and partially refract away, taking energy with them.}

\change{Finite $\beta$ introduces the slow mode wave and allows a coupling between the fast and slow modes.  \citet{McLaughlin:2006b} is the only other study we are aware of that attempted to quantify conversion near a null, so it is worth discussing their work in detail.  They studied a linear null and varied a parameter $\beta_0$ that set the size of the equipartition region surrounding their null.  As they point out, because the field of the linear null has no inherent length scale, by changing $\beta_0$ they effectively changed the distance between equipartition layer and the initial fast plane wave launched from the boundary.  When $\beta_0\ll1$, the equipartition region is small compared to the extent of the wave front (or size of the computational box), and this mimics the $\beta=0$ case: the wave refracts and the rays spiral inwards.  As $\beta_0$ increases, the size of the equipartition region increases, and the wave is initiated closer to it.  Once $\beta_0$ is large enough, $c_s>v_A$ everywhere in their domain, which corresponds to the wave being initiated inside the equipartition layer.}

\change{To estimate the amount of conversion, \citet{McLaughlin:2006b} initiated a strictly fast wave pulse in the $\beta>0$ region and noted where it completely separated into two pulses (fast and slow) within the $c_s=v_A$ contour.  They calculated the integral of the perpendicular velocity inside the slow wave pulse, which is a quantity related to the wave momentum (assuming the wave is completely polarized transverse to the field).  They found the ratio of this quantity to the integral of the perpendicular velocity over the initial pulse for different values of $\beta_0$.  The ratio is roughly proportional to $\beta_0$.  This is because increasing $\beta_0$ effectively initiates the wave closer to the null, so that a larger fraction of the wave refracts into the equipartition region where it can convert.  They could therefore identify the competing effects of refraction and conversion, but not study them further.}

\change{The most conversion \citet{McLaughlin:2006b} report is $\sim25\%$ for $\beta_0=10$, which gives the maximum size their equipartition region can be while still fitting inside the domain.  A peak conversion for this case is somewhat surprising, since conversion should begin roughly one scaleheight $H_e$ outside the equipartition boundary, and their initial pulse is inside that region.  It is possible that their use of the perpendicular velocity is no longer a good approximation for this case.  Their estimate is difficult to directly compare to our own because they used a momentum--related quantity and only calculated it when the waves entered the $c_s=v_A$ boundary, not when they exited.  In contrast, our estimate uses wave energy and relates the total incoming to total outgoing energy of each type.  We feel our approach is more physical, and allows us to quantify local dissipation.  Considering their results for different $\beta_0$ suggests that we should perform a set of our own experiments in which the height of the transition region is varied while the magnetic field and wavepacket parameters are held fixed.  This will change the size of the equipartition region at the null, but not the total initial wave energy, and can therefore be used to determine the relative importance of refraction and mode conversion for a given driver.}

\change{Refraction clearly dominates when $\beta<1$, and its dominance in much of the corona as has led multiple authors \citep{McClements:2004, McLaughlin:2009, Threlfall:2012}, including \citet{McLaughlin:2006b}, to argue that an azimuthally symmetric pulse will be a good approximation for waves approaching nulls.  However, we did not find this to be the case in our simulation, as can be seen around $t=4t_N$ in \figref{fig:en-time} and the animation of that figure.  This is because our wave front was of comparable size to the null's equipartition region, and because the density and temperature are not uniform around the null, so that the wave fronts do not simply spiral inward.  For nulls high in the corona where the scale height is large, and the equipartition contour is small, the approximation is more appropriate.  On the other hand, based on the statistics reviewed in \S\ref{sec:null-presence} we expect most nulls on the Sun to arise in locations where stratification is important.}

\change{From the discussion in this section, it seems that there are two important length scales.  (i) The first is global, and is caused by local extrema in the wave phase speeds.  This will determine the spatial portion of an arbitrary wave that will refract in towards a null.  (ii) The second is the size of the equipartition region around the null relative to the spatial extent of the wavefront that refracts towards the null.  The second scale will affect mode conversion, current accumulation, and dissipation at the null.   Combining our results with those of \citet{McLaughlin:2006b} shows that the combination of these two lengths will determine how much wave energy refracts towards a null, and how much of that mode converts at the null, for a given driver.  They can be combined to define an overall ``region of influence'' for nulls.}

\subsection{Wave energy guided along separatrices}
In our simulation, we found that slow mode energy leaving the null is concentrated near each separatrix field line.  \figref{fig:flux-null}(b) shows that $\sim30\%$ of the slow mode energy leaving the null travels upward, so the remaining $70\%$ propagates back to the photosphere and becomes increasingly concentrated around each of the other three separatrices.  This is a focusing effect due to the converging magnetic field which acts as a waveguide for the slow waves.  In total, we found that $8\%$ of the initial upward propagating wave packet becomes concentrated in the three downward propagating patches.  This study therefore shows that mode conversion near the null can take a distributed upward wave flux and create localized patches of downward wave flux, focused specifically on the separatrix foot points.

A similar effect has been reported by \citet{Russell:2013} for Alfv\'en waves in a zero $\beta$ plasma.  They introduced a transverse velocity perturbation at the apex of a model arcade in order to mimic a reconnection event.  The resulting pure Alfv\'en waves focus as they propagate downwards, mostly due to the converging field lines, although phase mixing also plays a role.  In contrast, when they introduced a fast wave above an arcade, they found that the energy density became more diffuse as it propagated towards their lower boundary.

In our case, the fast waves that convert near the null are not necessarily generated by the convective processes that we modeled.  For instance, a reconnection induced fast wave will be refracted towards any nulls in the chromosphere or corona.  Near the nulls, fast mode energy will partially convert to a set of slow modes, each of which will focus on null's separatrix foot points.  \change{The effect is especially important in the partially ionized chromosphere, which modifies the slow mode propagation and introduces frequency dependent damping \citep{Soler:2013}.  Focused Alfv\'en and slow waves could therefore contribute to the enhanced emission seen in flare ribbons.  This has been argued before \citep{Nakariakov:2011}, and our simulations at least show how focused wave energy could arrive at flare foot points.  The topic seems to deserve further consideration.}

\subsection{Comparison with simulations of stratified atmospheres}
We chose to isolate and analyze the dynamics of a single wave pulse.  This approach makes direct comparison with most other studies on atmospheric waves difficult, as they typically rely on time averaged properties of long lasting wave trains or stochastic fluctuations \citep{Fedun:2011a, Santamaria:2015, Rijs:2016}.  Some more recent efforts have included either short or instantaneous pulses \citep{Khomenko:2009b, Nutto:2012, Santamaria:2015, Rijs:2015, Moradi:2015}, but were analyzed by Fourier transform of, say, the velocity signal, which is still a time--averaged approach.  An exception is \citet{Shelyag:2016}, whose primary focus was dissipation due to ambipolar diffusion in a small, 3D flux tube.  They introduced a similar perturbation to ours and found that it converts to fast and slow waves, as well as Alfv\'enic waves due to the 3D geometry.  They found strong heating due to dissipation of currents associated with the Alfv\'enic waves, but did not attempt to quantify the conversion process.

In a complementary simulation to our own, \citet{Santamaria:2015} studied low frequency $(3.3-5\unit{mHz})$ waves in a stratified atmosphere spanning the upper convection zone to low corona which also included a null point.  They considered both horizontally and vertically driven waves.  For the vertically driven case they found little time--averaged Poynting flux above the height of the transition region, but did find significant low--$\beta$ acoustic flux confined to nearly vertical flux tubes at the edge of their domain.  That mechanism is distinct from the one we describe, where a fast mode propagates towards a null, converts at the equipartition region, and exits the null point as a slow acoustic wave confined to the separatrix field lines.  However, in their $\S3.3.1$ they appear to mention the mode conversion process near the null that we describe in much more detail.  In their case, conversion arises due to horizontal driving at the lower boundary.  Yet, by the time the wave reaches the photospheric level it contains substantial vertical oscillations in the regions of strongly inclined (near horizontal) field and therefore produces a similar type of driving to what we have implemented.  Indeed, one can see that some acoustic power is localized near the separatrices in their Figure 9, which would indicate the presence of slow mode waves generated by conversion near the null.  Their simulations therefore suggest that our results will carry over to lower frequency waves, closer to the Br\"unt--V\"ais\"al\"a frequency, but we caution that at those frequencies dispersion will play a strong role that we did not consider in our work.

\subsection{Comparison with 3D studies}
A limitation of the present study is the 2D geometry.  A 3D simulation will allow for the coupling of the Alfv\'en mode to the fast and slow modes discussed here, as \citet{Shelyag:2016} found.  \citet{Cally:2008} and \citet{Felipe:2012} demonstrated that the Alfv\'en mode coupling will show a second strong dependence on the relative orientation of the wave vector and the magnetic and gravitational fields.  These studies used slowly varying (or simply uniform) sunspot fields with no strong gradients.  

\change{For null point studies, \citet{McLaughlin:2008} performed a WKB analysis for a linear 3D null and found that fast wave energy accumulates at the null as a current density.  \citet{Thurgood:2012} performed a 3D numerical simulation and identified a coupling between the Alfv\'en and fast mode, though they did not attempt to quantify the amount of conversion.  Both of these 3D null point studies had $\beta=0$, so that the only form energy accumulation could take is an increase in current density as the wave contracts and the gradient across the wave front increases.  \citet{Pontin:2013} performed a $\beta\neq 0$, 3D simulation (see next section) and found that currents localized to a null and separatrix surface, but did not describe the dynamics in terms of waves.}

Determining the proportions of energy in each of the three modes is nontrivial in 3D.  Our 2D geometry allowed us to identify with relative ease the fast and slow modes based on their energy densities and propagation with respect to the field.  Distinguishing between the fast and Alfv\'en mode will require more care, and, we anticipate, either a greater reliance on the space--time diagrams like \figref{fig:enrays} or building up a statistical version of the local dispersion relation by studying oscillations of local parameters.  This latter approach has been used to analyze observational data \citep{Tomczyk:2009} but could be applied in three dimensions to simulation output.  A third approach is to project the energy density and flux terms onto characteristic directions of the magnetic field as \citet{Felipe:2012} has done for the 3D sunspot simulation, \change{\citet{Thurgood:2012} did for a 3D numerical simulation,} and \citet{McLaughlin:2016} did for a 3D null in a WKB ray--tracing investigation.  \change{\citet{Mumford:2015} had some success using this method to determine the energy flux associated with several modes of a 3D expanding flux tube, but the presence of multiple propagation velocities at a given height in their phase diagrams (see, e.g., their Figure 7(a)), suggests that the decomposition is not exact.}  It is clear that no single method can currently robustly distinguish between modes in 3D.  The best approach is probably to combine or compare multiple types of analysis, as we have done in two dimensions in the present investigation.

\subsection{Currents and reconnection}
In agreement with other studies, our simulations show that currents accumulate at topologically important locations: the null and the separatrices.  \citet{Pontin:2013} considered a 3D magnetic dome topology, similar to what would result if our initial magnetic field was spun around the axis passing through the null.  In that case, the two separatrices leaving our 2D null in the vertical direction become 3D spine lines, while the other two separatrices form a single dome--like separatrix surface.  \citet{Pontin:2013}'s initial condition had a uniform density and temperature, as opposed to our stratified atmosphere.  They applied an incompressible shear flow around the spine foot point inside the dome and observed an accumulation of current at the separatrix and around the null (see their Figure 9).  Other work on 3D nulls has also found that incompressible perturbations cause current accumulation \citep{Pontin:2007}, where the details depend on whether the area around the spine or the fan is perturbed, and if the perturbation is of rotational or shear type.  

\change{Our simulations show that current accumulation also occurs for compressive waves, even when $\beta>0$ and waves can travel through the null.  This situation is not well studied.  \citet{McLaughlin:2009} provide one of the most detailed analyses to date, but set $\beta=0$ in their initial condition.  They rely on current accumulation, shock formation, and the resulting Ohmic and shock dissipation to heat the plasma and raise $\beta$ above zero.  In their case, asymmetry in the heating about the null creates a pressure gradient which must be balanced by a Lorentz force, hence the persistence of the current density at the null after the wave has passed.  Mode conversion may play a role in their simulation, though they do not consider it in their paper.  \citet{McLaughlin:2006b} (discussed at length above) found that most current accumulates close to a linear null, but their simulation solved the linearized MHD equations, whereas the nonlinear effects are clearly important.  \citet{Afanasyev:2012} demonstrated that nonlinear effects substantially alter wave behavior near the null and allow fast waves to pass even for the $\beta=0$ case.  They studied both $\beta=0$ and finite $\beta$ cases, and used a modified WKB method that accounts for weak shocks.  Their method does not apply when one shock is downstream of another, e.g., when rays cross, which prevented them from studying current sheet formation.  However, we demonstrated in Figures \ref{fig:ray-paths}--\ref{fig:ray-ca} that this occurs frequently.  \citet{Gruszecki:2011} also studied the nonlinear evolution of a fast wave near a linear null and for an initially $\beta=0$ plasma.  They noted the shock formation and strong cospatial spikes in the current density, but stopped their simulations before the pulse reached the null in order to stay in the low $\beta$ regime.  They were thus unable to note if any current persisted at the null or along its separatrices.  So it is clear that more work on how currents localize to nulls, particularly for finite $\beta$, is necessary.}

\change{We found that the current sheet which forms at the null oscillates between $\pm45^\circ$, which is evidence for oscillatory reconnection (see \figref{fig:current-time}).  This finding is also not new \citep{Craig:1991}, but has recently received renewed attention, especially in connection with observations of quasi--periodic pulsations associated with flares \citep{vanDoorsselaere:2016}.  Oscillatory reconnection is found in the null point studies \citep{McLaughlin:2009, McLaughlin:2012a}, in simulations of flux emergence in a stratified atmosphere \citep{Murray:2009, McLaughlin:2012b}, and now in our simulation modeling convection--induced waves.  In short, oscillatory reconnection appears to be a rather robust feature of reconnection, at least in 2D, and it remains to determine its importance in 3D.}

\section{Conclusion}\label{sec:conclusion}
We have studied the propagation of an initially acoustic wavepacket through a stratified solar--type atmosphere with an inhomogeneous magnetic field.  We were able to quantify the energy to reach the null ($.155E_{input}$), mode conversion around a null ($0.12E_{input}$), and dissipation at the null and along separatrix field lines ($0.04E_{input}$), in terms of the initial energy of the wavepacket, $E_{input}$.  Some wave energy was able to escape into the corona $(0.11E_{input})$.  Most of the escaping fast mode energy ($\sim0.067E_{input}$) was due to refraction around the null, while escaping slow mode energy ($0.033E_{input}$) was due only to mode conversion at the null; the remainder of the escaped energy was due to fast mode energy leaving the null.

Several details and extensions of the present work have not been fully reported here, but are in preparation.  These include a discussion of the current accumulation at the null and along the separatrices (including substantially increasing the numerical resolution and thereby changing the effective Lundquist number), varying the initial location of the wavepacket, changing the height of the transition region and null point relative to each other, and varying the central frequency of the driver.  Finally, including the effect of partial ionization, already implemented for the LARE code as described in \citet{Leake:2013}, will bring the simulations into much closer agreement with the environment low in the solar atmosphere.

\appendix

\section{WKB}\label{sec:wkb}
WKB solutions provide insight into the very complex dynamics apparent in the numerical solutions.  The methods have been broadly applied and much discussed, so we will only give a brief description for reference, here.  We will justify the terms used in the dispersion relation Equation \eqref{eq:dispersion-relation} and write out the equations that were solved to produce the rays used throughout this work.  The general WKB theory and the derivation of the ray equations can be found in various forms in \citet{Lighthill:1978,Weinberg:1962,Stix:1992, Kulsrud:2005}.  \change{Analytic solutions to the ray equations can be found in \citet{Hansen:2009} and \citet{McLaughlin:2008}: the former is representative of the helioseismic literature and is applied to an isothermally stratified atmosphere with uniform, arbitrarily directed magnetic field; the latter is representative of the null point literature and is applied to a 3D linear magnetic null in a cold plasma.}

The wave equation \eqref{eq:wave-full} describes the evolution of a perturbation to the system.  Note that \eqref{eq:wave-full}, and the following discussion, are valid in 3D.  We will specialize to 2D at the end of this appendix.  As in the main text, we here assume the perturbation is of the form $\vect{v}_1=\vect{a}e^{i\Phi/\lambda}$, where now we have introduced a small parameter $\lambda$ that will shortly be absorbed into $\Phi$.  Taking the derivatives, we get
\begin{gather}
  -\frac{1}{\lambda^2}\vect{a}(\partial_t\Phi)^2 + \frac{1}{\lambda^2}c_s^2(\vect{a}\cdot\del\Phi)\del\Phi + \frac{1}{\lambda^2}v_A^2\Bigl[\hat{\vect{b}}\times(\del\Phi\times\hat{\vect{b}})(\vect{a}\cdot\del\Phi)-\hat{\vect{v}}\times(\del\Phi\times\vect{a})(\hat{\vect{v}}\cdot\del\Phi)\Bigr] + \mathcal{O}(\frac{1}{\lambda}) = \vect{0}.
\end{gather}
With $\lambda\ll 1$, the $1/\lambda^2$ terms dominate, and this justifies dropping the terms that explicitly involve gravity and derivatives of the background field from Equation \eqref{eq:wave-full}, which are $\mathcal{O}(\lambda^{-1})$ or $\mathcal{O}(1)$.  We now absorb $\lambda$ into the definition of $\Phi$.  Letting $\vect{k}\equiv\del\Phi$ and $\omega\equiv-\partial_t\Phi$, so that $\Phi = \vect{k}\cdot\vect{x}-\omega t$, the above equation can be rewritten in the form given by Equation \eqref{eq:dyad}, which is a set of coupled equations for the phase function $\Phi$.  It has nontrivial solution when the determinant of the coefficients is zero, given by Equation \eqref{eq:dispersion-relation}.  We rewrite the dispersion relation here in a more general form, allowing it to have explicit dependence on $t$ and $\Phi$:
\begin{equation}
  \mathcal{D}(\vect{x},t,\Phi,\vect{k},\omega)=0.
\end{equation}
This equation is solved by finding a characteristic curve $\vect{x}(\tau)$ and $t(\tau)$ which satisfies the four equations
\begin{gather}
  \label{eq:disp}\frac{d\vect{x}}{d\tau} = \frac{\partial \mathcal{D}}{\partial\vect{k}};\quad \frac{d t}{d\tau} = -\frac{\partial \mathcal{D}}{\partial\omega}.
\end{gather}
We want to determine how $\vect{k}$ and $\omega$ vary along this curve.  The method is to rewrite the derivatives of $\mathcal{D}$ along the characteristic curve.  The total derivatives of $\mathcal{D}$ are
\begin{gather}
  \frac{d\mathcal{D}}{d\vect{x}} = \frac{\partial \mathcal{D}}{\partial\vect{x}}+\frac{\partial \mathcal{D}}{\partial\Phi}\frac{\partial \Phi}{\partial\vect{x}}+\frac{\partial \mathcal{D}}{\partial\vect{k}}\cdot\frac{\partial \vect{k}}{\partial\vect{x}} + \frac{\partial \mathcal{D}}{\partial\omega}\frac{\partial \omega}{\partial\vect{x}} = 0\\
  \frac{d\mathcal{D}}{dt} = \frac{\partial \mathcal{D}}{\partial t}+\frac{\partial \mathcal{D}}{\partial\Phi}\frac{\partial \Phi}{\partial t}+\frac{\partial \mathcal{D}}{\partial\vect{k}}\frac{\partial \vect{k}}{\partial t} + \frac{\partial \mathcal{D}}{\partial\omega}\frac{\partial \omega}{\partial t} = 0.
\end{gather}
The dot product on the first line is between the two $\vect{k}$s, one in the denominator and one in the numerator.  But, $\vect{k}$ is a gradient so its curl is zero, $\frac{\partial k_i}{\partial x_j} = \frac{\partial k_j}{\partial x_i}$, and this allows us to rewrite the dot product as a contraction between $\vect{k}$ and $\vect{x}$, both in the denominator.  The space and time derivatives of $\Phi$ are just $\vect{k}$ and $-\omega$.  We substitute these expressions and rewrite the above equations as
\begin{gather}
  \frac{\partial \mathcal{D}}{\partial\vect{k}}\cdot\frac{\partial \vect{k}}{\partial\vect{x}} + \frac{\partial \mathcal{D}}{\partial\omega}\frac{\partial \omega}{\partial\vect{x}}=-\frac{\partial \mathcal{D}}{\partial \vect{x}} - \vect{k}\frac{\partial \mathcal{D}}{\partial\Phi}\\
  \frac{\partial \mathcal{D}}{\partial\vect{k}}\cdot\frac{\partial \vect{k}}{\partial t} + \frac{\partial \mathcal{D}}{\partial\omega}\frac{\partial \omega}{\partial t}=-\frac{\partial \mathcal{D}}{\partial t} + \omega\frac{\partial \mathcal{D}}{\partial\Phi}
\end{gather}
Next we substitute the definitions for the characteristic curve, Equation \eqref{eq:disp}, into the left hand side (LHS):
\begin{gather}
  \frac{d\vect{x}}{d\tau}\cdot\frac{\partial \vect{k}}{\partial\vect{x}}-\frac{dt}{d\tau}\frac{\partial \omega}{\partial\vect{x}} = -\frac{\partial \mathcal{D}}{\partial \vect{x}} - \vect{k}\frac{\partial \mathcal{D}}{\partial\Phi}\\
  \frac{d\vect{x}}{d\tau}\cdot\frac{\partial \vect{k}}{\partial t}-\frac{d t}{d\tau}\frac{\partial \omega}{\partial t} = -\frac{\partial \mathcal{D}}{\partial t} + \omega\frac{\partial \mathcal{D}}{\partial\Phi}
\end{gather}
The mixed second order derivatives of $\Phi$ must be equal: $\partial_\vect{x}\omega = \partial_\vect{x}(\-\partial_t\Phi) = -\partial_t(\partial_\vect{x}\Phi) = -\partial_t\vect{k}$.  The LHS of each equation is therefore a total derivative along the curve:
\begin{gather}
  \label{eq:ray-k}\frac{d\vect{k}}{d\tau} = -\frac{\partial\mathcal{D}}{\partial\vect{x}} - \vect{k}\frac{\partial\mathcal{D}}{\partial\Phi}\\
  \label{eq:ray-w}\frac{d\omega}{d\tau} = \frac{\partial\mathcal{D}}{\partial t} - \omega\frac{\partial\mathcal{D}}{\partial\Phi}.
\end{gather}
Equations \eqref{eq:disp}, \eqref{eq:ray-k}, and \eqref{eq:ray-w} fully describe the solution given an initial $\omega$ and $\vect{k}$ that satisfy the dispersion relation at location $\vect{x}$.  Our dispersion relation, Equation \eqref{eq:dispersion-relation}, has no explicit $\Phi$ or $t$ dependence, so those terms on the right hand side (RHS) of the equations are zero.  In particular, $d\omega/d\tau = 0$, and the frequency is constant along the ray.  Further, the second equation of \eqref{eq:disp} allows us to use the physical time $t$ as the parameter along the curve.  The result for $\vect{x}$ and $\vect{k}$ are
\begin{gather}
  \label{eq:hamilr}\frac{d\vect{x}}{d t} = - \frac{\partial_\vect{k}\mathcal{D}}{\partial_\omega\mathcal{D}}=\partial_\vect{k}\omega = \vect{v}_g\\
  \label{eq:hamilk_a}\frac{d\vect{k}}{d t} = \frac{\partial_\vect{x}\mathcal{D}}{\partial_\omega\mathcal{D}} = -\partial_\vect{x}\omega.
\end{gather}
The above equations are the ray equations, and demonstrate that $\vect{k}$ and $\vect{x}$, constructed along the rays, obey Hamilton's equations \citep{Goldstein:2002}, with $\omega$ taking the place of the Hamiltonian, $\vect{x}$ the generalized coordinates, and $\vect{k}$ their conjugate momentum densities.  

We now turn to our specific case of the dispersion relation for the fast and slow rays in 2D, Equation \eqref{eq:dispersion-relation}, which has a solution when
\begin{gather}
  \label{eq:vphase}\omega_\pm = \vert k\vert \Bigl[ \frac{1}{2} A\pm\frac{1}{2}\sqrt{A^2-4B\cos^2\theta}\Bigr]^{1/2} = \vert k\vert v_{\phi\pm},
\end{gather}
where $v_{\phi\pm}$ is the phase velocity.  $A = c_s^2+v_A^2$ and $B = c_s^2v_A^2$ depend only on space, while the ray direction $\theta$ depends on $\vect{k}$ through $\vert k\vert\cos\theta = \vect{k}\cdot\hat{\vect{b}}$.  Applying \eqref{eq:hamilk_a} to the dispersion relation gives
\begin{gather}
  \label{eq:hamilk}\frac{d\vect{k}}{dt} = -\vert k\vert \del v_{\phi\pm}
\end{gather}
The group velocity is found through the derivatives:
\begin{gather}
  \frac{\partial\omega_\pm}{\partial k_i}\hat{\vect{x}}_i = \frac{k_i}{\vert k\vert}v_{\phi\pm}\hat{\vect{x}}_i + \vert k\vert\frac{1}{2}\frac{1}{v_{\phi\pm}}\Bigl(\pm\frac{1}{2}\Bigr)\Bigl(\frac{1}{2}\Bigr)\Bigl[A^2 - 4B \cos^2\theta\Bigr]^{-\frac{1}{2}}(-4B)\Bigr(2\cos\theta\frac{\partial \cos\theta}{\partial k_i}\Bigl)\hat{\vect{x}}_i,\\
  \intertext{where the cosine term is}
  \frac{\partial \cos\theta}{\partial \vect{k}} = \frac{d}{d\vect{k}}\frac{\vect{k}\cdot\hat{\vect{b}}}{\sqrt{k^2}} = -\frac{\vect{k}}{\vert k\vert^3}(\vect{k}\cdot\hat{\vect{b}})+\frac{1}{\vert k\vert}\hat{\vect{b}} = \frac{1}{\vert k\vert}\bigl[\hat{\vect{b}}-\hat{\vect{k}}\cos\theta\bigr].
  \intertext{After substitution for $A$ and $B$, we write the group velocity as}
  \label{eq:vgroup}\vect{v}_{g\pm} = \frac{\partial\omega}{\partial\vect{k}} = v_{\phi\pm}\hat{\vect{k}}\mp\frac{c_s^2v_A^2\cos\theta}{v_{\phi\pm}\sqrt{(c_s^2+v_A^2)^2-4c_s^2v_A^2\cos^2\theta}}\bigl[\hat{\vect{b}}-\hat{\vect{k}}\cos\theta\bigr].
\end{gather}
Equivalently, we may determine the group velocity from the Jacobian:
\begin{gather}
  \vect{v}_{g\pm} = \frac{\partial\omega_\pm}{\partial\vect{k}} = \frac{\partial\omega_\pm}{\partial k}\hat{\vect{k}} + \frac{1}{k}\frac{\partial\omega_\pm}{\partial \theta}\hat{\pmb \theta}.
\end{gather}
With $\cos\theta\sin\theta\hat{\pmb \theta} = \cos\theta(\hat{\vect{b}} - \cos\theta\hat{\vect{k}})$, we see that the above equation is equivalent to \eqref{eq:vgroup}.

For each ray with initial values $\vect{x}_0$ and $\chi_0$, we solve the set of equations \eqref{eq:hamilr} and \eqref{eq:hamilk} for $\vect{x}(\chi_0,\vect{x}_0,\tau)$ and $\vect{k}(\chi_0,\vect{x}_0,\tau)$, together with the expressions for the phase \eqref{eq:vphase} and group \eqref{eq:vgroup} velocities, by numerical integration with an explicit update.  Note, for a particular ray we typically suppress the $\vect{x}_0$ and $\chi_0$ notation.  Values for each relevant quantity ($c_s, v_A, v_\phi,\vect{B},\del v_\phi$) are found at each LARE2D grid point and linearly interpolated to the ray point $\vect{x}(\tau)$.

\section{Wave energy and flux densities}\label{sec:waveflux}
Several authors have cited \citet{Bray:1974} for the derivation of conservation of wave energy density found at Equation \eqref{eq:wave-conservation}.  Their derivation results from taking the linearized equations, multiplying each by a first order quantity, and summing to obtain the desired conservation law.  However, in that case it is possible that some second order quantities have already been abandoned that may be important.  We show below that the expression may also be derived starting with the expression for total energy conservation and keeping all terms up to $\mathcal{O}(2)$ in perturbed quantities; c.f. \citet{Leroy:1985}, who discusses possible pitfalls for numerous ways of deriving a wave--energy conservation relation, and \citet{Bogdan:2003} \S4, who compares the total energy flux to wave energy flux determined in a simulation.

Total energy conservation for a system like the solar surface where plasma motions do not affect the gravitational field (labeled in this Appendix by a constant, uniform $\phi$) may be expressed as \citep{Kulsrud:2005}
\begin{gather}
  \label{eq:totalE}
  \partial_t\Bigl[\rho\epsilon + \frac{1}{2}\rho\abs{\vect{v}}^2+\frac{1}{2\mu_0}\abs{\vect{B}}^2+\rho\phi \Bigr] 
  + \del\cdot\Bigl[\rho\epsilon\vect{v} + \frac{1}{2}\rho\abs{\vect{v}}^2\vect{v}+P\vect{v}+\frac{\vect{B}\times(\vect{v}\times\vect{B})}{\mu_0}+\rho\phi\vect{v}\Bigr] = 0,
\end{gather}
where $\phi$ is the gravitational potential, $\vect{g}=-\del\phi$.  We begin by following \S65 of \citet{Landau:1987}.  We consider an adiabatic perturbation, so that $P$ is a function of $\rho$; through \eqref{eq:ideal}, $\epsilon$ is also a function of $\rho$.  We assume a stationary background with no flows, so that $\partial_t=0$ for all quantities, and $\vect{v}_0 = \vect{0}$.  First, we expand the internal energy to second order, at constant entropy
\begin{gather}
  \rho\epsilon = \rho\epsilon\Bigr\rvert_0 + \frac{\partial(\rho\epsilon)}{\partial\rho}\Bigr\rvert_0\rho_1 + \frac{1}{2}\frac{\partial^2(\rho\epsilon)}{\partial\rho^2}\Bigr\rvert_0(\rho_1)^2+\ldots
\end{gather}
\change{We have $\partial (\rho\epsilon)/\partial\rho = \epsilon + \rho\partial(\epsilon)/\partial\rho$.  The first law of thermodynamics in terms of specific energy $\epsilon$, specific entropy $s$, and specific volume $\zeta$ is $d\epsilon = Tds - pd\zeta$.  The specific volume is just $1/\rho$, so $d\zeta = -1/\rho^2 d\rho$, and $ds=0$ for an adiabatic process.  The derivative of $\epsilon$ with respect to $\rho$ is therefore $d\epsilon/d\rho = P/\rho^2$.  Thus, $\partial(\rho\epsilon)/\partial\rho = \epsilon + P/\rho$, and $\partial^2(\rho\epsilon)/\partial\rho^2 = 1/\rho\times(\partial P/\partial\rho)$, so that to second order}
\begin{gather}
\rho\epsilon= \rho_0\epsilon_0 + \frac{c_s^2}{\gamma-1}\rho_1 + \frac{1}{2}\frac{c_s^2}{\rho_0}\rho_1^2.
\end{gather}
The other terms inside the time derivative in Equation \eqref{eq:totalE} have straightforward expansions, and altogether, up to second order, are
\begin{equation}
  \label{eq:w2}
  \partial_t\Bigl[\frac{c_s^2}{\gamma-1}\rho_1 + \frac{c_s^2}{2\rho_0}\rho_1^2 + \frac{1}{2}\rho_0\abs{v_1}^2+\frac{\vect{B}_0\cdot\vect{b}}{\mu_0}+\frac{\vect{b}\cdot\vect{b}}{2\mu_0} + \rho_1\phi \Bigr].
\end{equation}
We expand the divergence term in the same way, keeping terms to second order:
\begin{equation}
  \Bigl[\rho_0\epsilon_0 + \frac{c_s^2}{\gamma-1}\rho_1\Bigr]\vect{v}_1 + P_0\vect{v}_1+P_1\vect{v}_1 + \frac{1}{\mu_0}(\vect{B}_0+\vect{b})\times\bigl[ \vect{v}_1\times (\vect{B}_0+\vect{b})\bigr]+\rho_0\phi\vect{v}_1+\rho_1\phi\vect{v}_1.
\end{equation}
Combining the first and third terms in the above expression through $\rho_0\epsilon_0 + P_0 = \gamma P_0/(\gamma-1)$, and dropping the $\vect{b}\times\vect{v}_1\times\vect{b}$ term because it is \bigo(3), we find that the full conservation equation up to second order is
\begin{multline}
  \label{eq:wave-cont-a}
  \partial_t\Bigl[\frac{c_s^2}{\gamma-1}\rho_1 + \frac{c_s^2}{2\rho_0}\rho_1^2 + \frac{1}{2}\rho_0\abs{v_1}^2+\frac{\vect{B}_0\cdot\vect{b}}{\mu_0}+\frac{\vect{b}\cdot\vect{b}}{2\mu_0}+\rho_1\phi\Bigr]\\
\shoveright{+\del\cdot\Bigl[
  \frac{\gamma P_0}{\gamma-1}\vect{v}_1+\frac{c_s^2\rho_1}{\gamma -1}\vect{v}_1 +P_1\vect{v}_1+\frac{\vect{B}_0\times(\vect{v}_1\times\vect{B}_0)}{\mu_0}+\frac{\vect{B}_0\times(\vect{v}_1\times\vect{b})}{\mu_0}+\frac{\vect{b}_0\times(\vect{v}_1\times\vect{B}_0)}{\mu_0} + \rho_0\phi\vect{v}_1+\rho_1\phi\vect{v}_1\Bigr] = 0.
}
\end{multline}
There are several first order quantities in the above expression.  The first terms under the time derivative and divergence cancel.  To see this, multiply the continuity equation by $c_s^2/(\gamma-1)$, and integrate by parts:
\begin{gather}
  \frac{c_s^2}{\gamma-1}\partial_t(\rho_0+\rho_1)+\frac{c_s^2}{\gamma-1}\del\cdot[(\rho_0+\rho_1)\vect{v}_1] = 0 \\
 \Rightarrow \partial_t\Bigl[\frac{c_s^2\rho_1}{\gamma-1}\Bigr]+\del\cdot\Bigl[\frac{c_s^2\rho_0}{\gamma-1}\vect{v}_1+\frac{c_s^2\rho_1}{\gamma-1}\vect{v_1}\Bigr]-\frac{\rho_0+\rho_1}{\gamma-1}\vect{v}_1\cdot\del c_s^2 = 0.
\end{gather}
Subtract this from Equation\eqref{eq:wave-cont-a}, and, noting that $c_s^2\rho_0=\gamma P_0$, we get:
\begin{multline}
  \label{eq:wave-cont-b}\partial_t\Bigl[\frac{c_s^2}{2\rho_0}\rho_1^2 + \frac{1}{2}\rho_0\abs{v_1}^2+\frac{\vect{B}_0\cdot\vect{b}}{\mu_0}+\frac{\vect{b}\cdot\vect{b}}{2\mu_0}+\rho_1\phi\Bigr]\\
  \shoveright{+\del\cdot\Bigl[
      P_1\vect{v}_1+\frac{\vect{B}_0\times(\vect{v}_1\times\vect{B}_0)}{\mu_0}+\frac{\vect{B}_0\times(\vect{v}_1\times\vect{b})}{\mu_0}+\frac{\vect{b}\times(\vect{v}_1\times\vect{B}_0)}{\mu_0} + \rho_0\phi\vect{v}_1+\rho_1\phi\vect{v}_1\Bigr]
  }\\
  \hfill\shoveright{
    +\frac{(\rho_0 +\rho_1)\vect{v}_1}{\gamma-1}\cdot\del c_s^2  = 0.
  }
\end{multline}
Next, we are assuming the gravitational field is unaffected by the plasma motions.  This allows us to write $\del\cdot(\rho\phi\vect{v}_1) = \rho\vect{v}_1\cdot\del\phi + \rho\del\cdot(\rho\vect{v}_1) = \rho\vect{v}_1\cdot\del\phi - \partial_t(\rho\phi)$ after using the continuity equation.  The last term here cancels the gravitational term under the time derivative of \eqref{eq:wave-cont-b}.  We have
\begin{multline}
  \partial_t\Bigl[\frac{c_s^2}{2\rho_0}\rho_1^2 + \frac{1}{2}\rho_0\abs{v_1}^2+\frac{\vect{B}_0\cdot\vect{b}}{\mu_0}+\frac{\vect{b}\cdot\vect{b}}{2\mu_0}\Bigr]
 +\del\cdot\Bigl[P_1\vect{v}_1+\frac{\vect{B}_0\times(\vect{v}_1\times\vect{B}_0)}{\mu_0}+\frac{\vect{B}_0\times(\vect{v}_1\times\vect{b})}{\mu_0}+\frac{\vect{b}\times(\vect{v}_1\times\vect{B}_0)}{\mu_0}\Bigr]
  \\
  \label{eq:wave-cont-c}\hfill\shoveright{
    +(\rho_0 +\rho_1)\vect{v}_1\cdot\del\Bigl[\phi +  \frac{c_s^2}{\gamma-1}\Bigr] = 0.
  }
\end{multline}
\change{To rewrite the gravitational term, we note that the derivative of the sound speed is 
\begin{equation}
\del c_s^2 = \frac{\gamma}{\rho_0}\del P_0 - \frac{\gamma P_0}{\rho_0^2}\del\rho_0 = \gamma\vect{g}-\frac{\gamma P_0}{\rho_0^2}\Bigl(-\frac{\vect{g}}{g}\partial_y\rho_0\Bigr)
\end{equation} 
after using the hydrostatic equation and the fact that $\rho_0$ varies only in the direction of gravity.  Next, we substitute for $\del\phi=-\vect{g}$ and combine terms to write the gradient from line two of \eqref{eq:wave-cont-c} as
\begin{equation}
\del[\phi+c_s^2/(\gamma-1)] = \vect{g}\Bigl[\frac{1}{\gamma-1}+\frac{c_s^2}{g}\partial_y\ln\rho_0\Bigr].
\end{equation}
Finally, we pull out a factor $c_s^2/(\gamma-1)g$ and recombine $\rho_0+\rho_1=\rho$ to write the full gravitational term from \eqref{eq:wave-cont-c} as}
\begin{equation}
  \frac{c_s^2}{(\gamma-1)g}\rho\vect{v}_1\cdot\vect{g}\Bigl[\frac{g}{c_s^2}+\partial_y\ln\rho_0\Bigr].
\end{equation}
The term in brackets is $-N^2/g$, where $N$ is the Br\"unt--V\"ais\"al\"a frequency, and is related to buoyant oscillations of the plasma.  

To deal with the magnetic terms under the divergence, we take the scalar product of the induction equation with $\vect{B}_0/\mu_0$,
\begin{gather}
  \label{eq:induction-a}\frac{\vect{B}_0}{\mu_0}\cdot \partial_t(\vect{B}_0+\vect{b}) - \frac{\vect{B}_0}{\mu_0} \cdot \del\times\Bigl[\vect{v}_1\times(\vect{B}_0+\vect{b})\Bigr] = 0.
  \intertext{Using $\del\cdot\Bigl[\vect{B}_0\times(\vect{v}_1\times\vect{B})\Bigr] = (\vect{v}_1\times\vect{B})\cdot\del\times\vect{B}_0-\vect{B}_0\times(\vect{v}\times\vect{B})$ to rewrite the triple product and then integrating by parts, we find}
  \label{eq:induction-b}\partial_t\Bigl(\frac{\vect{B}_0\cdot\vect{b}}{\mu_0}\Bigr) + \del\cdot\Bigl[
    \frac{\vect{B}_0}{\mu_0}\times(\vect{v}_1\times\vect{B_0})+\frac{\vect{B}_0}{\mu_0}\times(\vect{v}_1\times\vect{b})\Bigr] - (\vect{v}_1\times\vect{B})\cdot\cancelto{0}{\del\times\vect{B}_0} = 0,
\end{gather}
where in the last term we now make the assumption that our background state is curl free.  Subtract \eqref{eq:induction-b} from \eqref{eq:wave-cont-c} to obtain
\begin{equation}
  \label{eq:wave-cont-d}\frac{\partial}{\partial t}\biggl[\frac{c_s^2}{2\rho_0}\rho_1^2 + \frac{1}{2}\rho_0\abs{\vect{v}_1}^2 + \frac{\abs{\vect{b}}^2}{2\mu_0}\biggr] + \del\cdot\biggl[P_1\vect{v}_1 + \frac{\vect{b}\times(\vect{v}_1\times\vect{B}_0)}{\mu_0}\biggr] +  \frac{c_s^2}{(\gamma-1)g}\rho\vect{v}_1\cdot\vect{g}\Bigl[\frac{g}{c_s^2}+\partial_y\ln\rho_0\Bigr]= 0.
\end{equation}
Equation \eqref{eq:wave-cont-d} agrees with the second order conservation relation from \citet{Bray:1974}, but comes directly from the full equations expanded to second order.  The final term results from the stratification, and gives rise the gravitational term in \eqref{eq:wave-conservation} after writing $\vect{v}_1 = \partial_t\vect{X}$, where $\vect{X}=(X,Y)$ is the Lagrangian displacement of the fluid, and $\rho$ in terms of the displacement field, and then integrating by parts.  The gravitational terms are important for low frequencies, around and below the V\"ais\"al\"a frequency of $\approx 4.5\unit{mHz}$ in the solar atmosphere.  The derivation above requires a potential field, $\del\times\vect{B}_0 = \vect{0}$.  Bray and Loughhead assumed a uniform field, but this is actually not required by their argument, and a potential field is acceptable for their derivation, as well.  As a final note, dividing the flux term by the energy density term returns an expression for the group velocity of a wave, $\vect{v}_g$, which may be compared to the group velocity, for instance, from Appendix \ref{sec:wkb}.  This is most easily checked case by case for a specific wave with appropriate simplifications (e.g., a shear Alfv\'en wave, an acoustic wave in an isothermally stratified atmosphere, etc.).  

\acknowledgements
This work is supported by the the Chief of Naval Research.  JEL was supported by NASA’s HSR Program.  The simulations were performed under a grant of computer time from the DoD HPC program.  This research has made use of NASA's Astrophysics Data System.  
\software{LARE2D v.2.11 \citep{Arber:2001}}

\end{document}